\documentclass[english,numreferences]{kluwer}
\usepackage{babel}
\usepackage{amsmath}
\usepackage{latexsym}
\usepackage{epsfig}
\usepackage{theorem}
\usepackage{ifthen}

\newboolean{WITHAPPENDIX}
\setboolean{WITHAPPENDIX}{true}

\newboolean{PROOFSONLY}
\setboolean{PROOFSONLY}{true}

\ifthenelse{\boolean{PROOFSONLY}}{
\newcommand*{\ifcorr}[1]{\emph{See journal version of this paper.}}}{
\newcommand*{\ifcorr}[1]{#1}}
\ifthenelse{\boolean{PROOFSONLY}}{
\newcommand*{\ifcorrx}[1]{}}{
\newcommand*{\ifcorrx}[1]{#1}}

\newcommand{\fp}{\mathsf{fp}}

\renewcommand{\emptyset}{\mathord{\varnothing}}

\newcommand*{\sseq}{\subseteq}

\newcommand*{\sep}{\mathop{\star}\nolimits}
\newcommand*{\Out}{{\mathtt{out}}}
\newcommand*{\this}{{\mathtt{this}}}
\newcommand*{\wrt}{\textit{w.r.t.\ }}
\newcommand*{\ie}{\textit{i.e., }}
\newcommand*{\integer}{\mathord{\mathit{int}}}
\newcommand*{\Type}{\mathord{\mathit{Type}}}
\newcommand*{\Vars}{\mathord{\mathit{Vars}}}
\newcommand*{\Id}{\mathord{\mathit{Id}}}
\newcommand*{\nil}{\mathord{\mathit{null}}}
\newcommand*{\Typing}{\mathord{\mathit{TypEnv}}}
\newcommand*{\Fields}{\mathord{\mathit{Fields}}}
\newcommand*{\Methods}{\mathord{\mathit{Methods}}}
\newcommand*{\Pars}{\mathord{\mathit{Pars}}}
\newcommand*{\Frame}{\mathord{\mathit{Frame}}}
\newcommand*{\Obj}{\mathord{\mathit{Obj}}}
\newcommand*{\Memory}{\mathord{\mathit{Memory}}}
\newcommand*{\Loc}{\mathord{\mathit{Loc}}}
\newcommand*{\Value}{\mathord{\mathit{Value}}}
\newcommand*{\rs}{\mathord{\mathit{res}}}

\newcommand*{\er}{{\mathcal{ER}}}
\newcommand*{\e}{{\mathcal{E}}}
\newcommand*{\ee}{e}

\newcommand{\proofoperation}[1]{\newline\underline{$\mathsf{{#1}}$}}
\newcommand{\myproof}[3]{\begin{pf*}{Proof of {#1} \ref{#2} at page \pageref{#2}}{#3}\end{pf*}}
\newcommand{\myproofbis}[1]{\begin{pf}{#1}\hspace*{\fill} $\Box$\end{pf}}
\newcommand{\eref}[1]{(\ref{#1})}
\newcommand{\elabel}[1]{\label{i#1}}

\newcommand{\nat}{\mathbb{N}}
\newcommand{\integers}{\mathbb{Z}}

\newcommand{\domain}{\mathsf{dom}}
\newcommand{\codom}{\mathsf{rng}}
\newcommand{\init}{\Im}
\newenvironment{romanenumerate}{%
  \begin{enumerate}}{\end{enumerate}%
  }

\newcommand{\summary}[2]{\textrm{\textbf{\textup{#1}}} \textit{#2}}

\newtheorem{theorem}{Theorem}
\newtheorem{proposition}[theorem]{Proposition}
\newtheorem{lemma}[theorem]{Lemma}
\newtheorem{corollary}[theorem]{Corollary}
\newtheorem{definition}[theorem]{Definition}
\theoremstyle{plain}
\theoremheaderfont{\scshape}
{\theorembodyfont{\rmfamily} \newtheorem{example}[theorem]{Example}}

\begin{document}
\begin{article}
\begin{opening}

\ifthenelse{\boolean{PROOFSONLY}}{
\title{Deriving Escape Analysis by Abstract Interpretation: Proofs of results}
}{
\title{Deriving Escape Analysis by Abstract Interpretation}
}

\author{Patricia M.\ \surname{Hill}\email{hill@comp.leeds.ac.uk}}
\institute{University of Leeds, United Kingdom}
\author{Fausto \surname{Spoto}\email{fausto.spoto@univr.it}}
\institute{Universit\`a di Verona, Italy}
%
\begin{abstract}
Escape analysis of object-oriented languages
approximates the set of objects which do not \emph{escape}
from a given context. If we take a method as context,
the non-escaping objects can be allocated on its activation stack;
if we take a thread,
Java synchronisation locks on such objects are not needed.
In this paper, we formalise a basic escape domain $\e$
as an abstract interpretation of concrete states, which we then refine
into an abstract domain $\er$ which is more concrete than $\e$ and, hence,
leads to a more precise escape analysis than $\e$.
We provide optimality results for both $\e$ and $\er$, in the form of
Galois insertions from the concrete to the abstract domains and of
optimal abstract operations. The Galois insertion property is obtained
by restricting the abstract domains to those elements which do not contain
\emph{garbage}, by using an abstract garbage collector.
Our implementation of $\er$ is hence an implementation
of a formally correct escape analyser, able to
detect the stack allocatable creation points of Java (bytecode)
applications.
\ifthenelse{\boolean{PROOFSONLY}}{

This report contains the proofs of results of a paper with the same
title and authors and to be published in the Journal
\emph{Higher-Order Symbolic Computation}.
}{
}
\end{abstract}

\keywords{Abstract Interpretation, Denotational Semantics, Garbage Collection}




\end{opening}
\section{Introduction}\label{sec:introduction}
Escape analysis identifies, at compile-time, some
run-time data structures which do not \emph{escape}
from a given context,
in the sense that they are not reachable anymore from that context.
It has been studied for functional~\cite{ParkG92,Deutsch97,Blanchet98}
as well as for object-oriented
languages~\cite{RuggieriM88,Agrawal99,BogdaH99,WhaleyR99,GayS00,Ruf00,StreckenbachS00,RountevMR01,SalcianuR01,VivienR01,Blanchet03,ChoiGSSM03,WhaleyL04}.
It allows one to stack allocate dynamically created
data structures which would
normally be heap allocated. This is possible if these data structures
do not \emph{escape} from the method which created them.
Stack allocation reduces
garbage collection overhead at run-time \wrt heap allocation, since
stack allocated data structures are automatically deallocated when methods
terminate.
If, moreover, such data structures
do not occur in a loop and their size is statically determined,
they can be preallocated on the activation stack, which further
improves the efficiency of the code.
In the case of Java, which uses a mutual exclusion lock
for each object in order to synchronise accesses from different
threads of execution, escape analysis allows one also
to remove unnecessary synchronisations,
thereby making run-time accesses faster.
By removing the space for the mutual exclusion lock
associated with some of the objects,
escape analysis can also help with space constraints.
To this purpose, the analysis must
prove that an object is accessed by at most one thread.
This is possible if the object does not \emph{escape} its creating thread.

\ifcorrx{%
Consider for instance our running example in Figure~\ref{fig:program}.
This program defines geometric figures \texttt{Square} and
\texttt{Circle} that can be rotated.
The class \texttt{Scan} has two methods, \texttt{scan} and
\texttt{rotate}.  The method \texttt{scan} calls \texttt{rotate} on
each figure of a sequence \texttt{n} of figures passed as a parameter,
as well as on a new \texttt{Square} and a new \texttt{Circle}.
%
Each \texttt{new} statement is a creation point and has been decorated
with a label, such as $\pi_1$ or $\pi_2$. We often abuse notation
and call the label a creation point itself. Hence, we can say
that the creation points $\pi_2$ and $\pi_3$ can be stack allocated since
they create objects which are not reachable once the method
\texttt{scan} terminates.
On the contrary,
the creation point $\pi_4$ cannot be allocated in the activation stack of
the method \texttt{rotate}, since the \texttt{Angle} it creates is
actually stored inside the field \texttt{rotation} of the \texttt{Square}
object created at $\pi_2$, which is still reachable when
\texttt{rotate} terminates. However, if we created \texttt{Circle}s rather than
\texttt{Square}s at $\pi_2$, and
if we assumed that the \texttt{scan} method
is passed a list of \texttt{Circle}s as parameter, then
the creation point $\pi_4$ could be stack
allocated, since the virtual call \texttt{f.rot(a)} would always lead
to the method \texttt{rot} inside \texttt{Circle}, which does not store
its parameter in any field. The creation point $\pi_1$ cannot be stack
allocated since it creates objects that are stored in the \texttt{rotation}
field, and hence are still reachable when the method completes.
Note that we assume here that an object escapes from a method if it is
still \emph{reachable} when the method terminates.
Others~\cite{Ruf00,SalcianuR01,WhaleyL04} require that
that object is actually \emph{used} after the method terminates.
Our assumption is more conservative and hence leads to less precise analyses.
However, it lets us analyse libraries, whose calling contexts are not known
at analysis time,
so that it is undetermined whether an object is actually used after a library
method terminates or not.
}
\subsection{Contributions of Our Work}\label{subsec:work}
This paper presents two escape analyses for Java programs. The goal
of both analyses is to detect objects that do not escape (\ie are
unreachable from outside) a certain scope. This information can later
be used to stack-allocate captured (\ie non-escaping) objects.

Both analyses use the object allocation site model: all objects
allocated at a given program point (possibly in a loop) are
modelled by the same creation point. The first analysis, based
on the abstract domain $\e$,
expresses the information we need for our stack allocation. Namely,
for each program point, it provides
an over-approximation of the set of creation points
that escape because they are transitively reachable from a set
of escapability roots (\ie variables
including parameters, static fields, method
result). The domain $\e$
does not keep track of other information such as the creation points
pointed to by each individual variable or field.
\ifthenelse{\boolean{PROOFSONLY}}{
\addtocounter{figure}{1}
}{
\begin{figure}[ht]
{\small
$\begin{array}{l}
\mathtt{class\ Angle\ \{}\\
\mathtt{\ \ int\ degree;}\qquad\mathtt{int\ acute()\ \{\ return\ this.degree\ <\ 90;\ \}}\\
\mathtt{\}}\\
\mathtt{class\ Figure\ \{}\\
\mathtt{\ \ Figure\ next;}\\
\mathtt{\ \ void\ def()\ \{\}}\qquad\mathtt{void\ rot(Angle\ a)\ \{\}}
  \qquad\mathtt{void\ draw()\ \{\}}\\
\mathtt{\}}\\
\mathtt{class\ Square\ extends\ Figure\ \{}\\
\mathtt{\ \ int\ side,\ x,\ y;}\qquad\mathtt{Angle\ rotation;}\\
\mathtt{\ \ void\ def()\ \{}\\
\mathtt{\ \ \ \ this.side\ =\ 1;\ this.x\ =\ this.y\ =\ 0;}\\
\mathtt{\ \ \ \ this.rotation\ =\ new\ Angle();}\qquad\{\pi_1\}\\
\mathtt{\ \ \ \ this.rotation.degree\ =\ 0;}\\
\mathtt{\ \ \}}\\
\mathtt{\ \ void\ rot(Angle\ a)\ \{\ this.rotation\ =\ a;\ \}}\\
\mathtt{\ \ void\ draw()\ \{\ ...\ use\ this.rotation\ here\ ...\ \}}\\
\mathtt{\}}\\
\mathtt{class\ Circle\ extends\ Figure\ \{}\\
\mathtt{\ \ int\ radius,\ x,\ y;}\\
\mathtt{\ \ void\ def()\ \{}\\
\mathtt{\ \ \ \ this.radius\ =\ 1;\ this.x\ =\ this.y\ =\ 0;}\qquad\{w_0\}\\
\mathtt{\ \ \}}\\
\mathtt{\ \ void\ draw()\ \{\ ...\ \}}\\
\mathtt{\}}\\
\mathtt{class\ Scan\ \{}\\
\mathtt{\ \ void\ scan(Figure\ n)\ \{}\\
\mathtt{\ \ \ \ Figure\ f\ =\ new\ Square();}\qquad\{\pi_2\}\\
\mathtt{\ \ \ \ f.next\ =\ f;}\qquad\{w_1\}\\
\mathtt{\ \ \ \ f.def();\ rotate(f);}\\
\mathtt{\ \ \ \ f\ =\ new\ Circle();}\qquad\{\pi_3\}\\
\mathtt{\ \ \ \ f.def();}\qquad\{w_2\}\\
\mathtt{\ \ \ \ f.next\ =\ n;}\\
\mathtt{\ \ \ \ while\ (f\ !=\ null)\ \{\ rotate(f);\ f\ =\ f.next;\ \}}\\
\mathtt{\ \ \}}\\
\mathtt{\ \ void\ rotate(Figure\ f)\ \{}\\
\mathtt{\ \ \ \ Angle\ a\ =\ new\ Angle();}\qquad\{\pi_4\}\\
\mathtt{\ \ \ \ f.rot(a);\ a.degree\ =\ 0;}\\
\mathtt{\ \ \ \ while\ (a.degree\ <\ 360)\ \{\ a.degree+\!\!+;\ f.draw();\ \}}\\
\mathtt{\ \ \}}\\
\mathtt{\}}
\end{array}$
}
\caption{Running example.}\label{fig:program}
\end{figure}
}

Although $\e$ is the property neede for stack allocation,
a static analysis based on $\e$
is not sufficiently precise as it does not relate the
creation points with the variables and fields that point to them. We
therefore consider a refinement $\er$ of $\e$ that preserves this
information and also includes $\e$ so that $\er$ contains just the
minimum information needed for stack allocation.


Both analyses are developed in the abstract interpretation
framework~\cite{CousotC77,CousotC92},
and we present proofs that the associated transfer functions
are optimal with respect to the abstractions that are used by each
analysis \ie they make the best possible use of the abstract information
expressed by the abstract domains.

To increase the precision of the two analyses and to
get a Galois insertion, rather than a Galois connection, both
analyses use local variable scoping and type information.
Hence, the abstract domains contain no spurious element.
We achieve this goal through \emph{abstract garbage collectors}
which remove some elements from the abstract domains
whenever they reflect unreachable (and hence, for our analysis, irrelevant)
portions of the run-time heap,
as also~\cite{ChoiGSSM03} does, although~\cite{ChoiGSSM03} does not relate this
to the Galois insertion property.
Namely, the abstract domains are exactly the set of fixpoints of
their respective abstract garbage collectors and, hence, do not contain
spurious elements.

The contribution of this paper is a clean construction of an escape
analysis through abstract interpretation thus obtaining formal and
detailed proofs of correctness as well as optimality.
Optimality states that the abstract domains are related to the
concrete domain by a Galois \emph{insertion}, rather than just a
\emph{connection} and in the use of optimal abstract operations.
%
Precision and efficiency of the analysis are not the main
issues here, although we are pleased to see that our implementation
scales to relatively large applications and
compares well with some already existing and more precise escape
analyses (Section~\ref{sec:implementation}).
%
%
\subsection{The Basic Domain $\e$}\label{subsec:e}
Our work starts by defining a basic abstract domain $\e$ for escape analysis.
Its definition is guided by the observation that a creation point $\pi$
occurring
in a method $m$ can be stack allocated if the objects it creates are not
reachable at the end of $m$ from a set of
variables $E$ which includes $m$'s return value, the fields of the
objects bound to its formal parameters at call-time
(including the implicit $\this$ parameter)
and any exceptions thrown by $m$.
Note that we consider the
fields of the objects bound to the formal parameters at call-time
since they are aliases of the actual arguments, and hence still
reachable when the method returns.
For a language, such as Java, which allows static fields, $E$ also
includes the static fields.
Variables with integer type are not included in $E$ since no object
can be reached from an integer.
Moreover, local variables are also not included in $E$ since local variables
accessible inside a
method $m$ will disappear once $m$ terminates.
The basic abstract domain $\e$ is hence defined
as the collection of all sets of creation points. Each method
is decorated with an element of $\e$, which contains precisely the
creation points of the objects reachable from the variables in $E$
at the end of the method.
\begin{example}\label{ex:example_E}
\ifcorr{%
Consider for instance the method $\mathtt{scan}$ in Figure~\ref{fig:program}.
We assume that $\nil$
is passed to $\mathtt{scan}$ as a parameter, and that
its $\mathtt{this}$ object has been created at an external creation point
$\overline{\pi}$.
We have $\e=\wp(\{\overline{\pi},\pi_1,\pi_2,\pi_3,\pi_4\})$ and
$E=\{\mathtt{n}\}$
(the implicit parameter $\this$ has type $\mathtt{Scan}$ and hence no fields,
so that there is no need to consider it in $E$).
Then $\mathtt{scan}$ is decorated with $\emptyset$ since no
object can be reached at the end of $\mathtt{scan}$
from the variables in $E=\{\mathtt{n}\}$. Consequently, the creation points
$\pi_2$ and $\pi_3$ can be stack allocated since they do not belong
to $\emptyset$.
Note that, if $\mathtt{n}$ had been modified inside
the method $\mathtt{scan}$ then we would have used
$E=\{\mathtt{n}'\}$, where $\mathtt{n}'$ is a
\emph{shadow copy} of $\mathtt{n}$ which holds its \emph{initial} value
(we will see this technique in Example~\ref{ex:abstract_operations_er2}).
\qed
}
\end{example}

We still have to specify how this decoration is computed for
each method. We use abstract interpretation to propagate an input
set of creation points through the statements of each method, until its end
is reached. This is accomplished by defining a \emph{transfer function}
for every statement of the program which, in terms of abstract interpretation,
is called an \emph{abstract operation}
(see Section~\ref{sec:edomain} and Figure~\ref{fig:operations_e}).
The element of $\e$ resulting
at the end of each method is then \emph{restricted} to the appropriate set $E$
for that method
through an abstract operation called $\mathsf{restrict}$.
By applying the theory of abstract interpretation, we know that this restriction
is a conservative approximation of the actual decoration we need at the end of
each method.

\begin{example}\label{ex:e_imprecise}
\ifcorr{%
Consider again Example~\ref{ex:example_E}
and the method $\mathtt{scan}$.
In Figure~\ref{fig:e_imprecise}
we propagate the set $\{\overline{\pi}\}$ through $\mathtt{scan}$'s statements
by following the
translation of high-level statements into the \emph{bytecodes}
as specified in Figure~\ref{fig:operations_e}.
%
}
\ifthenelse{\boolean{PROOFSONLY}}{
\addtocounter{figure}{1}
}{
\begin{figure}
\[\begin{array}{l}
\mathtt{void}\ \mathtt{scan}(\mathtt{Figure}\ \mathtt{n})\ \{\\
\{\overline{\pi}\}\\
\ \ \mathtt{Figure\ f\ =\ new\ Square();}\ \quad\{\pi_2\}\\
\{\overline{\pi},\pi_2\}\\
\ \ \mathtt{f.next\ =\ f;}\ \quad\{w_1\}\\
\{\overline{\pi},\pi_2\}\\
\ \ \mathtt{f.def();}\\
\{\overline{\pi},\pi_1,\pi_2\}\\
\ \ \mathtt{rotate(f);}\\
\{\overline{\pi},\pi_1,\pi_2,\pi_4\}\\
\ \ \mathtt{f\ =\ new\ Circle();}\ \quad\{\pi_3\}\\
\{\overline{\pi},\pi_1,\pi_2,\pi_3,\pi_4\}\\
\ \ \mathtt{f.def();}\\
\{\overline{\pi},\pi_1,\pi_2,\pi_3,\pi_4\}\\
\ \ \mathtt{f.next\ =\ n;}\\
\{\overline{\pi},\pi_1,\pi_2,\pi_3,\pi_4\}\\
\ \ \mathtt{while(f\ !=\ null)\ \{}\\
\{\overline{\pi},\pi_1,\pi_2,\pi_3,\pi_4\}\\
\ \ \ \ \mathtt{rotate(f);}\\
\{\overline{\pi},\pi_1,\pi_2,\pi_3,\pi_4\}\\
\ \ \ \ \mathtt{f\ =\ f.next;}\\
\{\overline{\pi},\pi_1,\pi_2,\pi_3,\pi_4\}\\
\ \ \mathtt{\}}\\
\mathtt{\}}
\end{array}\]
\caption{The propagation of $\{\overline{\pi}\}$ through the method $\mathtt{scan}$ of the program in Figure~\ref{fig:program}.}
  \label{fig:e_imprecise}
\end{figure}
}
\ifcorrx{%
The restriction of the set $\{\overline{\pi},\pi_1,\pi_2,\pi_3,\pi_4\}$ to
$E=\{\mathtt{n}\}$
is $\{\pi_1,\pi_2,\pi_3,\pi_4\}$
(objects of class $\mathtt{Scan}$ created at $\overline{\pi}$ are
incompatible with the type of $\mathtt{n}$), which is a very
imprecise approximation of the desired result \ie $\emptyset$.
\qed
}
\end{example}

\ifthenelse{\boolean{PROOFSONLY}}{
\addtocounter{figure}{1}
}{
\begin{figure}[t]
\[\begin{array}{l}
\{\text{only $\mathtt{this}$ of type $\mathtt{Scan}$ is in scope here}\}\\
\mathtt{\{}\\
\ \ \ \mathtt{Figure\ f\ =\ new\ Square();}\quad\qquad\{\pi_s\}\\
\ \ \ \{p_1\!\}\\
\ \ \ \mathtt{f.def();}\qquad\{\text{this creates an object at $\pi_1$ (Figure
 ~\ref{fig:program})}\}\\
\ \ \ \{p_2\!\}\\
\mathtt{\}}\\
\{p_3\!\}
\end{array}\]
\caption{A program fragment where the set of variables in scope grows
  and shrinks.}\label{fig:abstract_garbage_collector}
\end{figure}
}

The problem here is that although the abstract domain $\e$
expresses the kind of
decoration we need for stack allocation, $\e$ has very poor
computational properties. In terms of abstract interpretation, it
induces very imprecise
abstract operations and, just as in the case of the basic domain
$\mathcal{G}$ for \emph{groundness analysis} of logic
programs~\cite{JS87}, it needs refining~\cite{GiacobazziR97,Scozzari00}.

\ifcorrx{%
Nevertheless, it must be observed that the abstract domain
$\e$ already contains some non-trivial information.
For instance, since $\overline{\pi}$ is a creation point
for objects of class $\mathtt{Scan}$
and $\pi_2$ is a creation point for objects of class
$\mathtt{Square}$, then the first $\mathtt{f.def()}$ virtual call
occurring in the method $\mathtt{scan}$ can only lead to
the method $\mathtt{def}$ inside $\mathtt{Square}$.
Hence we say that
\emph{our escape information contains information
on the run-time late-binding mechanism}, which can be exploited
to improve the precision of the analysis by refining the
call-graph.
This is what actually happens in Example~\ref{ex:e_imprecise}.
Note also that
the local scope may temporarily introduce new variables so that
at the end of the scope,
any creation points that can \emph{only} be reached from these variables
can be safely removed from the approximation. In
Figure~\ref{fig:abstract_garbage_collector}, the approximation computed at
$p_2$ is $\{\overline{\pi},\pi_s,\pi_1\}$, where $\overline{\pi}$
is the creation point of $\mathtt{this}$, but the approximation computed at
$p_3$ is $\{\overline{\pi}\}$, which is smaller.
For this reason, we say that
\emph{our escape information uses the static type information} to improve
the precision of the analysis.
That is, the possible approximations from $\e$ in a given program point,
are constrained by the (finite) set of
variables and their types that are in scope.
}

We formalise the fact that the approximation in $\e$ can shrink,
by means of an \emph{abstract garbage collector}
(Definition~\ref{def:delta})
\ie a garbage collector that works over sets of creation points
instead of concrete objects. When a variable's scope is
closed, the abstract garbage collector
removes from the approximation of the next statement all
creation points which can \emph{only} be reached \emph{from that variable}.
The name of abstract garbage collector is justified by the fact that
this conservatively maintains in the approximation the creation points of the
objects which \emph{might} be reachable in the concrete state, thus
modeling in the abstract domain a behaviour similar to that of a concrete
garbage collector. It must be noted, however, that our abstract garbage
collector only considers reachability from the variables in scope in the
current method, while
a concrete garbage collector would consider reachability from all variables
in the current activation stack.

\ifcorrx{%
The abstract garbage collector is of no use in
the propagation of $\bar{\pi}$ shown in Figure~\ref{fig:e_imprecise}
since, for instance,
after the creation point $\pi_3$, it is not possible to conclude
that the object $o$ created at $\pi_2$, and hence also those created at
$\pi_1$ and $\pi_4$ and stored in $o$'s $\mathtt{rotation}$ field,
are not reachable anymore.
This is because $\e$
does not distinguish between the
objects reachable from variables
\texttt{f} and \texttt{n}. Before $\pi_3$, the object $o$ created at
$\pi_2$ can be reached from \texttt{f} only,
but $\pi_3$ overwrites \texttt{f} so it cannot
be reached anymore. Because $\e$ does not make a distinction
between objects reachable from \texttt{f} and \texttt{n},
it cannot infer this, because
it considers that $o$ could be reachable from \texttt{n}.
The only safe choice is to
be conservative and assume that it cannot be garbage collected.
}
\subsection{The Refinement $\er$}\label{subsec:er}
The abstract domain $\e$ represents the information we need for
stack allocation, but it does not
include any other related information that may improve
the precision of the abstract operations,
such as explicit information
about the creation points of the objects bound to
\emph{a given} variable or field.
However, the ability to reason on a per variable basis is essential for
the precision of a static analysis of imperative languages, where assignment
to a given variable or field is the basic computational mechanism.
So we \emph{refine} $\e$ into a new abstract domain $\er$ which
splits the sets of creation points in $\e$ into
subsets, one for each variable or field.
We show that $\er$ strictly contains $\e$, justifying the name
of \emph{refinement}.

We perform a static analysis based on $\er$ exactly as for $\e$
but using the abstract operations for the domain $\er$
given in Section~\ref{sec:erdomain} (see Figure~\ref{fig:operations_er}).
\begin{example}\label{ex:er_precision}
\ifcorr{%
Consider again the method $\mathtt{scan}$ in Figure~\ref{fig:program}.
We start the analysis from the element $[\mathtt{this}\mapsto
\{\overline{\pi}\}]\sep[]$ of $\er$ which expresses the fact that the variable
$\mathtt{this}$ is initially bound to an object created at the
external creation point $\overline{\pi}$ and all other variables and
fields are initially bound to $\nil$ (if they have class type)
or to an integer (otherwise).
The operator $\sep$ is a pair-separator; its component
$[\mathtt{this}\mapsto\{\overline{\pi}\}]$ is the approximation for
the variables in scope and
its component $[]$ is the approximation for the fields.
The information is then propagated, as shown
in Figure~\ref{fig:er_precision}.
Then, just as for the domain $\e$, at the end of the method the result
$[\mathtt{this}\mapsto\{\overline{\pi}\}]\sep[]$ is restricted
to $E=\{\mathtt{n}\}$ and we get
$[]\sep[]$, which leads to $\emptyset$ which is a
much more precise approximation
than the set $\{\pi_1,\pi_2,\pi_3,\pi_4\}$
obtained in Example~\ref{ex:e_imprecise} with $\e$.
}
\ifthenelse{\boolean{PROOFSONLY}}{
\addtocounter{figure}{1}
}{
\begin{figure}
\[\begin{array}{l}
\mathtt{void\ scan(Figure\ n)\ \{}\\

[\mathtt{this}\mapsto\{\overline{\pi}\}]\sep[]\\
\ \ \mathtt{Figure\ f\ =\ new\ Square();}\ \quad\{\pi_2\}\\

[\mathtt{f}\mapsto\{\pi_2\},\mathtt{this}\mapsto\{\overline{\pi}\}]\sep[]\\
\ \ \mathtt{f.next\ =\ f;}\\

[\mathtt{f}\mapsto\{\pi_2\},\mathtt{this}\mapsto\{\overline{\pi}\}]
\sep[\mathtt{next}\mapsto\{\pi_2\}]\\
\ \ \mathtt{f.def();}\\

[\mathtt{f}\mapsto\{\pi_2\},\mathtt{this}\mapsto\{\overline{\pi}\}]
\sep[\mathtt{next}\mapsto\{\pi_2\},\mathtt{rotation\mapsto\{\pi_1\}}]\\
\ \ \mathtt{rotate(f);}\\

[\mathtt{f}\mapsto\{\pi_2\},\mathtt{this}\mapsto\{\overline{\pi}\}]
\sep[\mathtt{next}\mapsto\{\pi_2\},\mathtt{rotation\mapsto\{\pi_1,\pi_4\}}]\\
\ \ \mathtt{f\ =\ new\ Circle();}\ \quad\{\pi_3\}\\

[\mathtt{f}\mapsto\{\pi_3\},\mathtt{this}\mapsto\{\overline{\pi}\}]
\sep[\mathtt{next}\mapsto\{\pi_2\},\mathtt{rotation\mapsto\{\pi_1,\pi_4\}}]\\
\ \ \mathtt{f.def();}\\

[\mathtt{f}\mapsto\{\pi_3\},\mathtt{this}\mapsto\{\overline{\pi}\}]
\sep[\mathtt{next}\mapsto\{\pi_2\},\mathtt{rotation\mapsto\{\pi_1,\pi_4\}}]\\
\ \ \mathtt{f.next\ =\ n;}\\

[\mathtt{f}\mapsto\{\pi_3\},\mathtt{this}\mapsto\{\overline{\pi}\}]
\sep[\mathtt{next}\mapsto\{\pi_2\},\mathtt{rotation\mapsto\{\pi_1,\pi_4\}}]\\
\ \ \mathtt{while(f\ !=\ null)\ \{}\\

[\mathtt{f}\mapsto\{\pi_2,\pi_3\},\mathtt{this}\mapsto\{\overline{\pi}\}]
\sep[\mathtt{next}\mapsto\{\pi_2\},\mathtt{rotation\mapsto\{\pi_1,\pi_4\}}]\\
\ \ \ \ \mathtt{rotate(f);}\\

[\mathtt{f}\mapsto\{\pi_2,\pi_3\},\mathtt{this}\mapsto\{\overline{\pi}\}]
\sep[\mathtt{next}\mapsto\{\pi_2\},\mathtt{rotation\mapsto\{\pi_1,\pi_4\}}]\\
\ \ \ \ \mathtt{f\ =\ f.next;}\\

\ \ \mathtt{\}}\\

[\mathtt{this}\mapsto\{\overline{\pi}\}]\sep[]\\

\mathtt{\}}
\end{array}\]
\caption{The propagation of $[\mathtt{this}\mapsto\{\overline{\pi}\}]\sep[]$ through the method $\mathtt{scan}$ of the program in Figure~\ref{fig:program}.}
  \label{fig:er_precision}
\end{figure}
}

\ifcorrx{%
Note that at the end of the method
(when $\mathtt{f}$ and $\mathtt{n}$ go out of scope),
the approximation of the fields $\mathtt{next}$ and
$\mathtt{rotation}$ are reset to $\emptyset$.
The justification for this is that, at this point, it is no longer
possible to reach the $\mathtt{Square}$ object created at $\pi_2$
whose field $\mathtt{rotation}$ contained objects created at
$\pi_1$ or $\pi_4$, nor is it
possible to reach the $\mathtt{Circle}$ object created at
$\pi_3$ whose $\mathtt{next}$ field might have contained something
created at $\pi_2$. This is an example of the application
of our abstract garbage collector for $\er$ (Definition~\ref{def:xi}).
\qed
}
\end{example}

The domain $\er$ can hence be seen as the specification of a new escape
analysis, which includes $\e$ as its foundational kernel.
Example~\ref{ex:er_precision} shows that the abstract domain
$\er$ is actually more precise than $\e$.
Our implementation of $\er$ (Section~\ref{sec:implementation}) shows that
it can actually be
used to obtain non-trivial escape analysis information for Java bytecode.

%
%
\subsection{Structure of the Paper}\label{subsec:structure}
After a brief summary of our notation and terminology
in Section~\ref{sec:preliminaries}, we pass
in Section~\ref{sec:framework} to recall the framework
of~\cite{SpotoJ03} on which the analysis is based.
Then, in Section~\ref{sec:edomain}, we formalise our basic domain $\e$
and provide suitable abstract operations for its analysis.
We show that the analysis induced by $\e$ is very imprecise.
Hence, in Section~\ref{sec:erdomain} we refine the domain $\e$ into the
more precise domain $\er$ for escape analysis.
In Section~\ref{sec:implementation},
we discuss our prototype implementation and experimental results.
Section~\ref{sec:discussion} discusses related work.
Section~\ref{sec:conclusion} concludes the main part of the paper.
\ifthenelse{\boolean{WITHAPPENDIX}}{
} 
{Proofs not inlined in this paper are available in~\cite{corr-version}.
} 

Preliminary, partial versions of this paper appeared in~\cite{HillS02b}
and~\cite{HillS02}.
The current paper is a seamless
fusion of these papers, with the proofs of the theoretical results and
with a description and evaluation of the implementation of the escape analysis
over the domain $\er$.
\section{Preliminaries}\label{sec:preliminaries}
A total (partial) function $f$ is denoted by $\mapsto$ ($\to$).
The \emph{domain} (\emph{range}) of $f$ is $\domain(f)$ ($\codom(f)$).
We denote by $[v_1\!\mapsto\!t_1,\ldots,v_n\!\mapsto\!t_n]$ the function $f$
where $\domain(f) = \{v_1,\ldots,v_n\}$ and $f(v_i)=t_i$ for
$i=1,\ldots,n$.
Its \emph{update} is $f[w_1\mapsto d_1,\ldots,w_m\mapsto d_m]$,
where the domain may be enlarged.
By $f|_s$ ($f|_{-s}$) we denote the
\emph{restriction} of $f$ to $s\subseteq\domain(f)$
(to $\domain(f)\setminus s$).
If $f$ and $g$ are functions,
we denote by $fg$ the composition of $f$ and $g$,
such that $fg(x)=f(g(x))$.
If $f(x)=x$ then $x$ is a \emph{fixpoint} of $f$.
The set of fixpoints of $f$ is denoted by $\fp(f)$.

A \emph{pair} of elements is written $a\sep b$.
A definition of a pair $S$ such as
$S=a\sep b$, with $a$ and $b$ meta-variables,
silently defines the pair selectors $s.a$ and $s.b$ for $s\in S$.
The cardinality of a set $S$ is denoted by $\# S$.
The \emph{disjoint union} of two sets $S, T$ is denoted by $S + T$.
To simplify expressions, particulary when the set is used as a subscript,
we sometimes write a singleton set $\{x\}$ as $x$.
If $S$ is a set and $\le$ is a partial relation over $S$, we say that $S$
is a \emph{partial ordering} if it is reflexive ($s\le s$ for every
$s\in S$), transitive ($s_1\le s_2$ and $s_2\le s_3$ entail
$s_1\le s_2$ for every $s_1,s_2,s_3\in S$) and
anti-symmetric ($s_1\le s_2$ and $s_2\le s_1$ entail
$s_1=s_2$ for every $s_1,s_2\in S$).
If $S$ is a set and $\le$ a partial ordering on $S$, then
the pair $S\sep\le$ is a \emph{poset}.

A \emph{complete lattice} is a poset $\mathit{C}\sep\mathord\le$
where \emph{least upper bounds}
(lub) and \emph{greatest lower bounds} (glb) always exist. Let
$\mathit{C}\sep\mathord\le$ and $\mathit{A}\sep\mathord\preceq$ be posets and
$f:\mathit{C}\mapsto\mathit{A}$. We say that $f$ is \emph{monotonic}
if $c_1\le c_2$ entails $f(c_1)\preceq f(c_2)$. It is
\emph{(co-)additive} if it preserves lub's (glb's).
Let $f:\mathit{A}\mapsto\mathit{A}$.
The map $f$ is \emph{reductive} (respectively, \emph{extensive}) if
$f(a)\preceq a$ (respectively, $a\preceq f(a)$) for any $a\in\mathit{A}$.
It is \emph{idempotent} if $f(f(a))=f(a)$ for any $a\in\mathit{A}$.
It is a \emph{lower closure operator}
(\emph{lco}) if it is \emph{monotonic}, \emph{reductive}
and \emph{idempotent}.

We recall now the basics of abstract
interpretation~\cite{CousotC77,CousotC92}.
Let $\mathit{C}\sep\mathord\le$ and
$\mathit{A}\sep\mathord\preceq$ be two posets (the concrete and
the abstract domain). A \emph{Galois connection}
is a pair of monotonic maps $\alpha:\mathit{C}\mapsto\mathit{A}$ and
$\gamma:\mathit{A}\mapsto\mathit{C}$ such that
$\gamma\alpha$ is extensive and $\alpha\gamma$ is reductive. It is a
\emph{Galois insertion} when $\alpha\gamma$ is the identity map \ie
when the abstract domain does not contain \emph{useless} elements.
If $C$ and $A$ are complete lattices and $\alpha$ is strict and
additive, then $\alpha$ is the
abstraction map of a Galois connection.
If, moreover, $\alpha$ is onto or $\gamma$ is one-to-one, then
$\alpha$ is the abstraction map of a Galois insertion.
In a Galois connection, $\gamma$ can be defined in terms of $\alpha$ as
$\gamma(a)=\cup\{c\mid \alpha(c)\preceq a\}$, where $\cup$ is the
least upper bound operation over the concrete domain $C$.
Hence, it is enough to provide $\alpha$ to define a Galois connection.
An abstract operator $\hat{f}:\mathit{A}^n\mapsto\mathit{A}$ is
\emph{correct} \wrt $f:\mathit{C}^n\rightarrow\mathit{C}$
if $\alpha f\gamma\preceq\hat{f}$. For each operator $f$, there exists
an \emph{optimal} (most precise) correct abstract operator $\hat{f}$ defined
as $\hat{f}=\alpha f\gamma$.
This means that $\hat{f}$ does the best it can with the information
expressed by the abstract domain.
The composition of correct operators is correct.
The composition of optimal operators is not necessarily optimal.
The \emph{semantics} of a program is the fixpoint of a map
$f:\mathit{C}\mapsto\mathit{C}$, where $\mathit{C}$ is the
\emph{computational domain}. Its
\emph{collecting version}~\cite{CousotC77,CousotC92}
works over \emph{properties} of $\mathit{C}$ \ie over
$\wp(\mathit{C})$ and is the fixpoint of the powerset extension of $f$.
If $f$ is defined through suboperations, their powerset extensions
\emph{and $\cup$} (which merges the semantics of the branches of a conditional)
induce the extension of $f$.
\section{The Framework of Analysis}\label{sec:framework}
The framework presented here is for a simple typed object-oriented language
where the concrete states and operations are based on~\cite{SpotoJ03}.
It allows us to derive a compositional, denotational semantics,
which can be seen as an
analyser, from a specification of
a domain of abstract states and operations which work over them
(hence called \emph{state transformers}).
Then problems such as scoping, recursion and name clash can be ignored,
since these are already solved by the semantics.
Moreover, this framework relates the precision of the analysis to that
of its abstract domain so that traditional techniques for comparing the
precision of abstract domains can be
applied~\cite{CortesiFW98,CousotC77,CousotC92}.

The definition of a denotational semantics, in the style
of~\cite{Winskel93}, by using the state transformers
of this section can be found in~\cite{SpotoJ03}. Here we only want
to make clear some points:
\begin{itemize}
\item We allow expressions to have side-effects, such as method call
      expressions, which is not the case
      in~\cite{Winskel93}. As a consequence, the evaluation of an expression
      from an initial state yields both a final state \emph{and} the value of
      the expression. We use a special variable $\rs$ of the final
      state to hold this value;
\item The evaluation from an initial state $\sigma_1$ of a binary operation
      such as $e_1+e_2$, where $e_1$ and $e_2$ are
      expressions, first evaluates $e_1$ from $\sigma_1$, yielding
      an intermediate state $\sigma_2$, and then evaluates
      $e_2$ from $\sigma_2$, yielding a state $\sigma_3$.
      The value $v_1$ of $\rs$ in $\sigma_2$ is that of $e_1$, and the value
      $v_2$ of $\rs$ in $\sigma_3$ is that of $e_2$. We then modify
      $\sigma_3$ by storing in $\rs$ the sum $v_1+v_2$. This yields the
      final state.
      Note that the single variable $\rs$ is enough for this
      purpose. The complexity of this mechanism \wrt~a more
      standard approach~\cite{Winskel93} is,
      again, a consequence of the use of expressions with side-effects;
\item Our denotational semantics deals with method calls through
      \emph{interpretations}: an interpretation is the input/output
      behaviour of a method, and is used as its denotation whenever
      that method is called.
      As a nice consequence,
      our states contain only a single frame, rather than
      an activation stack of frames.
      This is standard in denotational semantics and has been used
      for years in logic programming~\cite{BossiGLM94}.
\item The computation of the semantics of a program starts from a
      bottom interpretation which maps every input state
      to an undefined final state and then
      updates this interpretation with the denotations of the methods body.
      This process is iterated until a fixpoint is reached as is done for
      logic programs~\cite{BossiGLM94}.
      The same technique can be applied
      to compute the abstract semantics of a program, but the computation is
      performed over the abstract domain. It is also possible to generate
      constraints which relate the abstract approximations at different
      program points, and then solve such constraints with a fixpoint engine.
      The latter is the technique that we use in
      Section~\ref{sec:implementation}.
\end{itemize}

%
\ifthenelse{\boolean{PROOFSONLY}}{
\addtocounter{figure}{1}
}{
\begin{figure}[t]
\begin{center}
{\small
\begin{gather}
\mathcal{K}=\left\{\begin{array}{l}
  \mathtt{Angle},\\
  \mathtt{Figure},\\
  \mathtt{Square},\\
  \mathtt{Circle},\\
  \mathtt{Scan}
\end{array}\right\}\quad
  \mathcal{M}=\left\{\begin{array}{l}
  \mathtt{Angle.acute},\\
  \mathtt{Figure.def},
    \mathtt{Figure.rot},\mathtt{Figure.draw},\\
  \mathtt{Square.def},
    \mathtt{Square.rot},\mathtt{Square.draw},\\
  \mathtt{Circle.def},
    \mathtt{Circle.draw},\\
  \mathtt{Scan.scan},\mathtt{Scan.rotate}
\end{array}\right\}\notag\\
\mathtt{Square}\le\mathtt{Figure},\quad\mathtt{Circle}\le\mathtt{Figure}
  \quad\text{and reflexive cases}\notag\\
\begin{align*}
F(\mathtt{Angle})&=[\mathtt{degree}\mapsto\integer]\qquad
  F(\mathtt{Scan})=[]\\
F(\mathtt{Figure})&=[\mathtt{next}\mapsto\mathtt{Figure}]\\
F(\mathtt{Square})&=\left[\begin{array}{l}
  \mathtt{side}\mapsto\integer,\mathtt{Square.x}\mapsto\integer,
    \mathtt{Square.y}\mapsto\integer,\\
  \mathtt{rotation}\mapsto\mathtt{Angle},\mathtt{next}\mapsto\mathtt{Figure}
  \end{array}\right]\\
F(\mathtt{Circle})&=\left[\begin{array}{l}
  \mathtt{radius}\mapsto\integer,\mathtt{Circle.x}\mapsto\integer,\\
  \mathtt{Circle.y}\mapsto\integer,\mathtt{next}\mapsto\mathtt{Figure}
\end{array}\right]
\end{align*}\\
\begin{align*}
M(\mathtt{Angle})&=[\mathtt{acute}\mapsto\mathtt{Angle.acute}]\\
M(\mathtt{Figure})&=\left[\begin{array}{l}
  \mathtt{def}\mapsto\mathtt{Figure.def},\mathtt{rot}\mapsto
    \mathtt{Figure.rot},\\
  \mathtt{draw}\mapsto\mathtt{Figure.draw}
\end{array}\right]\\
M(\mathtt{Square})&=
  \left[\begin{array}{l}
    \mathtt{def}\mapsto
      \mathtt{Square.def},\mathtt{rot}\mapsto
      \mathtt{Square.rot},\\
    \mathtt{draw}\mapsto\mathtt{Square.draw}
  \end{array}\right]\\
M(\mathtt{Circle})&=
  \left[\begin{array}{l}
    \mathtt{def}\mapsto\mathtt{Circle.def},\mathtt{rot}\mapsto
      \mathtt{Figure.rot},\\
    \mathtt{draw}\mapsto\mathtt{Circle.draw}
  \end{array}\right]\\
M(\mathtt{Scan})&=
  [\mathtt{scan}\mapsto\mathtt{Scan.scan},\mathtt{rotate}\mapsto
    \mathtt{Scan.rotate}]
\end{align*}\\
\begin{align*}
P(\mathtt{Angle.acute})&=[\Out\mapsto\integer,\mathtt{this}
  \mapsto\mathtt{Angle}]\\
P(\mathtt{Figure.rot})&=[\mathtt{a}\mapsto\mathtt{Angle},
  \Out\mapsto\integer,\mathtt{this}\mapsto\mathtt{Figure}]\\
P(\mathtt{Scan.rotate})&=[\mathtt{f}\mapsto\mathtt{Figure},
  \Out\mapsto\integer,\mathtt{this}\mapsto\mathtt{Scan}]\\
P(\mathtt{Figure.def})&=[\Out\mapsto\integer,\mathtt{this}
  \mapsto\mathtt{Figure}]\\
&\text{(the other cases of $P$ are as above)}
\end{align*}
\end{gather}}\normalsize
\end{center}
\caption{The static information of the program in Figure \ref{fig:program}.}
  \label{fig:static_information}
\end{figure}
}
\subsection{Programs and Creation Points}\label{subsec:creation_points}
We recall here the semantical framework of~\cite{SpotoJ03}.
\begin{definition}[Type Environment]\label{def:typing}
Each program in the language has a finite set of \emph{identifiers} $\Id$
such that $\Out, \this \in \Id$ and
a finite set of \emph{classes} $\mathcal{K}$ ordered by
a \emph{subclass relation} $\le$ such that
$\mathcal{K}\sep\mathord\le$ is a poset. 
Let $\Type = \{\integer\} \uplus \mathcal{K}$ and
 $\le$ be extended to $\Type$ by defining $\integer\le\integer$.
Let $\Vars\subseteq\Id$ be a set of \emph{variables} such that
$\{\Out,\this\} \subseteq \Vars$.
A \emph{type environment} for a program is any element of the set
\[
  \Typing=\left\{\tau:\Vars\to\Type\left|\begin{array}{l}
    \text{if }\this\in\domain(\tau)\text{ then }\tau(\this)
    \in\mathcal{K}\end{array}\right.\right\}.
\]
In the following, $\tau$ will implicitly stand for a type environment.
\end{definition}

A class contains local variables (\emph{fields})
and functions (\emph{methods}). A method has
a set of input/output variables called \emph{parameters}, including
\texttt{out}, which holds the result of the method, and
\texttt{this}, which is the object over which the method has been called
(the \emph{receiver} of the call).
Methods returning $\mathtt{void}$ are represented
as methods returning an $\integer$ of constant value $0$,
implicitly ignored by the caller of the method.

\begin{example}
\label{ex:type-environment}
\ifcorr{%
Consider the example program given in Figure~\ref{fig:program}.
Here $\Id$ includes, in addition to the identifiers
$\Out$ and $\this$, user-defined identifiers such as
\(
  \mathtt{rotation},
  \mathtt{def},
  \mathtt{x},
  \mathtt{y},
  \mathtt{rot}
\).
The set of classes is
\[
 \mathcal{K} =
   \{
      \mathtt{Angle},
      \mathtt{Figure},
      \mathtt{Square},
      \mathtt{Circle},
      \mathtt{Scan}
    \}
\]
where the ordering $\le$ is defined
 so that
$\mathtt{Square} \le \mathtt{Figure}$ and
$\mathtt{Circle} \le \mathtt{Figure}$.
Variables for this program include
$\mathtt{x}$, $\mathtt{y}$, $\mathtt{f}$, $\mathtt{n}$ and $\this$.
Variable $\Out$ is used to hold the return value of the methods,
so that the $\mathtt{return}\ e$ statement can be seen as syntactic sugar
for $\Out\ =\ e$ (with no following statements).
At points $w_0$ and $w_1$, the type environments are
\begin{align*}
\tau_{w_0} &=
  [\Out\mapsto\integer,
   \mathtt{this}\mapsto\mathtt{Circle}]
\\
\tau_{w_1} &=
  [\mathtt{f}\mapsto\mathtt{Figure},
     \mathtt{n}\mapsto\mathtt{Figure},
     \Out\mapsto\integer,\this\mapsto\mathtt{Scan}].
\end{align*}
\qed
}
\end{example}

$\Fields$ is a set of maps which bind each class to the type
environment of its fields.
The variable $\this$ cannot be a field.
$\Methods$ is a set of maps which bind each class to a map from identifiers to
methods. $\Pars$ is a set of maps which bind each method
to the type environment of its parameters (its signature).
\begin{definition}[Field, Method, Parameter]\label{def:static_information}
Let $\mathcal{M}$ be a finite set of \emph{methods}. We define
\begin{align*}
  \Fields&=\{F:\mathcal{K}\mapsto\Typing\mid\this\not\in\domain(F(\kappa))
    \text{ for every }\kappa\in\mathcal{K}\}\\
  \Methods&=\mathcal{K}\mapsto(\Id\to\mathcal{M})\\
  \Pars&=\{P:\mathcal{M}\mapsto\Typing\mid
    \{\Out,\this\}\subseteq\domain(P(\nu))
    \text{ for }\nu\in\mathcal{M}\}.
\end{align*}
\end{definition}

The \emph{static information} of a program is used by
the static analyser.
\begin{definition}[Static Information]\label{def:program}
The \emph{static information} of a program consists of a poset
$\mathcal{K}\sep\le$, a set of methods $\mathcal{M}$ and maps
$F\in\Fields$, $M\in\Methods$ and $P\in\Pars$.
\end{definition}

Fields in different classes but with the same name can be disambiguated
by using their \emph{fully qualified name} such as in
the Java Virtual Machine~\cite{LindholmY99}. For instance, we write
$\mathtt{Circle.x}$ for the field $\mathtt{x}$ of the class $\mathtt{Circle}$.

\begin{example}
\ifcorr{%
The static information of the program in Figure~\ref{fig:program}
is shown in Figure~\ref{fig:static_information}.
Note that the result of the method $\mathtt{Angle.acute}$
in Figure~\ref{fig:program}
becomes the type of $\Out$ in $P(\mathtt{Angle.acute})$
in Figure~\ref{fig:static_information}.
\qed
}
\end{example}

The only points in the program where
new objects can be created are the $\mathtt{new}$ statements.
We require that each of these statements
is identified by a unique label called its \emph{creation point}.
\begin{definition}[Creation Point]\label{def:creation_points}
Let $\Pi$ be a finite set of labels called \emph{creation points}.
A map $k:\Pi\mapsto\mathcal{K}$ relates every creation point $\pi\in\Pi$ with
the class $k(\pi)$ of the objects it creates.
\end{definition}
\begin{example}\label{ex:k-map}
\ifcorr{%
Consider again the program
in Figure~\ref{fig:program}. In that program
$\{\overline{\pi}, \pi_1, \pi_2, \pi_3, \pi_4\}$
is the set of creation points, where we assume that $\overline{\pi}$
decorates an external creation point for $\mathtt{Scan}$
(not shown in the figure). Then
\[
    k = [\overline{\pi} \mapsto \mathtt{Scan},
         \pi_1 \mapsto \mathtt{Angle},
         \pi_2 \mapsto \mathtt{Square},
         \pi_3 \mapsto \mathtt{Circle},
         \pi_4 \mapsto \mathtt{Angle}
        ].
\]
\qed
}
\end{example}

\subsection{Concrete States}\label{subsec:concrete_states}
To represent the concrete state of a computation at a
particular program point we need
to refer to the concrete values that may be assigned to the variables.
Apart from the integers and $\nil$, these values need to include
\emph{locations} which are the addresses of the memory cells
used at that point.
Then the concrete state of the computation
consists of a map that assigns type consistent values to variables
(\emph{frame})  and
a map from locations to objects (\emph{memory})
where an \emph{object} is characterised by its creation point
and the frame of its fields.
Hence the notion of object that we use here is more concrete
than that in~\cite{SpotoJ03}, which relates a \emph{class} rather
than a \emph{creation point} to each object.
A memory can be \emph{updated} by assigning new (type consistent)
values to the variables in its frames.
\begin{definition}[Location, Frame, Object, Memory]\label{def:domains2}
Let $\Loc$ be an infinite set of \emph{locations} and $\Value=\integers+\Loc
+\{\nil\}$. We define \emph{frames}, \emph{objects} and
\emph{memories} as
\begin{align*}
  \Frame_\tau&=\left\{\phi\in\domain(\tau)\mapsto\Value\left|
    \begin{array}{l}
      \text{for every }v\in\domain(\tau)\\
      \tau(v)=\integer\Rightarrow\phi(v)\in\integers\\
      \tau(v)\in\mathcal{K}
        \Rightarrow\phi(v)\in\{\nil\}\cup\Loc\\
    \end{array}
    \right.\right\}\\
  \Obj&=\{\pi\sep\phi\mid\pi\in\Pi,\ \phi\in\Frame_{F(k(\pi))}\}\\
  \Memory&=\{\mu\in\Loc\to\Obj\mid\domain(\mu)\text{ is finite}\}.
\end{align*}
Let $\mu_1,\mu_2\in\Memory$ and $L\subseteq\domain(\mu_1)$.
We say that $\mu_2$ is an $L$-\emph{update} of
$\mu_1$, written $\mu_1=_L\mu_2$, if $L\subseteq\domain(\mu_2)$ and
for every $l\in L$ we have $\mu_1(l).\pi=\mu_2(l).\pi$.

The initial value for a variable of a given type is used when we
add a variable in scope. It is defined as $\init(\integer)=0$,
$\init(\kappa)=\nil$ for $\kappa\in\mathcal{K}$. This function is extended
to type environments (Definition~\ref{def:typing}) as
$\init(\tau)(v)=\init(\tau(v))$ for every $v\in\domain(\tau)$.
\end{definition}
%
%
\ifthenelse{\boolean{PROOFSONLY}}{
\addtocounter{figure}{1}
}{
\begin{figure}
\begin{center}
\includegraphics{file=state.eps, width=8.0cm, height=7.5cm}
\begin{align*}
   k &= [\overline{\pi} \mapsto \mathtt{Scan},
         \pi_1 \mapsto \mathtt{Angle},
         \pi_2 \mapsto \mathtt{Square},
         \pi_3 \mapsto \mathtt{Circle},
         \pi_4 \mapsto \mathtt{Angle}
        ],\\
\tau_{w_1} &=[\mathtt{f}\mapsto\mathtt{Figure},\mathtt{n}\mapsto
\mathtt{Figure},\Out\mapsto\integer,
\this\mapsto\mathtt{Scan}],\\
  \phi_1&=[\mathtt{f}\mapsto l',\mathtt{n}\mapsto\nil,
\mathtt{out}\mapsto 2,\this\mapsto l],\\
  o_1&= \overline{\pi} \sep [],\\
  o_2&= \pi_2 \sep \left[\begin{array}{l}
    \mathtt{next}\mapsto l',\mathtt{rotation}\mapsto\nil,\\
    \mathtt{side}\mapsto 4,\mathtt{Square.x}\mapsto 3,
    \mathtt{Square.y}\mapsto -5
  \end{array}\right],\\
  o_3&=\pi_2 \sep \left[\begin{array}{l}
    \mathtt{next}\mapsto\nil,\mathtt{rotation}\mapsto l,\\
    \mathtt{side}\mapsto 4,\mathtt{Square.x}\mapsto 3,
    \mathtt{Square.y}\mapsto -5\\
  \end{array}\right],\\
  o_4&=\pi_3 \sep [\mathtt{Circle.x}\mapsto 4,
    \mathtt{Circle.y}\mapsto 3,\mathtt{next}\mapsto\nil,
    \mathtt{radius}\mapsto 3],\\
  \mu_1&=[l\mapsto o_1,l'\mapsto o_2,l''\mapsto o_4],\\
  \mu_2&=[l\mapsto o_2,l'\mapsto o_1],\\
  \mu_3&=[l\mapsto o_2,l'\mapsto o_3],\\
   \sigma_1 &= \phi_1 \sep \mu_1.
\end{align*}
\end{center}
\caption{%
The creation point map $k$ and, for program point $w_1$,
the type environment, a frame, objects and memories
for the program in Figure~\ref{fig:program}.%
}
  \label{fig:state}
\end{figure}
%
}
\begin{example}\label{ex:frames}\label{ex:memories}
\ifcorr{%
Consider again the program in Figure~\ref{fig:program}
and its static information in Figure~\ref{fig:static_information}.
The type environment $\tau_{w_1}$,
a frame $\phi_1$, memories $\mu_1, \mu_2, \mu_3$, objects
$o_1, o_2, o_3, o_4$ and a state $\sigma_1$
for this program at program point $w_1$
with locations
$l,l',l''\in\Loc$ are given in Figure~\ref{fig:state}.
Let also
\[
  \phi_2=[\mathtt{f}\mapsto 2,\mathtt{n}\mapsto l',
     \mathtt{out}\mapsto -2,\this\mapsto l].
\]
Then $\Frame_{\tau_{w_1}}$ contains $\phi_1$
but not $\phi_2$ because
$\mathtt{f}$ is bound to $2$ in $\phi_2$ (while it has class
$\mathtt{Figure}$ in $\tau_{w_1}$).
}

\ifcorrx{%
The object $o_1$ created at
$\overline{\pi}$ has class $k(\overline{\pi})=\mathtt{Scan}$
since $F(\mathtt{Scan})=[]$.
Objects created at $\pi_2$ have class $\mathtt{Square}$ so that
these could be $o_2$ and $o_3$.
Similarly, since $k(\pi_3)=\mathtt{Circle}$,
an object created at $\pi_3$ is $o_4$.
With these objects, $\Memory$ contains the maps $\mu_1, \mu_2, \mu_3$.
With these definitions of $\mu_1$, $\mu_2$ and $\mu_3$, the memory
$\mu_2$ is neither an $l$-update nor an $l'$-update of $\mu_1$ since
$\mu_1(l). \pi = \overline{\pi}$ whereas
$\mu_2(l). \pi = \pi_2$ and also
$\mu_1(l'). \pi = \pi_2$ whereas
$\mu_2(l'). \pi = \overline{\pi}$.
However, as
$\mu_3(l). \pi = \pi_2$ and $\mu_3(l'). \pi = \pi_2$,
we have $\mu_1 =_{l'} \mu_3$ and $\mu_2 =_{l} \mu_3$.
Also, letting $\mu_4 =[l\mapsto o_1,l'\mapsto o_3,l''\mapsto o_4]$,
then we have
$\mu_1 =_{\{l,l',l''\}} \mu_4$.
\qed
}
\end{example}

Type correctness and conservative garbage collection
guarantee that there are no dangling pointers and that
variables may only be bound to locations which contain objects
allowed by the type environment. This is a sensible constraint for
the memory allocated by strongly-typed languages such as
Java~\cite{ArnoldGH00}.
\begin{definition}[Weak Correctness]\label{def:weak_correctness}
Let $\phi\in\Frame_\tau$ and $\mu\in\mathit{Me}\-\mathit{mory}$. We say that
$\phi$ is \emph{weakly $\tau$-correct} \wrt $\mu$ if for every
$v\in\domain(\phi)$ such that $\phi(v)\in\Loc$ we have $\phi(v)\in\domain(\mu)$
and $k((\mu\phi(v)).\pi)\le\tau(v)$.
\end{definition}
We strengthen the correctness notion of Definition~\ref{def:weak_correctness}
by requiring that it also holds for the fields of the objects in memory.
\begin{definition}[$\tau$-Correctness]\label{def:proptotau}
Let $\phi\in\Frame_\tau$ and $\mu\in\Memory$. We say that
$\phi$ is \emph{$\tau$-correct} \wrt $\mu$ and write $\phi\sep\mu:\tau$, if
\begin{enumerate}
\item $\phi$ is weakly $\tau$-correct \wrt $\mu$ and,
\item for every $o\in\codom(\mu)$, $o.\phi$ is weakly
$F(k(o.\pi))$-correct \wrt $\mu$.
\end{enumerate}
\end{definition}
\begin{example}\label{ex:tau_correctness}
\ifcorr{%
Let $\tau_{w_1}$, $\phi_1$, $\mu_1$, $\mu_2$ and $\mu_3$
be as in Figure~\ref{fig:state}.
\begin{itemize}
\item $\phi_1 \sep \mu_1:\tau_{w_1}$.
Condition 1 of Definition~\ref{def:proptotau} holds because
\begin{gather*}
  \{v\in\domain(\phi_1)\mid\phi_1(v)\in\Loc\}
    = \{ \this,\mathtt{f} \},\\
  \{\phi_1(\this), \phi_1(\mathtt{f})\}
    = \{l,l'\} \subseteq \domain(\mu_1),\\
  k(\mu_1(l).\pi)
    = k(o_1.\pi)
    = k(\overline{\pi})
    = \mathtt{Scan}
    = \tau_{w_1}(\this),\\
  k(\mu_1(l').\pi)
    = k(o_2.\pi)
    = k(\pi_2)
    = \mathtt{Square}
    \le \mathtt{Figure}
    = \tau_{w_1}(\mathtt{f}).
\end{gather*}
 Condition 2 of Definition~\ref{def:proptotau} holds because
\begin{gather*}
  \codom(\mu_1) = \{o_1,o_2,o_4\},\\
  \codom(o_1.\phi) =
  \codom(o_4.\phi)\cap\Loc = \emptyset,\\
  \codom(o_2.\phi)\cap\Loc = \{l'\}
    \sseq \domain(\mu_1),\\
  k(\mu_1(l').\pi)
    = k(o_2.\pi)
    = \mathtt{Square}
    \le \mathtt{Figure}
    = F(o_2.\pi)(\mathtt{next}).
\end{gather*}
\item $\phi_1 \sep \mu_2:\tau_{w_1}$ does not hold, since
condition 1 of Definition~\ref{def:proptotau} does not hold. Namely,
$\tau_{w_1}(\this)=\mathtt{Scan}$,
$k((\mu_2\phi_1(\this)).\pi)=k(o_2.\pi)=k(\pi_2)=\mathtt{Square}$ and
$\mathtt{Square}\not\le\mathtt{Scan}$.
\item $\phi_1 \sep \mu_3:\tau_{w_1}$ does not hold, since
condition 2 of Definition~\ref{def:proptotau} does not hold. Namely,
$o_3\in\codom(\mu_3)$ and $o_3.\phi$ is not $F(k(o_3.\pi))$-correct
\wrt $\mu_3$, since we have that
$o_3.\phi(\mathtt{rotation})=l$, $\mathtt{Square}\not\le\mathtt{Angle}$
but $k(\mu_3(l).\pi)=k(o_2.\pi)=k(\pi_2)=\mathtt{Square}$ and moreover
$F(k(o_3.\pi))(\mathtt{rotation})
=F(\mathtt{Square})(\mathtt{rotation})=\mathtt{Angle}$.
\end{itemize}
\qed
}
\end{example}

Definition~\ref{def:concrete_states} defines
the state of the computation as a pair consisting of a frame and a memory.
The variable $\this$ in the domain of the frame
must be bound to an object. In particular, it cannot be $\nil$.
This condition could be relaxed in Definition~\ref{def:concrete_states}.
This would lead to simplifications in
the following sections (such as in Definition~\ref{def:delta}). However,
our condition is consistent with the specification of the Java programming
language~\cite{ArnoldGH00}. Note, however, that there is no such hypothesis
about the local variable number $0$ of the Java Virtual
Machine, which stores the $\mathtt{this}$ object~\cite{LindholmY99}.
\begin{definition}[State]\label{def:concrete_states}
If $\tau$ is a type environment associated with a program point,
the set of possible \emph{states} of a computation at that point is
any subset of
\[
  \Sigma_\tau=\left\{\phi\sep\mu\left|\begin{array}{l}
    \phi\in\Frame_\tau,\ \mu\in\Memory,\ \phi\sep\mu:\tau,\\
    \text{if }\this\in\domain(\tau)\text{ then }\phi(\this)
      \neq\nil\end{array}\right.\right\}.
\]
\end{definition}
\begin{example}\label{ex:states}
\ifcorr{%
Let $\tau_{w_1}$, $\phi_1$, $\mu_1$, $\mu_2$ and $\mu_3$
be as in Figure~\ref{fig:state}.
Then, in Example~\ref{ex:tau_correctness}, we have shown
that $\phi_1\sep\mu_1:\tau_{w_1}$
holds and that $\phi_1\sep\mu_2:\tau_{w_1}$ and $\phi_1\sep\mu_3:\tau_{w_1}$
do not hold.
Thus, at program point $w_1$,
we have $\phi_1\sep\mu_1\in\Sigma_{\tau_{w_1}}$,
$\phi_1\sep\mu_2\not\in\Sigma_{\tau_{w_1}}$
and $\phi_1\sep\mu_3\not\in\Sigma_{\tau_{w_1}}$.
\qed
}
\end{example}
The frame of an object $o$ in memory is
itself a state for the instance variables of $o$.
\begin{proposition}\label{prop:recursive}
Let $\phi\sep\mu\in\Sigma_\tau$ and
$o\in\codom(\mu)$. Then $(o.\phi)\sep\mu\in\Sigma_{F(k(o.\pi))}$.
\end{proposition}
\myproofbis{
Since $\phi\sep\mu\in\Sigma_\tau$, from Definition~\ref{def:concrete_states}
we have $\phi\sep\mu:\tau$. From
Definition~\ref{def:proptotau} we know that $o.\phi$ is weakly
$F(k(o.\pi))$-correct \wrt $\mu$ so that
$(o.\phi)\sep\mu:F(k(o.\pi))$. Since $\mathtt{this}\not\in\domain(F(k(o.\pi)))$
(Definition~\ref{def:static_information})
we conclude that $(o.\phi)\sep\mu\in\Sigma_{F(k(o.\pi))}$.}
\subsection{The Operations over the Concrete States}
  \label{subsec:concrete_operations}
\begin{figure}[ht]
{\small
\begin{center}
$\begin{array}{|rl|l|}
  \hline
  \multicolumn{2}{|c|}{\text{Operation}} &
    \multicolumn{1}{c|}{\text{Constraint ($\this\in\domain(\tau)$
    always)}}\\
  \hline\hline
  \mathsf{nop}_\tau&:\Sigma_\tau\mapsto \Sigma_\tau &\\\hline
  \mathsf{get\_int}^i_\tau&:\Sigma_\tau\mapsto\Sigma_{\tau[\rs\mapsto
    \integer]}
    &\rs\not\in\domain(\tau),\ i\in\integers\\\hline
  \mathsf{get\_null}^\kappa_\tau&:
    \Sigma_\tau\mapsto\Sigma_{\tau[\rs\mapsto\kappa]}
    &\rs\not\in\domain(\tau),\ \kappa\in\mathcal{K}\\\hline
  \mathsf{get\_var}^v_\tau&:\Sigma_\tau\mapsto \Sigma_{\tau[\rs\mapsto
    \tau(v)]}&\rs\not\in\domain(\tau),\ v\in\domain(\tau)\\\hline
  \mathsf{get\_field}^f_\tau&:
    \Sigma_\tau\to\Sigma_
    {\tau[\rs\mapsto i(f)]}&\rs\in\domain(\tau),\ \tau(\rs)
    \in\mathcal{K},\\
  && i=F\tau(res),\ f\in\domain(i)\\\hline
  \mathsf{put\_var}^v_\tau&:\Sigma_\tau\mapsto\Sigma_{\tau|_{-\rs}} &
    \rs\in\domain(\tau),\ v\in\domain(\tau),\\
  && v\neq\rs,\ \tau(\rs)
    \le\tau(v)\\\hline
  &&\rs\in\domain(\tau),\ \tau(\rs)\in\mathcal{K}\\
  \mathsf{put\_field}^f_{\tau,\tau'}&:\Sigma_\tau\mapsto\Sigma_{\tau'}\to
    \Sigma_{\tau|_{-\rs}}&f\in\domain(F\tau(\rs))\\
  &&\tau'=\tau[\rs\mapsto t]\text{ with }t\le(F\tau(\rs))(f)\\\hline
  \mathsf{=}_\tau,\mathsf{+}_\tau&:\Sigma_\tau\mapsto\Sigma_\tau\mapsto
    \Sigma_\tau & \rs\in\domain(\tau),\ \tau(\rs)=\integer\\\hline
  \mathsf{is\_null}_\tau&:\Sigma_\tau\mapsto\Sigma_{\tau[\rs\mapsto\integer]}
    & \rs\in\domain(\tau),\ \tau(\rs)\in\mathcal{K}\\\hline
  &&\rs\in\domain(\tau),\ \tau(\rs)\in\mathcal{K},\\
  && \{v_1,
    \ldots,v_n\}\subseteq\domain(\tau),\ \nu\in\mathcal{M}\\
  \mathsf{call}_\tau^{\nu,v_1,\ldots,v_n}\!\!\!&:
    \Sigma_\tau\mapsto\Sigma_{P(\nu)|_{-\Out}} &
    \domain(P(\nu))\!\setminus\!\{\Out,\!\this\}\!=\!
    \{\iota_1,\ldots,\iota_n\!\}\\
  &&\text{(alphabetically ordered)}\\
  &&\tau(\rs)\le P(\nu)(\this)\\
  && \tau(v_i)\le P(\nu)(\iota_i)\text{ for }i=1,\ldots,n\\\hline
  \mathsf{return}_\tau^\nu\!:\Sigma_\tau&\!\!\!\mapsto\Sigma_
    {p|_{\Out}}\!\!\!\to\Sigma_{\tau[\rs\mapsto p(\Out)]}
  &\rs\in\domain(\tau),\ \nu\in\mathcal{M},\ p=P(\nu)\\\hline
  \mathsf{restrict}_\tau^{\mathit{vs}}&:\Sigma_\tau\mapsto
    \Sigma_{\tau|_{-\mathit{vs}}}
  &\mathit{vs}\subseteq\domain(\tau)\\\hline
  \mathsf{expand}_\tau^{v:t}&:\Sigma_\tau\mapsto\Sigma_{\tau[v\mapsto t]}
  &v\in\Vars,\ v\not\in\domain(\tau),\ t\in\Type\\\hline
  \mathsf{new}_\tau^\pi&:\Sigma_\tau\mapsto\Sigma_{\tau[\rs\mapsto k(\pi)]}
  &\rs\not\in\domain(\tau),\ \pi\in\Pi\\\hline
  &&\rs\!\in\!\domain(\tau),\ \tau(\rs)\!\in\!\mathcal{K},\\
  && m\!\in\!\domain(M\tau(\rs)),\ \nu\!\in\!\mathcal{M}\\
  \mathsf{lookup}^{m,\nu}_\tau&:\Sigma_\tau\!\!\to\!
    \Sigma_{\tau[\rs\mapsto P(\nu)(\this)]} &
  \text{for every suitable $m$, $\sigma$ and $\tau$,}\\
  && \text{there is at most one $\nu$}\\
  &&\text{such that }
    \mathsf{lookup}_\tau^{m,\nu}(\sigma)\text{ is defined}\\\hline
  \mathsf{is\_true}_\tau&:\Sigma_\tau\to\Sigma_{\tau|_{-\rs}} &
    \rs\in\domain(\tau),\ \tau(\rs)=\integer,\\
  \mathsf{is\_false}_\tau&:\Sigma_\tau\to\Sigma_{\tau|_{-\rs}} &
    \domain(\mathsf{is\_true}_\tau)\cap
      \domain(\mathsf{is\_false}_\tau)=\emptyset\\
  &&\domain(\mathsf{is\_true}_\tau)
    \cup\domain(\mathsf{is\_false}_\tau)=\Sigma_\tau\\\hline
\end{array}$
\end{center}
}
\caption{The signature of the
  operations over the states.}\label{fig:signatures}
\end{figure}

\begin{figure}[t]
{\scriptsize
\begin{gather}
\begin{align*}
  \mathsf{nop}_\tau(\phi\sep\mu)&=\phi\sep\mu\\
  \mathsf{get\_int}_\tau^i(\phi\sep\mu)&=
      \phi[\rs\mapsto i]\sep\mu\\
  \mathsf{get\_null}^\kappa_\tau
    (\phi\sep\mu)&=\phi[\rs\mapsto\nil]\sep\mu\\
  \mathsf{get\_var}_\tau^v
    (\phi\sep\mu)&=\phi[\rs\mapsto\phi(v)]\sep\mu\\
  \mathsf{restrict}_\tau^{\mathit{vs}}(\phi\sep\mu)&=
    \phi|_{-\mathit{vs}}\sep\mu\\ 
  \mathsf{expand}_\tau^{v:t}(\phi\sep\mu)&=
    \phi[v\mapsto\init(t)]\sep\mu\\
  \mathsf{put\_var}_\tau^v(\phi\sep\mu)
      &=\phi[v\mapsto\phi(\rs)]|_{-\rs}\sep\mu\\
  \mathsf{get\_field}_\tau^f(\phi'\sep\mu)&=
    \begin{cases}
      \phi'[res\mapsto((\mu\phi'(\rs)).\phi)(f)]\sep\mu &
        \text{if $\phi'(\rs)\neq \nil$}\\
      \text{undefined} & \text{otherwise}
    \end{cases}\\
   \begin{array}{c}
     \mathsf{put\_field}_{\tau,\tau'}^f\\
     (\phi_1\sep\mu_1)(\phi_2\sep\mu_2)
   \end{array}&=
    \begin{cases}
      \phi_2|_{-\rs}\sep\mu_2[l\mapsto\mu_2(l).\pi\sep\mu_2(l).
        \phi[f\mapsto\phi_2(\rs)]] &\\
      \qquad\text{if $l=\phi_1(\rs), l\neq \nil$ and $\mu_1=_l\mu_2$}\\
      \text{undefined}\quad\text{otherwise}&
    \end{cases}\\
  \mathsf{=}_\tau(\phi_1\sep\mu_1)
    (\phi_2\sep\mu_2)&=
    \begin{cases}
      \phi_2[\rs\mapsto 1]\sep\mu_2 & \text{if $\phi_1(\rs)=\phi_2(\rs)$}\\
      \phi_2[\rs\mapsto -1]\sep\mu_2 &
        \text{if $\phi_1(\rs)\neq \phi_2(\rs)$}
    \end{cases}\\
  \mathsf{+}_\tau(\phi_1\sep\mu_1)(\phi_2\sep\mu_2)&=
      \phi_2[\rs\mapsto\phi_1(\rs)+\phi_2(\rs)]\sep\mu_2\\
  \mathsf{is\_null}_\tau(\phi\sep\mu)&=
    \begin{cases}
      \phi[\rs\mapsto 1]\sep\mu & \text{if $\phi(\rs)=\nil$}\\
      \phi[\rs\mapsto -1]\sep\mu & \text{otherwise}
    \end{cases}\\
  \mathsf{call}_\tau^{\nu,v_1,\ldots,v_n}
    (\phi\sep\mu)&=
      [\iota_1\mapsto\phi(v_1),\ldots,\iota_n\mapsto\phi(v_n),
        \this\mapsto\phi(\rs)]\sep\mu\\
  \text{where $\{\iota_1,\ldots,\iota_n\}$}&=\text{$P(\nu)\setminus
    \{\Out,\this\}$ (alphabetically ordered)}\\
  \begin{array}{c}
    \mathsf{return}_\tau^\nu\\
    (\phi_1\sep\mu_1)(\phi_2\sep\mu_2)
  \end{array}&=
    \begin{cases}
      \phi_1[\rs\mapsto\phi_2(\Out)]\sep\mu_2\\
      \quad \text{if $L = \codom(\phi_1)|_{-\rs}\cap\Loc$ and
              $\mu_1=_L \mu_2$}\\
      \mbox{}\\
      \text{undefined}\quad\text{otherwise}
    \end{cases}\\
  \mathsf{new}_\tau^\pi(\phi\sep\mu)&=
    \phi[\rs\mapsto l]\sep\mu[l\mapsto
    \pi\sep\init(F(k(\pi)))],\ l\in\Loc\setminus\domain(\mu)\\
  \mathsf{lookup}^{m,\nu}_\tau(\phi\sep\mu)&=\begin{cases}
    \phi\sep\mu\\
    \quad\text{if $\phi(\rs)\neq \nil$ and
      $M(k((\mu\phi(\rs)).\pi))(m)=\nu$}\\
    \mbox{}\\
    \text{undefined}\quad\text{otherwise}
  \end{cases}\\
  \mathsf{is\_true}_\tau(\phi\sep\mu)&=\begin{cases}
    \phi|_{-\rs}\sep\mu & \text{if $\phi(\rs)\ge 0$}\\
    \text{undefined} & \text{otherwise}
  \end{cases}\\
  \mathsf{is\_false}_\tau(\phi\sep\mu)&=\begin{cases}
    \phi|_{-\rs}\sep\mu & \text{ if $\phi(\rs)<0$}\\
    \text{undefined} & \text{otherwise.}
  \end{cases}
\end{align*}\end{gather}}\normalsize
\caption{The operations over concrete states.}\label{fig:concrete_states}
\end{figure}

Figures~\ref{fig:signatures} and~\ref{fig:concrete_states} show the
signatures and the definitions, respectively, of a set of operations over
the concrete states for a type environment $\tau$.
The variable $\rs$ holds intermediate results, as we said at the beginning of
this section.
We briefly introduce these operations.

\begin{itemize}
\item
The $\mathsf{nop}$ operation does nothing.
\item
A $\mathsf{get}$ operation loads into $\rs$ a constant,
the value of another variable or the value of the field of an object.
In the last case ($\mathsf{get\_field}$),
that object is assumed to be stored in $\rs$
\emph{before} the $\mathsf{get}$ operation.
Then $(\mu\phi'(\rs))$ is the object whose field $f$ must be read,
$(\mu\phi'(\rs)).\phi$ are its fields and $(\mu\phi'(\rs)).\phi(f)$ is the
value of the field named $f$.
\item
A $\mathsf{put}$ operation stores in $v$ the value of $\rs$ or
of a field of an object pointed to by $\rs$.
Note that, in the second case,
$\mathsf{put\_field}$ is a binary operation since the
evaluation of $e_1.f=e_2$ from an initial state $\sigma_1$
works by first evaluating $e_1$ from $\sigma_1$, yielding an intermediate
state $\sigma_2$, and then evaluating $e_2$ from $\sigma_2$, yielding a
state $\sigma_3$. The final state is then
$\mathsf{put\_field}(\sigma_2)(\sigma_3)$~\cite{SpotoJ03}, where the
variable $\rs$ of $\sigma_2$ holds the value of $e_1$ and the variable
$\rs$ of $\sigma_3$ holds the value of $e_2$.
The object whose field is modified
must still exist in the memory of $\sigma_3$.
This is expressed by the update relation (Definition~\ref{def:domains2}).
As there is no result, $\rs$ is removed.
Providing two states \ie two frames and
two heaps for $\mathsf{put\_field}$ and, more generally, for
binary operations, may look like an overkill and it might be expected that
a single state and a single frame would be enough. However,
our decision to have two states has been dictated by the intended
use of this semantics
\ie abstract interpretation. By \emph{only} using operations over states,
we have exactly one concrete domain, which can be abstracted into just one
abstract domain. Hybrid operations, working on states and frames,
would only complicate the abstraction.
%
\item
For every binary operation such as $=$ and $+$ over values, there is an
operation on states. Note that (in the case of $\mathsf{=}$)
Booleans are implemented by means of
integers (every non-negative integer means true).
We have already explained why we use two states for binary operations.
\item
The operation $\mathsf{is\_null}$ checks that $\rs$ points
to $\nil$.
\item
The operation $\mathsf{call}$ is used before, and the operation
$\mathsf{return}$ is used after, a call to a method $\nu$.
While $\mathsf{call}^\nu$ creates a new state in which $\nu$ can execute,
the operation $\mathsf{return}^\nu$ restores
the state $\sigma$ which was current
before the call to $\nu$, and stores in $\rs$ the result of the call.
As said in (the beginning of) Section~\ref{sec:framework},
the denotation of the method is taken from an \emph{interpretation},
in a denotational fashion~\cite{BossiGLM94}. Hence the execution
from an initial state $\sigma_1$ of
a method call denoted, in the current interpretation,
by $d:\Sigma\to\Sigma$, yields the final state
$\mathsf{return}(\sigma_1)(d(\mathsf{call}(\sigma_1)))$.
Note that $\mathsf{return}$ is a binary operation whose first argument is
the state of the caller at call-time and whose second argument is
the state of the callee at return-time. Its definition in
Figure~\ref{fig:concrete_states} restores the state of the caller but
stores in $\rs$ the return value of the callee. By using
a binary operation we can define our semantics in terms of states
rather than in terms of activation stacks. This is a useful simplification when
passing to abstraction, since states must be abstracted rather than stacks.
Note that the update relation (Definition~\ref{def:domains2}) requires that the
variables of the caller have not been changed during the execution of the
method (although the fields of the objects bound to those
variables may be changed).
\item
The operation $\mathsf{expand}$ ($\mathsf{restrict}$) adds (removes)
variables.
\item
The operation $\mathsf{new}^\pi$ creates a new object $o$
of creation point $\pi$.
A pointer to $o$ is put in $\rs$. Its fields
are initialised to default values.
\item
The operation $\mathsf{lookup}^{m,\nu}$ checks if, by calling the method
identified by $m$ of the object $o$
pointed to by $\rs$, the method $\nu$ is run.
This depends on the class $k(o.\pi)$ of $o=\mu\phi(\rs)$.
\item
The operation $\mathsf{is\_true}$ ($\mathsf{is\_false}$)
checks if $\rs$ contains true (false).
\end{itemize}
\begin{example}\label{ex:concrete_operations}
\ifcorr{%
Consider again the example in Figure~\ref{fig:program}.
Let $\tau = \tau_{w_1}$, $\phi_1$, $\mu_1$ and $\sigma_1 = \phi_1\sep\mu_1$
be as in Figure~\ref{fig:state}
so that, as indicated in Example~\ref{ex:states},
state $\sigma_1$ could be the current state at program point $w_1$.
The computation continues as follows~\cite{SpotoJ03}.
\[
  \sigma_2=\mathsf{get\_var}_\tau^\mathtt{f}(\sigma_1) \qquad
    \text{read $\mathtt{f}$.}
\]
Let $o_1, o_2, o_3, o_4$ be as in Figure~\ref{fig:state}. Then
$\sigma_2=\phi_1 \left[\rs\mapsto \phi_1(f)\right]\sep \mu_1
           =\phi_1 \left[\rs\mapsto l'\right]\sep \mu_1
          =[
    \mathtt{f}\mapsto l',\mathtt{n}\mapsto\nil,
    \Out\mapsto 2,\rs\mapsto l',\this\mapsto l]\sep \mu_1$.
The $\mathsf{lookup}$ operations determine which is the target
of the first virtual call $\mathtt{f.def()}$ in Figure~\ref{fig:program}.
As a result only one of the following blocks of code is run
depending on which $\mathsf{lookup}$ check is defined.
\[
\left.\begin{array}{rll}
  \sigma_3'=&\mathsf{lookup}_{\tau[\rs\mapsto\mathtt{Figure}]}
    ^{\mathtt{def},\mathtt{Figure.def}}
    (\sigma_2) & \text{select $\mathtt{Figure.def}$}\\
  \sigma_4'=&\mathsf{call}_{\tau[\rs\mapsto P(\mathtt{Figure.def})
    (\this)]}^\mathtt{Figure.def}
    (\sigma_3') & \text{initialise the $\mathtt{Figure}$}\\
  \sigma_5'=&\text{the final state of $\mathtt{Figure.def}$ from $\sigma_4'$}\\
  \sigma_6'=&\mathsf{return}_{\tau[\rs\mapsto P(\mathtt{Figure.def})
    (\this)]}^\mathtt{Figure.def}(\sigma_3')(\sigma_5')
    & \text{return to the caller}\\
  \mbox{}\\
  \sigma_3''=&\mathsf{lookup}_{\tau[\rs\mapsto\mathtt{Figure}]}
    ^{\mathtt{def},\mathtt{Square.def}}
    (\sigma_2) & \text{select $\mathtt{Square.def}$}\\
  \sigma_4''=&\mathsf{call}_{\tau[\rs\mapsto P(\mathtt{Square.def})
    (\this)]}^\mathtt{Square.def}
    (\sigma_3'') & \text{initialise the $\mathtt{Square}$}\\
  \sigma_5''=&\text{the final state of $\mathtt{Square.def}$
    from $\sigma_4''$}\\
  \sigma_6''=&\mathsf{return}_{\tau[\rs\mapsto P(\mathtt{Square.def})
    (\this)]}^\mathtt{Square.def}(\sigma_3'')(\sigma_5'')
    & \text{return to the caller}\\
  \mbox{}\\
  \sigma_3'''=&\mathsf{lookup}_{\tau[\rs\mapsto\mathtt{Figure}]}
    ^{\mathtt{def},\mathtt{Circle.def}}
    (\sigma_2) & \text{select $\mathtt{Circle.def}$}\\
  \sigma_4'''=&\mathsf{call}_{\tau[\rs\mapsto P(\mathtt{Circle.def})
    (\this)]}^\mathtt{Circle.def}
    (\sigma_3''') & \text{initialise the $\mathtt{Circle}$}\\
  \sigma_5'''=&\text{the final state of $\mathtt{Circle.def}$
    from $\sigma_4'''$}\\
  \sigma_6'''=&\mathsf{return}_{\tau[\rs\mapsto P(\mathtt{Circle.def})
    (\this)]}^\mathtt{Circle.def}(\sigma_3''')(\sigma_5''')
    & \text{return to the caller.}
\end{array}\right.
\]
The states $\sigma_5'$, $\sigma_5''$ and $\sigma_5'''$ are computed
from the current intepretation for the methods.
For each $\mathsf{lookup}$ operation, we have
$(\sigma_2.\phi)(\rs)=l'\neq \nil$ and $(\sigma_2.\mu)(l')=o_2$; then the
method of $o_2$ identified by $\mathtt{def}$ is called.
Now $o_2.\pi\!=\!\pi_2$ and $k(\pi_2)\!=\!\mathtt{Square}$. Moreover
$M(\mathtt{Square})(\mathtt{def})=\mathtt{Square.def}$
(Figure~\ref{fig:static_information}).
So the only \emph{defined} $\mathsf{lookup}$
operation is that for $\mathtt{Square.def}$. This means
that $\mathtt{Square.def}$ is called and $\sigma_3''=\sigma_2$,
while $\sigma_3'$ and $\sigma_3'''$ are undefined.
}

\ifcorrx{%
We obtain $\sigma_4''=[\this\mapsto l']\sep \mu_1$.
Note that the object $o_2$ 
is now the $\this$ object of this instantiation of the method
$\mathtt{Square.def}$. To compute $\sigma_5''$, we should execute
the operations which implement $\mathtt{Square.def}$, starting from the
state $\sigma_4''$. For simplicity,
we report
only
the final state of this execution which is
\[
  \sigma_5''=[\mathtt{out}\mapsto 0]\sep
     \underbrace{[l\mapsto o_1,l'\mapsto o_5,l''\mapsto o_4]}_{\mu_5},
\]
where
\begin{align*}
   o_5&=\pi_2\sep\left[\begin{array}{l}
    \mathtt{next}\mapsto l',\mathtt{side}\mapsto 1,
      \mathtt{Square.x}\mapsto 0,\\
    \mathtt{Square.y}\mapsto 0,\mathtt{rotation}\mapsto o_6
   \end{array}\right],\\
   o_6&=\pi_1\sep[\mathtt{degree}\mapsto 0].
\end{align*}
The $\mathsf{return}$ operation returns the control to the caller method.
This means that the frame will be that of the caller, but the return value
of the callee is copied into the $\rs$ variable of the caller. Namely,
\[
  \sigma_6''=\left[\begin{array}{l}
    \mathtt{f}\mapsto l',\mathtt{n}\mapsto\nil,\\
    \Out\mapsto 2,\rs\mapsto 0,\this\mapsto l
  \end{array}\right]\sep\mu_5.
\]
\qed
}
\end{example}
\subsection{The Collecting Semantics}\label{subsec:collecting}
The operations of Figure~\ref{fig:concrete_states} can be used to define
the transition function from states to states, or \emph{denotation},
of a piece of code $c$, as shown in
Example~\ref{ex:concrete_operations}.
By use of $\mathsf{call}$ and $\mathsf{return}$,
there is a denotation for each method called in $c$;
thus, by adding $\mathsf{call}$ and $\mathsf{return}$,
we can plug the method's denotation in the calling points inside $c$
(as shown in Subsection~\ref{subsec:concrete_operations} and in
Example~\ref{ex:concrete_operations}).
A function $I$ binding each method $\mathtt{m}$
in a program $P$ to its denotation
$I(\mathtt{m})$ is called an \emph{interpretation} of $P$.
Given an interpretation $I$, we are hence able to define
the denotation $T_P(I)(\mathtt{m})$ of the body of a method $\mathtt{m}$,
so that we are able to transform $I$ into a new interpretation $T_P(I)$.
This leads to the definition of the \emph{denotational semantics} of $P$
as the minimal (\ie less defined) interpretation which is a fixpoint
of $T_P$.
This way of defining the concrete semantics in a denotational way
through interpretations,
is useful for a subsequent abstraction~\cite{CousotC92}.
The technique, which has been extensively used
in the logic programming tradition~\cite{BossiGLM94},
has been adapted in~\cite{SpotoJ03} for object-oriented
imperative programs by adding the mechanism for dynamic dispatch
through the $\mathsf{lookup}$ operation in Figure~\ref{fig:concrete_states}.
Note that the fixpoint of $T_P$ is
not finitely computable in general, but it does exist as a consequence
of Tarski's theorem and it is the limit of the ascending chain of
interpretations $I_0$, $T_P(I_0)$, $T_P(T_P(I_0))$, \ldots,
where, for every method $\mathtt{m}$, the denotation
$I_0(\mathtt{m})$ is always undefined~\cite{Ta55}.

The concrete semantics described above denotes each method with a map on
states \ie a function from $\Sigma$ to $\Sigma$.
However, abstract interpretation is interested in
\emph{properties} of states; so that each property of interest,
is identified with the set of all the states satisfying that property.
This leads to the definition of a \emph{collecting}
semantics~\cite{CousotC77,CousotC92} \ie
a concrete semantics working over the powerset $\wp(\Sigma)$.
The operations of this collecting semantics are the powerset
extension of the operations in Figure~\ref{fig:concrete_states}.
For instance, $\mathsf{get\_int}_\tau^i$ is extended into
\[
  \mathsf{get\_int}_\tau^i(S)=\{\mathsf{get\_int}_\tau^i(\sigma)\mid
    \sigma\in S\}
\]
for every $S\in\wp(\Sigma_\tau)$.
Note that dealing with powersets means that the semantics becomes
non-determi\-ni\-stic. For instance, in Example~\ref{ex:concrete_operations}
more than one target of the $\mathtt{f.def()}$ virtual call could
be selected at the same time and more than one of the blocks of code
could be executed. Hence we need a $\cup$ operation over sets
of states which merges different threads of execution at the end of a virtual
call (or, for similar motivations, at the end of a conditional).
The notion of denotation now becomes a map over $\wp(\Sigma_\tau)$.
Interpretations and the transformer on interpretations are defined exactly
as above.
We will assume the result, proved in~\cite{SpotoJ03}, that every abstraction of
$\wp(\Sigma_\tau)$, $\mathord{\cup}$ and of the powerset extension
of the operations in Figure~\ref{fig:concrete_states}
induces an abstraction of the concrete collecting semantics.
This is an application to object-oriented imperative programs of the
\emph{fixpoint transfer} Proposition~27 in~\cite{CousotC92}.
Two such abstractions will be described in Sections~\ref{sec:edomain}
and~\ref{sec:erdomain}.
\section{The Basic Domain $\e$}\label{sec:edomain}
We define here a basic abstract domain $\e$ as a
property of the concrete states of Definition~\ref{def:concrete_states}.
Its definition is guided by our goal to
\emph{over}approximate, for every program point $p$,
the set of creation points of objects reachable at $p$
from some variable or field in scope.
Thus an element of the abstract domain $\e$ which decorates a program point $p$
is simply
a set of creation points of objects that may be reached at $p$.
The choice of an \emph{over}approximation
follows from the typical use of the information provided by
an escape analysis. For instance, an object can be stack allocated
if it does not escape the method
which creates it \ie if it does not belong to a superset of the objects
reachable at its end.
Moreover, our goal is to stack allocate specific creation points.
Hence, we are not interested in the identity of the objects but in
their creation points.

Although, at the end of this section, we will see that $\e$ induces
rather imprecise abstract operations, its definition is important since
$\e$ comprises exactly the information needed to implement our escape
analysis. Even though its abstract operations lose
precision, we still need $\e$ as a basis for
comparison and as a minimum requirement for new, improved domains
for escape analysis. Namely, in Section~\ref{sec:erdomain}
we will define a more precise abstract domain $\er$
for escape analysis, and we will prove
(Proposition~\ref{prop:inclusion}) that it strictly contains $\e$.
This situation is similar to that of the abstract domain $\mathcal{G}$
for groundness analysis of logic programs~\cite{Sondergaard86} which,
although imprecise, expresses the property looked for by the analysis,
and is the basis of all the other abstract domains for groundness
analysis, derived as \emph{refinements} of $\mathcal{G}$~\cite{Scozzari00}.
The definition of more precise abstract
domains as refinements of simpler ones is actually standard methodology
in abstract interpretation nowadays~\cite{GiacobazziR97}.
Another example is strictness analysis of functional programs,
where a first simple domain is subsequently enriched to express more precise
information~\cite{Jensen97}.
A similar idea has also been applied to model-checking, through a
sequence of refinements of a simple abstract domain~\cite{Dams96}.
A \emph{refinement}, in this context, is just an operation that transforms
a simpler domain into a richer one \ie one containing more abstract elements.
There are many standard refinements operations.
One of this is \emph{reduced product}, which allows one to compose
two abstract domains in order to express the composition of the
properties expressed by the two domains, and \emph{disjunctive
completion}, which enriches an abstract domain with the ability
to express disjunctive information about the properties expressed
by the domain~\cite{CousotC79}. Another example is the
\emph{linear refinement} of a domain \wrt another, which expresses
the dependencies of the abstract properties expressed by the two
domains~\cite{GiacobazziS98}.
In Section~\ref{sec:erdomain}
we use a refinement which is significant for imperative programs, where
assignments to program variables are the pervasive operation.
Hence, a variable-based approximation often yields improved precision
\wrt a global approximation of the state, such
as expressed by $\e$.
This same refinement is used, for instance, when passing from
rapid type analysis to a variable-based \emph{class analysis} of
object-oriented imperative programs in~\cite{SpotoJ03}.

We show an example now that clarifies the idea of \emph{reachability}
for objects
at a program point.
\begin{example}\label{ex:escape_analysis}
\ifcorr{%
Consider the program in Figure~\ref{fig:program}
and the type environment $\tau_{w_1}$ for program point $w_1$
given in Figure~\ref{fig:state}.
We show that no objects created at $\pi_4$
will be reachable at program point $w_1$.
The type environment
at $w_1$, which is $\tau_{w_1}$,
shows that we cannot reach any object from $\Out$,
since $\Out$ can only contain integers.
The variable $\this$ has class $\mathtt{Scan}$ which has no fields.
Since in $\pi_4$ we create objects of class $\mathtt{Angle}$,
they cannot be reached from $\this$.
The variables $\mathtt{f}$ and $\mathtt{n}$ have class $\mathtt{Figure}$
whose only field has type $\mathtt{Figure}$ itself.
Reasoning as for $\this$, we could falsely conclude that no object
created at $\pi_4$ can be reached from $\mathtt{f}$ or $\mathtt{n}$.
This conclusion is false
since, as $\mathtt{Square} \le \mathtt{Figure}$, the call
$\mathtt{rotate(f)}$ could result in a call
in the class $\mathtt{Square}$ to the method $\mathtt{f.rot(a)}$
which stores $\mathtt{a}$,
created at $\pi_4$, in the field $\mathtt{rotation}$;
and $\mathtt{rotation}$
is still accessible to other methods such as $\mathtt{f.draw()}$.
\qed
}
\end{example}

The reasoning in Example~\ref{ex:escape_analysis} leads to
the notion of \emph{reachability} in Definition~\ref{def:reachability}
where we use
the actual fields of the objects instead of those of the declared class
of the variables.
\begin{definition}[Reachability]\label{def:reachability}
Let $\sigma=\phi\sep\mu\in\Sigma_\tau$ and
$S\subseteq\Sigma_\tau$. The set of the objects \emph{reachable} in $\sigma$ is
$O_\tau(\sigma)=\cup\{O_\tau^i(\sigma)\mid i\ge 0\}$ where
\begin{align*}
  O_\tau^0(S)&=\emptyset\\
  O_\tau^{i+1}(S)&=\bigcup\left\{\{o\}\cup O^i_{F(k(o.\pi))}(o.\phi\sep\mu)
    \left|\begin{array}{l}
      \phi\sep\mu\in S,\ v\in\domain(\tau)\\
      \phi(v)\in\Loc,\ o=\mu\phi(v)
    \end{array}\right.\right\}.
\end{align*}
The maps $O_\tau^i$ are extended to $\wp(\Sigma_\tau)$ as
$O_\tau^i(S)=\cup\{O_\tau^i(\sigma)\mid\sigma\in S\}$.
\end{definition}
Proposition~\ref{prop:recursive}
provides a guarantee that Definition~\ref{def:reachability}
is well-defined.
Observe that variables and fields of type $\integer$
do not contribute to $O_\tau$.
We can now define the abstraction map for $\e$.
It selects the creation points of the reachable objects.
\begin{definition}[Abstraction Map for $\e$]\label{def:alpha}
Let $S\subseteq\Sigma_\tau$.
The \emph{abstraction map} for $\e$ is
\[
   \alpha_\tau^\e(S)=\{o.\pi\mid\sigma\in S\text{ and }
     o\in O_\tau(\sigma)\}\subseteq\Pi.
\]
\end{definition}
\begin{example}\label{ex:e_domain}
\ifcorr{%
Let $\phi_1$, $\mu_1$, $\sigma_1 = \phi_1 \sep \mu_1$, $o_1$ and
$o_2$ be as defined in Figure~\ref{fig:state}. Then
$\{v \in \domain(\tau_{w_1}) \mid \phi_1(v) \in \Loc\} = \{\mathtt{f},\this\}$,
$\mu_1\phi_1(\this) = o_1$ and  $\mu_1\phi_1(\mathtt{f}) = o_2$
so that we have
\(
   O_\tau^1(\sigma_1)
    = \{o_1,o_2\}.
\)
However $o_1.\pi = \overline{\pi}$ and $o_2.\pi = \pi_2$ so that,
by using the static information in Figure~\ref{fig:static_information},
we have $F(k(o_1.\pi)) = \emptyset$ and
$\domain(F(k(o_2.\pi)))
  = \{\mathtt{next},
      \mathtt{rotation}, \mathtt{side},\linebreak
      \mathtt{Square.x},
     \mathtt{Square.y}\}$.
From Figure~\ref{fig:state} we conclude that
\(
\{\mu_1(o_2.\phi(f))\mid \ f\in\domain(F(k(o_2.\pi)))
    ,
    o_2.\phi(f)\in\Loc\}
   = \{o_2\}
\)
and therefore
\[
  O^1_{F(k(o_2.\pi))}(o_2.\phi\sep\mu_1)=
    \left\{\mu(o_2.\phi(f))\left|\begin{array}{l}
      f\in\domain(F(k(o_2.\pi)))\\
      o_2.\phi(f)\in\Loc
    \end{array}\right.\right\}=\{o_2\}
\]
so that
\(
   O_\tau^2(\sigma_1)
    = \{o_1, o_2\} \cup \{o_2\} = \{o_1, o_2\}
    = O_\tau^1(\sigma_1)
\).
Thus we have a fixpoint and $O_\tau(\sigma_1) = \{o_1,o_2\}$.
Note that $o_4\not\in O_\tau(\sigma_1)$ \ie it is \emph{garbage}.
}

\ifcorrx{%
As $o_1. \pi = \overline{\pi}$ and $o_2. \pi = \pi_2$, we have
$\alpha_{\tau_{w_1}}^\e(\sigma_1)=\{\overline{\pi},\pi_2\}$.
This corresponds with the approximation
we used in Example~\ref{ex:e_imprecise}
to decorate program point $w_1$.
\qed
}
\end{example}

\subsection{The Domain $\e$ in the Presence of Type Information}
  \label{subsec:types_escape}
Definition~\ref{def:alpha} seems to suggest that $\codom(\alpha^\e_\tau)
=\wp(\Pi)$ \ie that every set of creation points is a legal approximation
in each given program point.
However, this is not true if type information is taken into account.
\begin{example}\label{ex:rho_not_onto}
\ifcorr{%
Consider the program point $w_0$ in Figure~\ref{fig:program} and its type
environment $\tau_{w_0}=[\Out\mapsto\integer,\mathtt{this}\mapsto
\mathtt{Circle}]$.
Then $\alpha^\e_{\tau_{w_0}}(\sigma)\not=\{\pi_4\}$ for every
$\sigma=\phi\sep\mu\in\Sigma_\tau$. This is because
\[
  \alpha^\e_{\tau_{w_0}}(\sigma)=\bigcup\left\{\left.
    \{o.\pi\}\cup\alpha^\e_{F(k(o.\pi))}((o.\phi)\sep\mu)\right|
    \begin{array}{l}
      v\in\{\mathtt{this}\},\ \phi(v)\in\Loc\\
      o=\mu\phi(v)
    \end{array}\right\}.
\]
By Definition~\ref{def:weak_correctness} we know that if
$\phi(v)\in\Loc$ then $k(o.\pi)=\mathtt{Circle}$. Hence
$o.\pi=\pi_3$. We conclude that either $\phi(v)=\nil$
and $\alpha_{\tau_{w_0}}^\e(\sigma)=\emptyset$, or $\phi(v)\in\Loc$
and $\pi_3\in\alpha_{\tau_{w_0}}^\e(\sigma)$. In both cases it is not
possible that $\alpha^\e_{\tau_{w_0}}(\sigma)=\{\pi_4\}$.
\qed
}
\end{example}%
Example~\ref{ex:rho_not_onto} shows that
\emph{static type information provides escape information} by
indicating which subsets of creation points are not the abstraction of any
concrete states. We should therefore characterise which are the \emph{good} or
meaningful elements of $\wp(\Pi)$.
This is important because it reduces the size of the abstract domain
and removes useless creation points during the analysis through the use
of an \emph{abstract garbage collector} $\delta_\tau$
(Definition~\ref{def:delta}).

Let $\ee\in\wp(\Pi)$. Then $\delta_\tau(\ee)$ is defined as the largest subset
of $\ee$ which contains only those
creation points deemed useful by the type environment $\tau$.
This set is computed first by collecting the creation points
that create objects compatible with the types in $\tau$.
For each of these points, this check is reiterated
for each of the fields of the object it creates until a fixpoint is reached.
Note that if there are no possible creation points for \texttt{this},
all creation points are useless.
%
\begin{definition}[Abstract Garbage Collector $\delta$]\label{def:delta}
Let $\ee\subseteq\Pi$. We define
$\delta_\tau(\ee)=\cup\{\delta_\tau^i(\ee)\mid i\ge 0\}$ with
\begin{align*}
  \delta_\tau^0(\ee)&=\emptyset\\
  \delta_\tau^{i+1}(\ee)&=\begin{cases}
    \emptyset\\
    \quad\text{if $\mathtt{this}\in\domain(\tau)$ and
      no $\pi\in \ee$ is
      s.t.\ $k(\pi)\le\tau(\mathtt{this})$}\\
    \mbox{}\\
    \cup\bigl\{\,
       \{\pi\}\cup\delta^i_{F(\pi)}(\ee)\bigm|
      \kappa\in\codom(\tau)\cap\mathcal{K},\ \pi\in \ee,\ k(\pi)\le\kappa
    \,\bigr\}\\
    \quad\text{otherwise.}
    \end{cases}
\end{align*}
\end{definition}
It follows from Definition~\ref{def:delta}
that $\delta_\tau^i\subseteq\delta_\tau^{i+1}$ and
hence $\delta_\tau=\delta_\tau^{\#\Pi}$.
Note that in Definition~\ref{def:delta} we consider all
subclasses of $\kappa$ (Example~\ref{ex:escape_analysis}).
\begin{example}\label{ex:delta1}
\ifcorr{%
Let us look at the program in Figure~\ref{fig:program}.
}

\ifcorrx{%
Consider first the program point $w_0$ and the type environment
\(
\tau_{w_0} \!=
  [\Out\!\mapsto\integer,
   \mathtt{this}\!\mapsto\mathtt{Circle}]
\)
at program point  $w_0$.
Let $\ee\!=\!\{\overline{\pi},\pi_1,\pi_3,\pi_4\}$.
Then, it can be seen in Figure~\ref{fig:static_information} that
$\codom(F(\mathtt{Circle})) \cap \mathcal{K} = \{\mathtt{Figure}\}$.
Note that $\kappa(\pi_3) = \mathtt{Circle} = \tau_{w_0}(\this)$.
Thus, for every $i\in\nat$, we have
\begin{align*}
  \delta^1_{F(\mathtt{Circle})}(\ee)&=
    \cup \bigl\{\, \{\pi\}\cup\delta^0_{F(k(\pi))}(\ee) \bigm|
    \pi\in \ee,\ k(\pi)\le\mathtt{Figure} \,\bigr\}=\{\pi_3\}\\
  \delta^{i+1}_{F(\mathtt{Circle})}(\ee)&=
    \cup \bigl\{\, \{\pi\}\cup\delta^i_{F(k(\pi))}(\ee) \bigm|
    \pi\in \ee,\ k(\pi)\le\mathtt{Figure} \,\bigr\}\\
  &=\{\pi_3\}\cup\delta^i_{F(\mathtt{Circle})(\ee)}~,
\end{align*}
which is enough to prove, by induction, that
$\delta^i_{F(\mathtt{Circle})}(\ee)=\{\pi_3\}$ for every $i\ge 1$.
Then we have
\begin{align*}
  \delta^{i+2}_{\tau_{w_0}}(\ee)&=\cup \bigl\{\,
  \{\pi\}\cup\delta^{i+1}_{F(k(\pi))}(\ee) \bigm|
    \kappa\in\codom(\tau)\cap\mathcal{K},\ \pi\in \ee,\ k(\pi)\le\kappa
   \,\bigr\}\\
  &=\cup \bigl\{\, \{\pi\}\cup\delta^{i+1}_{F(k(\pi))}(\ee) \bigm|
    \pi\in \ee,\ k(\pi)\le\mathtt{Circle}
   \,\bigr\}\\
  &=\{\pi_3\}\cup\delta^{i+1}_{F(\mathtt{Circle})}(\ee)=\{\pi_3\}.
\end{align*}
}

\ifcorrx{%
Consider now the program point $w_1$ and
the type environment
$\tau_{w_1}$ at program point  $w_1$ as given in Figure~\ref{fig:state}.
Let $\ee=\{\overline{\pi},\pi_1,\pi_3,\pi_4\}$.
Suppose that $i > 0$. As $\codom(F(\mathtt{Scan}))= \emptyset$, we have
$\delta^i_{F(\mathtt{Scan})}(\ee)=\emptyset$;
in the previous paragraph we have shown that
$\delta^i_{F(\mathtt{Circle})}(\ee)=\{\pi_3\}$;
similarly, it can be seen that
$\delta^i_{F(\mathtt{Square})}(\ee)=\{\pi_1,\pi_3,\pi_4\}$.
We therefore can conclude that, for all $i > 0$,
\begin{align*}
  \delta_{\tau_{w_1}}(\ee)
  &=\delta^{i+1}_{\tau_{w_1}}(\ee)\\
  &=\cup \bigl\{\{\pi\}\cup\delta^i_{F(k(\pi))}(\ee) \bigm|
    \kappa\in\{\mathtt{Figure},\mathtt{Scan}\},\ \pi\in \ee,\ k(\pi)\le\kappa
    \,\bigr\}\\
  &=\{\overline{\pi},\pi_3\}\cup\delta^{i+1}_{F(\mathtt{Square})}(\ee)
    \cup\delta^i_{F(\mathtt{Circle})}(\ee)
    \cup\delta^i_{F(\mathtt{Scan})}(\ee)\\
  &=\{\overline{\pi},\pi_3\}\cup\{\pi_1,\pi_3,\pi_4\}\cup\{\pi_3\}\cup
    \emptyset\\
  &=\{\overline{\pi},\pi_1,\pi_3,\pi_4\}.
\end{align*}
Then all the creation points in $\ee$
are useful in $w_1$ (compare this with Example~\ref{ex:escape_analysis}).
\qed
}
\end{example}

Proposition~\ref{prop:delta_lco} states
that the abstract garbage collector $\delta_\tau$ is a lower closure operator
so that it possesses the properties of monotonicity,
reductivity and idempotence that would be expected in a garbage collector.
\begin{proposition}\label{prop:delta_lco}
Let $i\in\nat$. The abstract garbage collectors $\delta_\tau^i$ and
$\delta_\tau$ are lco's.
\end{proposition}
The following result proves that
$\delta_\tau$ can be used to define $\codom(\alpha^\e_\tau)$. Namely,
the \emph{useful} elements of $\wp(\Pi)$ are those that do not contain
any garbage. The proof of Proposition~\ref{prop:delta_fixpoints}
relies on the explicit
construction, for every $\ee\subseteq\Pi$,
of a set of concrete states $X$ such
that $\alpha_\tau(X)=\delta_\tau(\ee)$, which is a fixpoint of $\delta_\tau$
by a well-known property of lco's.
\begin{proposition}\label{prop:delta_fixpoints}
Let $\delta(\tau)$ be an abstract garbage collector.
We have that $\fp(\delta_\tau)=\codom(\alpha^\e_\tau)$ and
$\emptyset\in\fp(\delta_\tau)$. Moreover, if
$\mathtt{this}\in\domain(\tau)$,
then for every $X\subseteq\Sigma_\tau$ we have
$\alpha^\e_\tau(X)=\emptyset$ if and only if $X=\emptyset$.
\end{proposition}

Proposition~\ref{prop:delta_fixpoints} lets us assume that
$\alpha^\e_\tau:\wp(\Sigma_\tau)\mapsto\fp(\delta_\tau)$. Moreover, it
justifies the following definition of our domain
$\e$ for escape analysis.
Proposition~\ref{prop:delta_fixpoints} can be used to
compute the possible
approximations from $\e$ at a given program point. However,
it does not specify which of these is best. This is
the goal of an escape analysis (Subsection~\ref{subsec:static_analysis_e}).
\begin{definition}[Abstract Domain $\e$]\label{def:property_e}
Our basic domain for escape analysis is
$\e_\tau=\fp(\delta_\tau)$, ordered by set inclusion.
\end{definition}
\begin{example}\label{ex:ew}
\ifcorr{%
Let $\tau_{w_0}$
and $\tau_{w_1}$ be as given in Figure~\ref{fig:state}. Then
\begin{align*}
  \e_{\tau_{w_0}}&=\{\emptyset\}\cup\{\ee\in\wp(\Pi)\mid
      \pi_3\in \ee\text{ and }
      \bigl(\{\pi_1,\pi_4\}\cap \ee \neq \emptyset
        \text{ entails }\pi_2\in \ee\bigr)\}\\
  \e_{\tau_{w_1}}&=\{\emptyset\}\cup\{\ee\in\wp(\Pi)\mid
    \overline{\pi}\in \ee\text{ and }
      \bigl(\{\pi_1,\pi_4\}\cap \ee \neq \emptyset
        \text{ entails }\pi_2\in \ee\bigr)\}.
\end{align*}
The constraints say that there must be a creation point for
the $\mathtt{this}$ variable and that to reach
an $\mathtt{Angle}$ (created at $\pi_1$ or at $\pi_4$)
from the variables in $\domain(\tau_{w_0})$ or
$\domain(\tau_{w_1})$, we must be able to reach
a $\mathtt{Square}$ (created at $\pi_2$).
\qed
}
\end{example}
By Definition~\ref{def:alpha}, we know that $\alpha^\e_\tau$
is strict and additive and, by Proposition~\ref{prop:delta_fixpoints},
onto $\e_\tau$. Thus,
by a general result of
abstract interpretation~\cite{CousotC77,CousotC92}
(Section~\ref{sec:preliminaries}),
we have the following proposition.
\begin{proposition}\label{prop:e_insertion}
The map $\alpha^\e_\tau$ (Definition~\ref{def:alpha}) is the abstraction
map of a Galois insertion from $\wp(\Sigma_\tau)$ to $\e_\tau$.
\end{proposition}

Note that if, in Definition~\ref{def:property_e},
we had defined $\e_\tau$ as $\wp(\Pi)$, the map $\alpha^\e_\tau$
would induce just a Galois connection instead of a Galois insertion, as a
consequence of Proposition~\ref{prop:delta_fixpoints}.

The domain $\e$
induces optimal abstract operations
which can be used for an actual escape analysis.
We discuss this in the next subsection.
\subsection{Static Analysis over $\e$}\label{subsec:static_analysis_e}
Figure~\ref{fig:operations_e} defines the abstract counterparts of the
concrete operations in Figure~\ref{fig:concrete_states}.
Proposition~\ref{prop:operations_e} states that they are correct and optimal,
in the sense of abstract interpretation (Section~\ref{sec:preliminaries}).
Optimality is proved by showing that each operation in
Figure~\ref{fig:operations_e} coincides with the optimal operation
$\alpha^\e\circ\mathit{op}\circ\gamma^\e$, where $\mathit{op}$ is the
corresponding concrete operation in Figure~\ref{fig:concrete_states},
as required by the
abstract interpretation framework. Note that the map $\gamma^\e$ is
induced by $\alpha^\e$ (Section~\ref{sec:preliminaries}). 
\begin{proposition}\label{prop:operations_e}
The operations in Figure~\ref{fig:operations_e} are the optimal counterparts
induced by $\alpha^\e$
of the operations in Figure~\ref{fig:concrete_states} and
of $\cup$.
They are implicitly strict on $\emptyset$, except for
$\mathsf{return}$, which is strict in its first argument only,
and for $\cup$.
\end{proposition}

\begin{figure}[t]
{\small
\begin{gather}
\begin{align*}
  \mathsf{nop}_\tau(\ee)&=\ee & \mathsf{get\_int}_\tau^i(\ee)&=\ee\\
  \mathsf{get\_null}^\kappa_\tau(\ee)&=\ee & \mathsf{get\_var}_\tau^v(\ee)&=\ee\\
  \mathsf{is\_true}_\tau(\ee)&=\ee & \mathsf{is\_false}_\tau(\ee)&=\ee\\
  \mathsf{put\_var}_\tau^v(\ee)&=\delta_{\tau|_{-v}}(\ee) &
    \mathsf{is\_null}_\tau(\ee)&=\delta_{\tau|_{-\rs}}(\ee)\\
  \mathsf{new}_\tau^\pi(\ee)&=\ee\cup\{\pi\} &
    \mathsf{=}_\tau(\ee_1)(\ee_2)&=\mathsf{+}_\tau(\ee_1)(\ee_2)=\ee_2\\
  \mathsf{expand}_\tau^{v:t}(\ee)&=\ee &
    \mathsf{restrict}_\tau^{\mathit{vs}}(\ee)&=\delta_{\tau_{-\mathit{vs}}}(\ee)\\
  \mathsf{call}_\tau^{\nu,v_1,\ldots,v_n}(\ee)&=\delta_{\tau|_{\{v_1,\ldots
    v_n,\rs\}}}(\ee) & \cup_\tau(\ee_1)(\ee_2)&=\ee_1\cup \ee_2
\end{align*}\\
\begin{align*}
  \mathsf{get\_field}_\tau^f(\ee)&=\begin{cases}
    \emptyset & \text{if $\{\pi\in \ee\mid k(\pi)\le\tau(\rs)\}=\emptyset$}\\
    \delta_{\tau[\rs\mapsto F(\tau(\rs))(f)]}(\ee) & \text{otherwise}
  \end{cases}\\
  \mathsf{put\_field}_{\tau,\tau'}^f(\ee_1)(\ee_2)&=\begin{cases}
    \emptyset & \text{if $\{\pi\in \ee_1\mid k(\pi)\le\tau(\rs)\}=
      \emptyset$}\\
    \delta_{\tau|_{-\rs}}(\ee_2) & \text{otherwise}
  \end{cases}\\
  \mathsf{return}_\tau^\nu(\ee_1)(\ee_2)&=\cup\left\{
    \{\pi\}\cup\delta_{F(k(\pi))}(\Pi)\left|
      \begin{array}{l}
        \kappa\in\codom(\tau|_{-\rs})\cap\mathcal{K}\\
        \pi\in\ee_1,\ k(\pi)\le\kappa
      \end{array}\right.\right\}\cup\ee_2\\
  \mathsf{lookup}^{m,\nu}_\tau(\ee)&=\begin{cases}
    \emptyset\qquad\text{if $\ee'=\left\{\pi\in \ee\left|
      \begin{array}{l}
        k(\pi)\le\tau(\rs)\\
        M(k(\pi))(m)=\nu
      \end{array}\right.\right\}=\emptyset$}\\
    \delta_{\tau|_{-\rs}}(\ee)\cup\left(\bigcup
    \{\{\pi\}\cup\delta_{F(k(\pi))}(\ee)\mid\pi\in \ee'\}\right)\quad
      \text{otherwise.}
  \end{cases}
\end{align*}
\end{gather}}
\caption{The optimal abstract operations over $\e$.}
  \label{fig:operations_e}
\end{figure}

Many operations in Figure~\ref{fig:operations_e} coincide with the
identity map.
This is a sign of the computational imprecision conveyed by the
domain $\e$. Other operations call the $\delta$ garbage collector quite often
to remove creation points of objects which might become unreachable
since some variable has disappeared from the scope. For instance, as the
concrete $\mathsf{put\_var}$ operation removes variable $v$ from
the scope (Figure~\ref{fig:concrete_states}), its abstract counterpart
in Figure~\ref{fig:operations_e} calls the garbage collector.
The same happens for $\mathsf{restrict}$ which, however, removes a \emph{set}
of variables from the scope.
There are also some operations ($\mathsf{is\_null}$,
$\mathsf{put\_field}$, $\mathsf{lookup}$) that use $\rs$ as a
temporary variable and one operation ($\mathsf{get\_field}$) that
changes the type of $\rs$.  Hence these abstract operations
also need to call the garbage collector.
Note that the definitions of the
$\mathsf{get\_field}$, $\mathsf{put\_field}$ and
$\mathsf{lookup}$ operations also consider, separately, the unusual situation
when we read a field, respectively, write a field or call a method and
the receiver is \emph{always} $\nil$. In this case, the concrete computation
always stops so that the best approximation of the (empty) set of subsequent
states is $\emptyset$.
The garbage collector is also called by
$\mathsf{call}$ since it creates a scope for the callee where
only some of the variables of the caller (namely, the parameters of
the callee) are addressable.
The $\mathsf{new}$ operation adds its creation point to the
approximation, since its concrete counterpart creates an object and
binds it to the temporary variable $\rs$.
The $\cup$ operation computes the union of the creation points
reachable from at least one of the two branches of a conditional.
The $\mathsf{return}$ operation states that all fields of the objects bound
to the
variables in scope before the call might have been modified by the
call.  This is reflected by the use of $\delta_{F(k(\pi))}(\Pi)$ in
$\mathsf{return}$, which plays the role of a worst-case assumption on
the content of the fields.  After
Example~\ref{ex:abstract_operations_e} we discuss how to cope with the
possible imprecision of this definition.
The $\mathsf{lookup}$ operation computes first the set $e'$ of the
creation points of objects that may be receivers of the virtual
call. If this set is not empty, the variable $\rs$ (which holds the
receiver of the call) is required to be bound to an object created at
some creation point in $e'$. This further constrains the creation
points reachable from $\rs$ and this is why we call the garbage
collector $\delta_{F(k(\pi))}$ for each $\pi\in e'$.

The definitions of $\mathsf{return}$ and $\mathsf{lookup}$ are quite complex;
this is a consequence of our quest for \emph{optimal} abstract operations.
It is possible to replace their definitions in  Figure~\ref{fig:operations_e}
by the less precise but simpler definitions:
\[
  \mathsf{return}_\tau^\nu(e_1)(e_2)=\delta_\tau(\Pi)\cup e_2\qquad
    \mathsf{lookup}_\tau^{m,\nu}(e)=e.
\]
Note though that, in practice, the results with the simpler definitions will
often be the same.
\begin{example}\label{ex:abstract_operations_e}
\ifcorr{%
Let us mimic, in $\e$, the concrete computation
of Example~\ref{ex:concrete_operations}.
Let the type environment $\tau = \tau_{w_1}$
and the concrete states $\sigma_1$ and $\sigma_2$
be as given in Figure~\ref{fig:state}.
We start by constructing elements $\ee_1$ and $\ee_2$ of $\e$
corresponding to $\sigma_1$ and $\sigma_2$.
The abstract state $\ee_1$ is obtained by abstracting $\sigma_1$
(see Example~\ref{ex:e_domain}):
\begin{align*}
  \ee_1&=\alpha^\e_\tau(\{\sigma_1\})=\alpha^\e_\tau(\sigma_1)
    =\{\overline{\pi},\pi_2\}\\
  \ee_2&=\mathsf{get\_var}_\tau^\mathtt{f}(\ee_1)
    =\{\overline{\pi},\pi_2\}.
\end{align*}
There are three abstract $\mathsf{lookup}$ operations corresponding to the
concrete ones and hence we construct
for $i=3,\ldots,6$,  elements $\ee_i'$, $\ee_i''$, $\ee_i'''$ of $\e$
corresponding to the concrete states $\sigma_i'$, $\sigma_i''$, $\sigma_i'''$,
respectively.
\begin{align*}
  \ee_3'&=\mathsf{lookup}_{\tau[\rs\mapsto\mathtt{Figure}]}
    ^{\mathtt{def},\mathtt{Figure.def}}(\ee_2)=\emptyset\\
    &\text{since $\{\pi\in \ee_2\mid k(\pi)\le
    \mathtt{Figure}\text{ and }M(\pi)(\mathtt{def})=
    \mathtt{Figure.def}\}=\emptyset$}\\
  \ee_3''&=\mathsf{lookup}_{\tau[\rs\mapsto\mathtt{Figure}]}
    ^{\mathtt{def},\mathtt{Square.def}}(\ee_2)\\
    &=\delta_\tau(\ee_2)\cup\left(\bigcup\left\{\left.\{\pi\}\cup
      \delta_{F(k(\pi))}(\ee_2)\right|\pi\in \ee''\right\}\right)
      =\{\overline{\pi},\pi_2\}\cup\{\pi_2\}\\
    &=\{\overline{\pi},\pi_2\}~,\\
    &\text{since $\ee''=\left\{\pi\in \ee_2\left|\begin{array}{l}
      k(\pi)\le\mathtt{Figure}\\
      M(\pi)(\mathtt{def})=\mathtt{Square.def}
    \end{array}\right.\right\}=\{\pi_2\}$}
\end{align*}
\begin{align*}
  \ee_3'''&=\mathsf{lookup}_{\tau[\rs\mapsto\mathtt{Figure}]}
    ^{\mathtt{def},\mathtt{Circle.def}}(\ee_2)=\emptyset\\
    &\text{since $\ee'''=\left\{\pi\in \ee_2\left|
      \begin{array}{l}
        k(\pi)\le\mathtt{Figure}\\
        M(\pi)(\mathtt{def})=\mathtt{Circle.def}
      \end{array}\right.\right\}=\emptyset$.}
\end{align*}
The $\mathsf{lookup}$ operations for
$\mathtt{Figure}$ and $\mathtt{Circle}$ return $\emptyset$ so that,
as the abstract operations over $\e$ are strict on $\emptyset$
(Proposition~\ref{prop:operations_e}), $\ee_4' = \ee_5' = \ee_6' =
\ee_4'''=\ee_5'''=\ee_6'''= \emptyset$. This is because the analysis is able
to guess the target of the virtual call $\mathtt{f.def()}$, since the only
objects reachable there are
a $\mathtt{Square}$ (created in $\pi_2$)
and a $\mathtt{Scan}$ which, however, is not compatible with the declared type
of the receiver of the call.
Hence we only have to consider the case when a
$\mathtt{Square}$ is selected:
\begin{align*}
  \ee_4''&=\mathsf{call}^\mathtt{Square.def}_{\tau[\rs\mapsto\mathtt{Square}]}
    (\ee_3'')=\delta_{[\rs\mapsto\mathtt{Square}]}(\ee_3'')=\{\pi_2\}.
\end{align*}
Since $P(\mathtt{Square.def})|_{\Out}
=[\Out\mapsto\integer]$ and $\e_{[\Out\mapsto\integer]}=\{\emptyset\}$,
we do not need to execute the method $\mathtt{Square.def}$ to conclude that
\[
  \ee_5''=\emptyset.
\]
Hence
\begin{align*}
  \ee_6''&=\mathsf{return}^\mathtt{Square.def}
    _{\tau[\rs\mapsto\mathtt{Square}]}(\ee_3'')(\ee_5'')\\
  &=\cup\{\{\pi\}\cup\delta_{F(k(\pi))}(\Pi)\mid\pi\in
    \{\overline{\pi},\pi_2\}\}=
    \{\overline{\pi},\pi_1,\pi_2,\pi_3,\pi_4\}.
\end{align*}
Since the abstract semantics is non-deterministic, we merge the results
of every thread of execution through the $\cup$ operation. Hence the abstract
state after the execution of the call $\mathtt{f.def()}$ in Figure
\ref{fig:program} is
\[
  \ee_6=\ee_6'\cup \ee_6''\cup \ee_6'''=\emptyset\cup
    \{\overline{\pi},\pi_1,\pi_2,\pi_3,\pi_4\}\cup\emptyset=
  \{\overline{\pi},\pi_1,\pi_2,\pi_3,\pi_4\}.
\]
\qed
}
\end{example}

\ifcorrx{%
In Example~\ref{ex:abstract_operations_e}, the imprecision of the analysis
induced by $\e$ is largely due to the abstract operation $\mathsf{return}$
used to compute $\ee_6''$. The creation points $\pi_1$ and
$\pi_4$ for $\mathtt{Angle}$s need to be added
because in the execution of the methods $\mathtt{Square.def}$
any field of the object bound to $\mathtt{this}$ \emph{could} be
modified. For instance, an $\mathtt{Angle}$
could be bound to the field $\mathtt{rotation}$ of
the object bound to $\mathtt{this}$.
This is what actually happens
for $\pi_1$ in the method $\mathtt{Square.def}$
(Figure~\ref{fig:program}), while the introduction of the creation point
$\pi_4$ is an imprecise (but correct) assumption.
This is a consequence of the definition
of $\e$ as the set of sets of \emph{reachable} creation points.
At the end of a method, we only have rather weak information about
the set $\ee$ of creation points of the objects reachable from $\Out$.
For instance, if $\Out$ has type $\integer$, such as in
Example~\ref{ex:abstract_operations_e}, we can only have $\ee=\emptyset$.
When we return to the caller, the actual parameters return into scope.
The definition of $\mathsf{return}$ in Figure~\ref{fig:concrete_states} shows
that these parameters are unchanged (because of the condition
$\mu_1=_L\mu_2$, where $L = \codom(\phi_1)|_{-\rs}\cap\Loc$,
on the concrete operation).
This is reflected by the condition $\pi\in e_1$
in the abstract $\mathsf{return}$ operation in Figure~\ref{fig:operations_e}.
However, we do not know anything about their \emph{fields}.
Without such information,
only a pessimistic assumption can be made, which is
expressed by the use of $\Pi$ in the abstract $\mathsf{return}$ operation.
This problem can be solved
by including a \emph{shadow copy} of the actual parameters among
the variables in scope inside a method.
An example of the use of this
technique will be given later (see Example~\ref{ex:abstract_operations_er2}).
By using this technique, we can actually improve the precision of the
computation in Example~\ref{ex:abstract_operations_e}.  As reported in
Example~\ref{ex:e_imprecise}, we get the more precise approximation
$\{\overline{\pi},\pi_1,\pi_2\}$ after the first call to $\mathtt{f.def()}$.
}

There is, however, another problem related with the domain $\e$.
It is exemplified below.
\begin{example}\label{ex:abstract_operations_e2}
\ifcorr{%
Let the approximation provided by $\e$ before the statement
$\mathtt{f\ =\ new\ Circle()}$ in Figure~\ref{fig:program}
be $\{\overline{\pi},\pi_1,\pi_2,\pi_4\}$ (Example~\ref{ex:e_imprecise}).
We can compute the approximation \emph{after}
that statement by executing two abstract operations.
Since the object created in $\pi_3$ gets stored
inside the variable $\mathtt{f}$, we would expect the creation point $\pi_2$
of the old value of $\mathtt{f}$ to disappear. But this does not happen:
\begin{align*}
  \mathsf{new}_\tau^{\pi_3}(\{\overline{\pi},\pi_1,\pi_2,\pi_4\})
    &=\{\overline{\pi},\pi_1,\pi_2,\pi_3,\pi_4\}\\
  \mathsf{put\_var}^\mathtt{f}_{\tau[\rs\mapsto\mathtt{Circle}]}
    (\{\overline{\pi},\pi_1,\pi_2,\pi_3,\pi_4\})&=
    \{\overline{\pi},\pi_1,\pi_2,\pi_3,\pi_4\}.
\end{align*}
\qed
}
\end{example}
\ifcorrx{%
This time, the imprecision is a consequence of the fact that
$\mathtt{n}\in\domain(\tau)$, $\tau(\mathtt{n})=\mathtt{Figure}$ and
$k(\pi_2)=\mathtt{Square}\le\mathtt{Figure}$. Since the abstract domain
$\e$ does not allow one to know the creation points of the objects
bound to \emph{a given} variable,
but only provides global information on the creation points of the objects
bound to variables and fields \emph{as a whole},
we do not know whether $\pi_2$ is the creation point of an object bound
to $\mathtt{f}$ (and in such a case it disappears) or to
$\mathtt{n}$ instead (and in such a case it must \emph{not} disappear). Hence
a correct $\mathsf{put\_var}$ operation cannot make $\pi_2$ disappear.
We solve this problem in Section~\ref{sec:erdomain} by introducing this
missing information into a new, more precise, abstract domain $\er$.
}
\section{The Refined Domain $\er$}\label{sec:erdomain}
We define here a \emph{refinement} $\er$ of the domain $\e$ of
Section~\ref{sec:edomain}, in the sense that $\er$ is a concretisation of
$\e$ (Proposition~\ref{prop:inclusion}). The idea underlying the
definition of $\er$ is that the precision of $\e$ can be improved
if we can speak about the creation points of the objects bound
to \emph{a given} variable or field (see the problem
highlighted in Example~\ref{ex:abstract_operations_e2}).
The construction of $\er$ is very similar to that of $\e$.
\subsection{The Domain}\label{subsec:erdomain}
Definition~\ref{def:domains2} defines concrete values.
The domain $\er$ we are going to define approximates
every concrete value with an \emph{abstract value}.
An abstract value is either $*$, which approximates the integers, or
a set $\ee\subseteq\Pi$, which approximates $\nil$ and all locations
containing an object created in some creation point in $\ee$.
An abstract frame maps variables to
abstract values consistent with their type.
%
\begin{definition}[Abstract Values and Frames]\label{def:frames1}
Let the \emph{abstract values} be
$\Value^{\er}=\{*\}\cup\wp(\Pi)$. We define
\[
  \Frame_\tau^{\er}=\left\{\phi\in\domain(\tau)\mapsto\Value^{\er}\left|
    \begin{array}{l}
      \text{for every }v\in\domain(\tau)\\
      \text{ if }\tau(v)=\integer\text{ then }\phi(v)=*\\
      \text{ if }\tau(v)\in\mathcal{K}\text{ and }\pi\in\phi(v)\\
      \ \ \text{ then }k(\pi)\le\tau(v)
    \end{array}
    \right.\right\}.
\]
The set $\Frame_\tau^\er$ is ordered by pointwise set-inclusion.
\end{definition}
\begin{example}\label{ex:frames_er}
\ifcorr{%
Let $\tau_{w_1}$ be as defined in Figure~\ref{fig:state}.
Then we have
\begin{align*}
  [\mathtt{f}\mapsto\{\pi_2\},\mathtt{n}\mapsto\{\pi_2,\pi_3\},
    \Out\mapsto *,\mathtt{this}\mapsto\{\overline{\pi}\}]
    &\in\Frame_{\tau_{w_1}}^{\er}\\
  [\mathtt{f}\mapsto\{\overline{\pi},\pi_2\},\mathtt{n}\mapsto\{\pi_2,\pi_3\},
    \Out\mapsto *,\mathtt{this}\mapsto\{\overline{\pi}\}]
    &\not\in\Frame_{\tau_{w_1}}^{\er}~,
\end{align*}
since $k(\overline{\pi})=\mathtt{Scan}$,
$\tau_{w_1}(\mathtt{f})=\mathtt{Figure}$ and
$\mathtt{Scan}\not\le\mathtt{Figure}$.
\qed
}
\end{example}

The map $\varepsilon$ \emph{extracts} the creation points of the objects
bound to the variables.
\begin{definition}[Extraction Map]\label{def:extractor}
The map $\varepsilon_\tau:\wp(\Sigma_\tau)\mapsto\Frame^\er_\tau$ is such that,
for every $S\subseteq\Sigma_\tau$ and $v\in\domain(\tau)$,
\[
  \varepsilon_\tau(S)(v)=
  \begin{cases}
    * & \text{if $\tau(v)=\integer$}\\
    \{(\mu\phi(v)).\pi\mid\phi\sep\mu\in S
      \text{ and }\phi(v)\in\Loc\} & \text{if $\tau(v)\in\mathcal{K}$.}
  \end{cases}
\]
\end{definition}
\begin{example}\label{ex:extractor}
\ifcorr{%
Consider the state $\sigma_1$ in Figure~\ref{fig:state}. Then
\[
  \varepsilon_{\tau_{w_1}}(\sigma_1)=
   [
    \mathtt{f}\mapsto\{\pi_2\},\mathtt{n}\mapsto\emptyset,
    \Out\mapsto *,\mathtt{this}\mapsto\{\overline{\pi}\}
   ].
\]
\qed
}
\end{example}

Since it is assumed that all the fields are uniquely identified by their
fully qualified name, the type environment $\widetilde{\tau}$
\emph{of all the fields} introduced by the program is well-defined.
\begin{definition}[Type Environment of All Fields]\label{def:all_fields}
\hspace*{-1ex}We define the \emph{type environment of all fields} as
$\widetilde{\tau}=\cup\{F(\kappa)\mid\kappa\in\mathcal{K}\}$.
Let $\tau\in\Typing$
be such that $\domain(\tau)\subseteq\domain(\widetilde{\tau})$
and $\phi\in Frame_\tau$. Its \emph{extension}
$\widetilde{\phi}\in\Frame_{\widetilde{\tau}}$ is such that, for every
$v\in\domain(\widetilde{\tau})$,
\[
  \widetilde{\phi}(v)=\begin{cases}
    \phi(v) & \text{if $v\in\domain(\tau)$}\\
    \init(\widetilde{\tau}(v)) &
      \text{otherwise (Definition~\ref{def:domains2}).}
  \end{cases}
\]
\end{definition}
\begin{example}\label{ex:tauoverlined}
\ifcorr{%
Consider the map $F$ in Figure~\ref{fig:static_information}
for the program in Figure~\ref{fig:program}. Then
\[
  \widetilde{\tau}=\left[\begin{array}{l}
    \mathtt{Circle.x}\mapsto\integer,
      \mathtt{Circle.y}\mapsto\integer,\mathtt{degree}\mapsto\integer\\
    \mathtt{next}\mapsto\mathtt{Figure},\mathtt{radius}\mapsto\integer,
      \mathtt{rotation}\mapsto\mathtt{Angle}\\
    \mathtt{side}\mapsto\integer,
      \mathtt{Square.x}\mapsto\integer,\mathtt{Square.y}\mapsto\integer
  \end{array}\right].
\]
Let $\phi=[\mathtt{Circle.x}\mapsto 12,\mathtt{Circle.y}\mapsto 5,
\mathtt{next}\mapsto l,\mathtt{radius}\mapsto 5]\in F(\mathtt{Circle})$,
with $l\in\Loc$. We have
\[
  \widetilde{\phi}=\left[\begin{array}{l}
    \mathtt{Circle.x}\mapsto 12,
      \mathtt{Circle.y}\mapsto 5,\mathtt{degree}\mapsto 0\\
  \mathtt{next}\mapsto l,\mathtt{radius}\mapsto 5,
    \mathtt{rotation}\mapsto\nil,
    \mathtt{side}\mapsto 0\\
  \mathtt{Square.x}\mapsto 0,\mathtt{Square.y}\mapsto 0
  \end{array}\right].
\]
}
\qed\end{example}

An abstract memory is an abstract frame for $\widetilde{\tau}$.
The abstraction map computes the abstract memory by extracting
the creation points of the fields of the reachable objects
of the concrete memory (Definition~\ref{def:reachability}).
\begin{definition}[Abstract Map for $\er$]\label{def:er_abstraction}
Let the set of abstract memories be
$\Memory^\er=\Frame_{\widetilde{\tau}}^\er$. We define the map
\[
   \alpha_\tau^\er:\wp(\Sigma_\tau)\mapsto\{\bot\}\cup
      (\Frame_\tau^\er\times\Memory^\er)
\]
 such that, for $S\subseteq\Sigma_\tau$,
\[
  \alpha_\tau^\er(S)=\begin{cases}
    \bot & \text{if $S=\emptyset$}\\
    \varepsilon_\tau(S)\sep\varepsilon_{\widetilde{\tau}}(
    \{\widetilde{o.\phi} \sep \sigma.\mu \mid
    \sigma\in S\text{ and }o\in O_\tau(\sigma)\}) & \text{otherwise.}
  \end{cases}
\]
\end{definition}
\begin{example}\label{ex:er_domain}
\ifcorr{%
Consider the state $\sigma_1$ in Figure~\ref{fig:state}.
Let $\tau = \tau_{w_1}$ be as given in Figure~\ref{fig:state}.
In Example~\ref{ex:e_domain}
we have shown that $O_\tau(\sigma_1) = \{o_1, o_2\}$
and in Example~\ref{ex:extractor}
we have computed the value of $\varepsilon_\tau(\sigma_1)$. We have
\begin{align*}
  \widetilde{o_1.\phi}&=\left[\begin{array}{l}
    \mathtt{Circle.x}\mapsto 0,
      \mathtt{Circle.y}\mapsto 0,\mathtt{degree}\mapsto 0\\
    \mathtt{next}\mapsto\nil,\mathtt{radius}\mapsto 0,
      \mathtt{rotation}\mapsto\nil\\
    \mathtt{side}\mapsto 0,
      \mathtt{Square.x}\mapsto 0,\mathtt{Square.y}\mapsto 0
  \end{array}\right],\\
  \widetilde{o_2.\phi}&=\left[\begin{array}{l}
    \mathtt{Circle.x}\mapsto 0,
      \mathtt{Circle.y}\mapsto 0,\mathtt{degree}\mapsto 0\\
    \mathtt{next}\mapsto l',\mathtt{radius}\mapsto 0,
      \mathtt{rotation}\mapsto\nil\\
    \mathtt{side}\mapsto 4,
      \mathtt{Square.x}\mapsto 3,\mathtt{Square.y}\mapsto -5
  \end{array}\right]
\end{align*}
%
Then (fields not represented are implicitly bound to $*$)
\begin{align*}
  \alpha_\tau^\er(\sigma_1)
  &=\alpha_\tau^\er(\phi_1 \sep \mu_1)\\
  &=\varepsilon_\tau(\sigma_1)\sep\varepsilon_{\widetilde{\tau}}(
    \{\widetilde{o_1.\phi}\sep\mu_1,\widetilde{o_2.\phi}\sep\mu_1\})\\
  &=\varepsilon_\tau(\sigma_1)\sep\varepsilon_{\widetilde{\tau}}(
    \{\widetilde{o_2.\phi}\sep\mu_1\})\\
  &=\varepsilon_\tau(\sigma_1)\sep
    [\mathtt{next}\mapsto\{\mu_1(l').\pi\},
    \mathtt{rotation}\mapsto\emptyset,\ldots]\\
  &=[\mathtt{f}\mapsto\{\pi_2\},\mathtt{n}\mapsto\emptyset,
    \Out\mapsto *,\mathtt{this}\mapsto\{\overline{\pi}\}]\\
  &\qquad\qquad\qquad \sep [\mathtt{next}\mapsto\{\pi_2\},\mathtt{rotation}\mapsto\emptyset,\ldots].
\end{align*}
\qed
}
\end{example}
Compare Examples~\ref{ex:er_domain} and~\ref{ex:e_domain}. You can see
that $\er$ distributes over the variables and fields the same creation points
observed by $\e$.

As a notational simplification, we often assume that each field not
reported in the approximation of the memory is implicitly bound to
$\emptyset$, if it has class type, and bound to $*$, if it has
$\integer$ type.

Just as for $\alpha_\tau^\e$ (Example~\ref{ex:rho_not_onto}), the
following example shows that the map $\alpha_\tau^\er$ is not
necessarily onto.
\begin{example}\label{ex:alpha_er_not_onto}
\ifcorr{%
Let $\tau=[\mathtt{c}\mapsto\mathtt{Circle}]$. A $\mathtt{Circle}$ has no
field called $\mathtt{rotation}$. Then there is no state
$\sigma\in\Sigma_\tau$ such that its abstraction is
$\alpha_\tau^\er(\sigma)=[\mathtt{c}\mapsto\{\pi_3\}]\sep
[\mathtt{next}\mapsto\emptyset,\mathtt{rotation}\mapsto\{\pi_1\},\ldots]$,
since only a $\mathtt{Circle}$ created at $\pi_3$ is reachable from the
variables.
\qed
}
\end{example}
Hence, we define
a map $\xi$ which forces to $\emptyset$ the fields of type class
of the objects which have no reachable creation points.
Just as for the garbage collector $\delta$ for $\e$,
the map $\xi$ can be seen as an abstract garbage collector for $\er$.
This $\xi$ uses an auxiliary map $\rho$ to compute
the set of creation points $r$ reachable from
the variables in scope.
The approximations of the fields of the objects created at $r$ are not
garbage collected by $\xi$. The approximations of the other fields are
garbage collected instead.
\begin{definition}[Abstract Garbage Collector $\xi$]\label{def:xi}
We define $\rho_\tau:\Frame_\tau^\er
\times\Memory^\er\mapsto\wp(\Pi)$ and
$\xi_\tau:\{\bot\}\cup(\Frame_\tau^\er\times\Memory^\er)\linebreak
\mapsto\{\bot\}\cup(\Frame_\tau^\er\times\Memory^\er)$
as $\rho_\tau(s)=\cup\{\rho_\tau^i(s)\mid i\ge 0\}$, where
\begin{align*}
  \rho_\tau^0(\phi\sep\mu)&=\emptyset\\
  \rho_\tau^{i+1}(\phi\sep\mu)&=\bigcup\left\{\left.\{\pi\}\cup
    \rho^i_{F(k(\pi))}
    (\mu|_{\domain(F(k(\pi)))}\sep\mu)\right|\begin{array}{l}
      v\in\domain(\tau)\\
      \pi\in\phi(v)
    \end{array}\right\}
\end{align*}
and
\begin{align*}
  \xi_\tau(\bot)&=\bot\\
  \xi_\tau(\phi\sep\mu)&=\begin{cases}
    \bot\qquad\text{if $\mathtt{this}\in\domain(\tau)$ and $\phi(\mathtt{this})
      =\emptyset$}\\
    \phi\sep\left(\cup\{\mu|_{\domain(F(k(\pi)))}\mid
    \pi\in\rho_\tau(\phi \sep \mu)\}
    \right)\quad\text{otherwise.}
  \end{cases}
\end{align*}
\end{definition}
\begin{example}
\ifcorr{%
Let $s=[\mathtt{c}\mapsto\{\pi_3\}]
\sep[\mathtt{next}\mapsto\emptyset,\mathtt{rotation}\mapsto\{\pi_1\},\ldots]$.
We have $\rho_\tau(s)=\{\pi_3\}$ and hence
\[
  \xi_\tau(s)=[\mathtt{c}\mapsto\{\pi_3\}]\sep
    [\mathtt{next}\mapsto\emptyset,\mathtt{rotation}\mapsto\emptyset,\ldots]
\]
\ie the abstract garbage collector $\xi$ has recognised $\pi_1$ as garbage
(compare with Example~\ref{ex:alpha_er_not_onto}).
\qed
}
\end{example}
The following property is expected to hold for a garbage collector.
Compare Propositions~\ref{prop:delta_lco} and~\ref{prop:xi_lco}.
\begin{proposition}\label{prop:xi_lco}
The abstract garbage collector $\xi_\tau$ is an lco.
\end{proposition}
The garbage collector $\xi_\tau$ can be used
to define $\codom(\alpha_\tau^\er)$.
Namely, the \emph{useful} elements of $\Frame^\er_\tau\times\Memory^\er$
are exactly those that do not contain any garbage.
Compare Propositions~\ref{prop:delta_fixpoints} and~\ref{prop:xi_fixpoints}.
\begin{proposition}\label{prop:xi_fixpoints}
Let $\xi_\tau$ be the abstract garbage collector of
Definition~\ref{def:xi}. Then $\fp(\xi_\tau)=\codom(\alpha_\tau^\er)$.
\end{proposition}
Proposition~\ref{prop:xi_fixpoints}
 allows us to assume that $\alpha_\tau^\er:
\wp(\Sigma_\tau)\mapsto\fp(\xi_\tau)$
and justifies the following definition.
\begin{definition}[Abstract Domain $\er$]\label{def:er_domain}
We define $\er_\tau=\fp(\xi_\tau)$, ordered by
pointwise set-inclusion (with the assumption that $*\subseteq *$ and
$\bot\subseteq s$ for every $s\in\er_\tau$).
\end{definition}
By Definitions~\ref{def:extractor} and~\ref{def:er_abstraction} we know
that the map $\alpha_\tau^\er$ is strict and additive.
By Proposition~\ref{prop:xi_fixpoints} we know that it is onto.
Thus we have the following result corresponding to
Proposition~\ref{prop:e_insertion} for the domain $\e$.
\begin{proposition}\label{prop:er_insertion}
The map $\alpha_\tau^\er$ is
the abstraction map of a Galois insertion from
$\wp(\Sigma_\tau)$ to $\er_\tau$.
\end{proposition}
\subsection{Static Analysis over $\er$}\label{subsec:static_analysis_er}
In order to use the domain $\er$ for an escape analysis, we need to provide
the abstract counterparts over $\er$ of the concrete operations in
Figure~\ref{fig:concrete_states}. Since $\er$ approximates every
variable and field with an abstract value, those abstract
operations are similar to those of the Palsberg and Schwartzbach's domain
for \emph{class analysis} in~\cite{PalsbergS91} as formulated
in~\cite{SpotoJ03}. However, $\er$ observes the fields of just
the reachable objects (Definition~\ref{def:er_abstraction}), while
Palsberg and Schwartzbach's domain observes the fields of all objects
in memory.

Figure~\ref{fig:operations_er} reports the abstract counterparts on $\er$ of
the concrete operations in Figure~\ref{fig:concrete_states}.
These operations are implicitly strict on $\bot$ except for $\cup$.
In this case, we define
$\bot \cup (\phi\sep\mu) = (\phi\sep\mu) \cup \bot = \phi\sep\mu$.
Their optimality is proved by showing that each operation in
Figure~\ref{fig:operations_er} coincides with the optimal operation
$\alpha^\er\circ\mathit{op}\circ\gamma^\er$, where $\mathit{op}$ is the
corresponding concrete operation in Figure~\ref{fig:concrete_states},
as required by the
abstract interpretation framework. Note that the map $\gamma^\er$ is
induced by $\alpha^\er$ (Section~\ref{sec:preliminaries}).
\begin{proposition}\label{prop:operations_er}
The operations in Figure~\ref{fig:operations_er} are the optimal counterparts
induced by $\alpha^\er$ of the operations in Figure~\ref{fig:concrete_states}
and of $\cup$.
\end{proposition}
\begin{figure}[ht]
{\small
\begin{gather}
\begin{align*}
  \mathsf{nop}_\tau(\phi\sep\mu)&=\phi\sep\mu\\
  \mathsf{get\_int}_\tau^i(\phi\sep\mu)&=\phi[\rs\mapsto *]\sep\mu\\
  \mathsf{get\_null}^\kappa_\tau(\phi\sep\mu)&=\phi[\rs\mapsto\emptyset]\sep\mu\\
  \mathsf{get\_var}_\tau^v(\phi\sep\mu)&=\phi[\rs\mapsto\phi(v)]\sep\mu\\
  \mathsf{is\_true}_\tau(\phi\sep\mu)&=\phi\sep\mu\\
  \mathsf{is\_false}_\tau(\phi\sep\mu)&=\phi\sep\mu\\
  \cup_\tau(\phi_1\sep\mu_1)(\phi_2\sep\mu_2)&=(\phi_1\cup\phi_2)\sep
    (\mu_1\cup\mu_2)\\
  \mathsf{is\_null}_\tau(\phi\sep\mu)&=\xi_{\tau[\rs\mapsto\integer]}
    (\phi[\rs\mapsto *]\sep\mu)\\
  \mathsf{new}_\tau^\pi(\phi\sep\mu)&=\phi[\rs\mapsto\{\pi\}]\sep\mu\\
  \mathsf{put\_var}_\tau^v(\phi\sep\mu)&=\xi_{\tau|_{-\rs}}
    (\phi[v\mapsto\phi(\rs)]|_{-\rs}\sep\mu)\\
  \mathsf{restrict}_\tau^{\mathit{vs}}(\phi\sep\mu)
    &=\xi_{\tau|_{-\mathit{vs}}}(\phi|_{-\mathit{vs}}\sep\mu)\\
  \mathsf{expand}_\tau^{v:t}(\phi\sep\mu)&=\begin{cases}
    \phi[v\mapsto *]\sep\mu & \text{if $t=\integer$}\\
    \phi[v\mapsto\emptyset]\sep\mu & \text{otherwise}
  \end{cases}\\
  \mathsf{=}_\tau(\phi_1\sep\mu_1)(\phi_2\sep\mu_2)&=
    \mathsf{+}_\tau(\phi_1\sep\mu_1)(\phi_2\sep\mu_2)=\phi_2\sep\mu_2
  \end{align*}\\
  \begin{align*}
  \mathsf{get\_field}_\tau^f(\phi\sep\mu)&=\begin{cases}
    \bot\quad\text{if $\phi(\rs)=\emptyset$}\\
    \xi_{\tau[\rs\mapsto F(\tau(\rs))(f)]}(\phi[\rs\mapsto\mu(f)]\sep\mu)
      \quad\text{else}
  \end{cases}\\
  \begin{array}{c}
    \mathsf{put\_field}_{\tau,\tau'}^f\\
    (\phi_1\sep\mu_1)(\phi_2\sep\mu_2)
  \end{array}&=
  \begin{cases}
    \bot\quad\text{if $\phi_1(\rs)=\emptyset$}\\
    \xi_{\tau|_{-\rs}}(\phi_2|_{-\rs}\sep\mu_2) \\
    \quad\text{else, if no
       $\pi\in\phi_1(\rs)$ occurs in $\phi_2|_{-\rs}\sep\mu_2$}\\
    \xi_{\tau|_{-\rs}}(\phi_2|_{-\rs}\sep\mu_2[f\mapsto\mu_2(f)\cup
      \phi_2(\rs)])\\
    \quad\text{otherwise}
  \end{cases}\\
  \mathsf{call}_\tau^{\nu,v_1,\ldots,v_n}(\phi\sep\mu)&=
    \xi_{P(\nu)|_{-\Out}}
    \left(\left[\begin{array}{l}
      \iota_1\mapsto\phi(v_1),\ldots,\iota_n\mapsto\phi(v_n)\\
      \mathtt{this}\mapsto\phi(\rs)
    \end{array}\right]\sep\mu\right)\\
  \begin{array}{c}
    \mathsf{return}_\tau^\nu\\
    (\phi_1\sep\mu_1)(\phi_2\sep\mu_2)
  \end{array}&=
    \xi_{\tau|_{-\rs}}(\phi_1|_{-\rs}\sep\mu^\top)
    \cup([\rs\mapsto\phi_2(\Out)]\sep\mu_2)\\
    &\quad\text{where $\mu^\top$ is the top of $\Memory^\er$}\\
  \mathsf{lookup}^{m,\nu}_\tau(\phi\sep\mu)&=\begin{cases}
    \bot\quad\text{if $\ee\!=\!\{\pi\in\phi(\rs)\mid
      M(\pi)(m)\!=\!\nu\}\!=\!\emptyset$}\\
    \xi_\tau(\phi[\rs\mapsto \ee]\sep\mu)\quad\text{otherwise.}
  \end{cases}
\end{align*}
\end{gather}}
\caption{The abstract operations over $\er$.}\label{fig:operations_er}
\end{figure}
Let us consider each of the abstract operations.
The operation $\mathsf{nop}$ leaves the
state unchanged. The same happens for the operations working
with integer values only, such as $\mathsf{is\_true}$,
$\mathsf{is\_false}$, $\mathsf{=}$ and $\mathsf{+}$, since the domain
$\er$ ignores variables with integer values.
The concrete operation $\mathsf{get\_int}$ loads an integer
into $\rs$. Hence, its abstract
counterpart loads $*$ into $\rs$, since $*$ is the approximation for
integer values (Definition~\ref{def:frames1}). The concrete operation
$\mathsf{get\_null}$ loads $\nil$ into $\rs$ and hence its
abstract counterpart approximates $\rs$ with $\emptyset$.
The operation $\mathsf{get\_var}^v$ copies the creation points of $v$
into those of $\rs$.
The $\cup$ operation merges the creation points of the objects bound to each
given variable or field in one of the two branches of a conditional.
The concrete $\mathsf{is\_null}$ operation checks if $\rs$ contains
$\nil$ or not, and loads $1$ or $-1$ in $\rs$ accordingly.
Hence its abstract counterpart loads $*$ into $\rs$.
Since the old value of $\rs$ may no longer be reachable,
we apply the abstract garbage collector $\xi$.
The $\mathsf{new}^\pi$ operation binds $\rs$ to an object created at $\pi$.
The $\mathsf{put\_var}^v$ operation copies the value of $\rs$ into $v$,
and removes $\rs$. Since the old value of $v$ may be lost, we apply the
abstract garbage collector $\xi$. The $\mathsf{restrict}$ operation removes
some variables from the scope and, hence, calls $\xi$.
The $\mathsf{expand}^v$ operation adds the variable $v$ in scope.
Its initial value is approximated with $*$, if it is $0$, and with
$\emptyset$, if it is $\nil$. The $\mathsf{get\_field}^f$ operation
returns $\bot$ if it is \emph{always} applied to states where the receiver
$\rs$ is $\nil$. This is because $\bot$ is the best approximation
of the empty set of final states.
If, instead, the receiver is not necessarily
$\nil$, the creation points of the field $f$ are copied from the
approximation $\mu(f)$ into the approximation of $\rs$. Since this operation
changes the value of $\rs$, possibly making
some object unreachable, it needs to call $\xi$.
For the $\mathsf{put\_field}^f$ operation, we first check if the
receiver is always $\nil$, in which case the abstract operation returns $\bot$.
Then we consider the case in which the evaluation of what is going to be
put inside the field makes the receiver unreachable. This (pathological) case
happens in a situation such as $\mathtt{a.g.f=m(a)}$ where the method call
$\mathtt{m(a)}$ sets to $\nil$ the field $\mathtt{g}$ of the object
bound to $\mathtt{a}$.
Since we assume that the left-hand side is evaluated before
the right-hand side, the receiver is not necessarily $\nil$,
but the field updates might not be observable if $\mathtt{a.g.f}$ is only
reachable from $\mathtt{a}$. In the third and final
case for $\mathsf{put\_field}$ we consider the standard situation
when we write into a reachable field of a non-$\nil$ receiver.
The creation points of
the right-hand side are added to those already approximating the objects stored
in $f$.
The $\mathsf{call}$ operation restricts the scope to the parameters
passed to a method and hence $\xi$ is used.
The $\mathsf{return}$ operation copies into $\rs$ the return value of the
method which is held in $\Out$. The local variables of the caller
are put back into scope, but the approximation of their fields is provided
through a worst-case assumption $\mu^\top$ since they may be modified by
the call. This loss of precision
can be overcome by means of shadow copies of the variables, just as for $\e$
(see Example~\ref{ex:abstract_operations_er2}).
The $\mathsf{lookup}^m$ operation first computes the subset $e$ of the
approximation of the receiver of the call only containing the creation
points whose class leads to a call to the method $m$.  If $e =
\emptyset$, a call to $m$ is impossible and the result of the
operation is $\bot$.  Otherwise, $e$ becomes the approximation of the
receiver $\rs$, so that some creation points can disappear
and we need to call $\xi$.
\begin{example}\label{ex:abstract_operations_er}
\ifcorr{%
As in Example~\ref{ex:abstract_operations_e} for $\e$,
let us mimic, in $\er$, the concrete computation
of Example~\ref{ex:concrete_operations}.
We start from the abstraction (Definition~\ref{def:er_abstraction})
of $\sigma_1$, given in Example~\ref{ex:er_domain}.
Variables and fields not shown are implicitly bound to $\emptyset$ if they have
class type and to $*$ if they have type $\integer$.
\begin{align*}
  s_1&=\alpha_\tau^\er(\sigma_1)=\left[\begin{array}{l}
      \mathtt{f}\mapsto\{\pi_2\},\mathtt{n}\mapsto\emptyset\\
      \mathtt{this}\mapsto\{\overline{\pi}\}
    \end{array}\right]\sep
    \left[\begin{array}{l}
      \mathtt{next}\mapsto\{\pi_2\}\\
      \mathtt{rotation}\mapsto\emptyset
    \end{array}\right]\\
  s_2&=\mathsf{get\_var}_\tau^\mathtt{f}(s_1)=\left[\begin{array}{l}
      \mathtt{f}\mapsto\{\pi_2\},\mathtt{n}\mapsto\emptyset\\
      \rs\mapsto\{\pi_2\},\mathtt{this}\mapsto\{\overline{\pi}\}
    \end{array}\right]\sep
    \left[\begin{array}{l}
      \mathtt{next}\mapsto\{\pi_2\}\\
      \mathtt{rotation}\mapsto\emptyset
    \end{array}\right].
\end{align*}
There are three abstract $\mathsf{lookup}$ operations corresponding to the
concrete ones and hence we construct
for $i=3,\ldots,6$,  elements $s_i'$, $s_i''$, $s_i'''$ of $\er$
corresponding to the concrete states $\sigma_i'$, $\sigma_i''$, $\sigma_i'''$,
respectively.
\begin{align*}
  s_3'&=\mathsf{lookup}_{\tau[\rs\mapsto\mathtt{Figure}]}
    ^{\mathtt{def},\mathtt{Figure.def}}(s_2)=\bot\\
    &\text{since $e'=\{\pi\in\{\pi_2\}\mid M(\pi)(\mathtt{def})=
    \mathtt{Figure.def}\}=\emptyset$},\\
  s_3''&=\mathsf{lookup}_{\tau[\rs\mapsto\mathtt{Figure}]}
    ^{\mathtt{def},\mathtt{Square.def}}(s_2)=\xi_\tau(s_2)=s_2\\
    &\text{since $e''=\{\pi\in\{\pi_2\}\mid M(\pi)(\mathtt{def})=
    \mathtt{Square.def}\}=\{\pi_2\}$},\\
  s_3'''&=\mathsf{lookup}_{\tau[\rs\mapsto\mathtt{Figure}]}
    ^{\mathtt{def},\mathtt{Circle.def}}(s_2)=\bot\\
    &\text{since $e'''=\{\pi\in\{\pi_2\}\mid
      M(\pi)(\mathtt{def})=\mathtt{Circle.def}\}=\emptyset$.}
\end{align*}
The $\mathsf{lookup}$ operations for
$\mathtt{Figure}$ and $\mathtt{Circle}$ return $\bot$ so that,
as the abstract operations over $\er$ are strict on $\bot$
(Proposition~\ref{prop:operations_er}), $s_4' = s_5' = s_6' =
s_4'''=s_5'''=s_6'''= \bot$. This is because the analysis is able
to guess the target of the virtual call $\mathtt{f.def()}$, since the only
creation point for the receiver $\mathtt{f}$ is $\pi_2$, which creates
$\mathtt{Square}$s. Hence we only have to consider the case when a
$\mathtt{Square}$ is selected:
\begin{align*}
  s_4''&=\mathsf{call}^\mathtt{Square.def}_{\tau[\rs\mapsto\mathtt{Square}]}
    (s_3'')\\
  &=\xi_{[\rs\mapsto\mathtt{Square}]}\left(
    [\mathtt{this}\mapsto\{\pi_2\}]
    \sep\left[\begin{array}{l}
      \mathtt{next}\mapsto\{\pi_2\}\\
      \mathtt{rotation}\mapsto\emptyset
    \end{array}\right]\right)\\
  &=[\mathtt{this}\mapsto\{\pi_2\}]
    \sep[\mathtt{next}\mapsto\{\pi_2\},\mathtt{rotation}\mapsto\emptyset].
\end{align*}
Since $P(\mathtt{Square.def})|_{\Out}
=[\Out\mapsto\integer]$ and $\er_{[\Out\mapsto\integer]}=\{\bot,
[\Out\mapsto *]\sep[\mathtt{next}\mapsto\emptyset
,\ \mathtt{rotation}\mapsto\emptyset]\}$,
we can just observe that the method
$\mathtt{Square.def}$ does not diverge to conclude that
\[
  s_5''=[\Out\mapsto *]\sep[\mathtt{next}\mapsto\emptyset
    ,\ \mathtt{rotation}\mapsto\emptyset].
\]
Hence, by letting $\mu^\top$ denote the top element of $\Memory^\er$
so that
\[
   \mu^\top =
    \left[
      \mathtt{next}\mapsto\{\pi_2,\pi_3\},
      \mathtt{rotation}\mapsto\{\pi_1,\pi_4\}
   \right],
\]
we have
\begin{align*}
  s_6''
    &=\mathsf{return}^\mathtt{Square.def}
      _{\tau[\rs\mapsto\mathtt{Square}]}(s_3'')(s_5'')\\
    &=\mathsf{return}_{\tau[\rs\mapsto\mathtt{Square}]}^\mathtt{Square.def}
      (s_2)(s_5'')\\
    &=\xi_{\left[
    \begin{smallmatrix}
      \mathtt{f}\mapsto\mathtt{Figure},\ \mathtt{n}\mapsto\mathtt{Figure},\\
      \Out\mapsto\integer,\ \mathtt{this}\mapsto\mathtt{Scan}
    \end{smallmatrix}
          \right]}
          \left(\left[
      \mathtt{f}\mapsto\{\pi_2\},\mathtt{n}\mapsto\emptyset,
      \mathtt{this}\mapsto\{\overline{\pi}\}
    \right]\sep \mu^{\top} \right)
       \cup\\
  &\qquad\cup([\rs\mapsto *]\sep[\mathtt{next}\mapsto\emptyset,
    \ \mathtt{rotation}\mapsto\emptyset])\\
  &=\left[
      \mathtt{f}\mapsto\{\pi_2\},\mathtt{n}\mapsto\emptyset,
      \mathtt{this}\mapsto\{\overline{\pi}\}
    \right]\sep \mu^{\top}.
\end{align*}
Since the abstract semantics is non-deterministic, we merge the results
of every thread of execution through the $\cup$ operation. Hence the abstract
state after the execution of the call $\mathtt{f.def()}$ in Figure
\ref{fig:program} is
\[
  s_6=s_6'\cup s_6''\cup s_6'''=\bot\cup s_6''\cup\bot=s_6''.
\]
\qed
}
\end{example}

The abstract state $s_6''$ shows that the imprecision problem of $\e$,
related to the $\mathsf{return}$ operation, is still present in $\er$.
By comparing $s_2$ with $s_6''$, it can be seen that
the $\mathsf{return}$ operation makes a very pessimistic
assumption about the possible creation points for the $\mathtt{next}$
and $\mathtt{rotation}$ fields.
In particular, from $s_6''$ it seems that creation points
$\pi_3$ and $\pi_4$ are reachable (they belong to $\mu^\top$),
which is not the case in the concrete state
(compare this with $\sigma_6''$ in Example~\ref{ex:concrete_operations}).
As for the domain $\e$,
this problem can be solved by including, in the state of the callee,
\emph{shadow copies} of the parameters of the caller.
This is implemented through a preprocessing of the bodies of the
methods which prepend statements of the form
$v'\mathtt{:=}v$ for each parameter $v$, where $v'$ is the shadow
copy of $v$. Since shadow copies are fresh new variables, not
already occurring in the method's body, their value is never changed.
In this way, at the end of the method we know which creation points are
reachable from the fields of the objects bound to such parameters.
%
\begin{example}\label{ex:abstract_operations_er2}
\ifcorr{%
Let us reexecute the abstract computation of
Example~\ref{ex:abstract_operations_er}, but including shadow copies of the
parameters in the abstract states. We denote by $p'$ the shadow copy of the
parameter $p$. We assume that the method $\mathtt{scan}$ was called
with an actual parameter $\nil$ for the formal parameter $\mathtt{n}$.
The abstract state $s_1$ contains now two shadow copies
\[
  s_1=\left[\begin{array}{l}
      \mathtt{f}\mapsto\{\pi_2\},
        \mathtt{n}\mapsto\emptyset,\mathtt{n}'\mapsto\emptyset\\
      \mathtt{this}\mapsto\{\overline{\pi}\},
      \mathtt{this}'\mapsto\{\overline{\pi}\}
    \end{array}\right]\sep
    \left[\begin{array}{l}
      \mathtt{next}\mapsto\{\pi_2\}\\
      \mathtt{rotation}\mapsto\emptyset
    \end{array}\right].
\]
The same change applies to the abstract states $s_2$ and
$s_3''$ in Example~\ref{ex:abstract_operations_er}. The abstract state
$s_4''$ uses a new shadow copy for the actual parameter
of the method $\mathtt{Square.def}$ \ie
\[
  s_4''=[\mathtt{this}\mapsto\{\pi_2\},\mathtt{this}'\mapsto
    \{\pi_2\}]\sep[
    \mathtt{next}\mapsto\{\pi_2\},\ \mathtt{rotation}\mapsto\emptyset].
\]
The static analysis of the method $\mathtt{Square.def}$ easily concludes that
no object has been created. Hence now we have
\[
  s_5''=[\Out\mapsto *,\mathtt{this}'\mapsto\{\pi_2\}]\sep[
    \mathtt{next}\mapsto\{\pi_2\},\ \mathtt{rotation}\mapsto\{\pi_1\}].
\]
Note that the abstract memory is not empty now (compare with
Example~\ref{ex:abstract_operations_er}). This is because the shadow variable
$\mathtt{this}'$ prevents the abstract garbage collector from deleting the
creation point $\pi_2$ from $\mathtt{next}$ and the creation point
$\pi_1$ from $\mathtt{rotation}$.
Moreover, we know what is reachable at the end of
the execution of the $\mathtt{Square.def}$ method from its
parameters (through their shadow copies). Therefore we do not need to apply any
pessimistic assumption at $\mathsf{return}$ time, and we can define the
abstract $\mathsf{return}$ operation
in such a way that it just transfers
the result of the method call into the $\rs$ variable:
\[
  \mathsf{return}_\tau^\nu(\phi_1\sep\mu_1)(\phi_2\sep\mu_2)
    =\xi_{\tau[\rs\mapsto P(\nu)(\Out)]}
    (\phi_1[\rs\mapsto\phi_2(\Out)]\sep\mu_2)
\]
so that we have
\[
  s_6''=\left[\begin{array}{l}
      \mathtt{f}\mapsto\{\pi_2\},
        \mathtt{n}\mapsto\emptyset,\mathtt{n}'\mapsto\emptyset\\
      \mathtt{this}\mapsto\{\overline{\pi}\},
      \mathtt{this}'\mapsto\{\overline{\pi}\}
    \end{array}\right]\sep
    \left[\begin{array}{l}
      \mathtt{next}\mapsto\{\pi_2\}\\
      \mathtt{rotation}\mapsto\{\pi_1\}
    \end{array}\right].
\]
Note that creation points $\pi_3$ and $\pi_4$ are no longer
reachable (compare with Example~\ref{ex:abstract_operations_er}).
\qed
}
\end{example}
As previously noted in Subsection~\ref{subsec:e}, shadow copies of the
parameters are also useful for dealing with methods that modify their
formal parameters.

There was another problem with $\e$, related to the fact that $\e$ does not
distinguish between different variables
(see end of Section~\ref{sec:edomain}). It is not surprising that
$\er$ solves that problem, as shown below.
\begin{example}\label{ex:abstract_operations_er3}
\ifcorr{%
Let the approximation provided by $\er$ before the statement
$\mathtt{f\ =\ new\ Circle()}$ in Figure~\ref{fig:program} be
\[
  s=[\mathtt{f}\mapsto\{\pi_2\},\mathtt{this}\mapsto\{\overline{\pi}\}]
    \sep[\mathtt{next}\mapsto\{\pi_2\},\mathtt{rotation}\mapsto\{\pi_1,\pi_4\}]
\]
(Example~\ref{ex:er_precision}). We can compute the approximation \emph{after}
that statement by executing two abstract operations.
Since the object created at $\pi_3$ gets stored
inside the variable $\mathtt{f}$, we expect the creation point $\pi_2$
of the old value of $\mathtt{f}$ to disappear from the approximation
of $\mathtt{f}$, which is what actually happens:
\begin{align*}
  &s'=\mathsf{new}_\tau^{\pi_3}(s)\\
  &=\left[\begin{array}{l}
      \mathtt{f}\mapsto\{\pi_2\},\rs\mapsto\{\pi_3\},\\
      \mathtt{this}\mapsto\{\overline{\pi}\}
    \end{array}\right]
    \sep\left[\begin{array}{l}
      \mathtt{next}\mapsto\{\pi_2\},\\
      \mathtt{rotation}\mapsto\{\pi_1,\pi_4\}\end{array}\right],\\
  &\mathsf{put\_var}^\mathtt{f}_{\tau[\rs\mapsto\mathtt{Circle}]}(s')\\
  &=\xi_\tau\left([\mathtt{f}\mapsto\{\pi_3\},
    \mathtt{this}\mapsto\{\overline{\pi}\}]
    \sep\left[\begin{array}{l}
      \mathtt{next}\mapsto\{\pi_2\}\\
      \mathtt{rotation}\mapsto\{\pi_1,\pi_4\}
    \end{array}\right]\right)\\
  &=[\mathtt{f}\mapsto\{\pi_3\},
    \mathtt{this}\mapsto\{\overline{\pi}\}]
        \sep\left[\begin{array}{l}
      \mathtt{next}\mapsto\{\pi_2\}\\
      \mathtt{rotation}\mapsto\{\pi_1,\pi_4\}
    \end{array}\right].
\end{align*}
Note that the creation points for $\mathtt{rotation}$ do not disappear,
since from the $\mathtt{Circle}$ bound to $\mathtt{f}$ it might be
possible to reach a $\mathtt{Square}$ through its $\mathtt{next}$ field,
and a $\mathtt{Square}$ has a $\mathtt{rotation}$ field.
\qed
}
\end{example}
\subsection{$\er$ is a Refinement of $\e$}\label{subsec:refinement}
We have called $\er$ a \emph{refinement} of $\e$. In order to give
this word a formal justification, we show here that
$\er$ actually includes the elements of $\e$.
Namely, we show how every element $\ee\in\e$ can be \emph{embedded} into
an element $\theta(\ee)$ of $\er$, such that $\ee$ and $\theta(\ee)$
have the same concretisation \ie they represent the same property
of concrete states.
The idea, formalised in Definition \ref{def:implementation},
is that every variable or field must be bound in $\er$ to all those
creation points in $\ee$ compatible with its type.
\begin{definition}[Embedding of $\e$ into $\er$]
  \label{def:implementation}
Let $s\subseteq\Pi$. We define
$\vartheta_\tau(s)\in\Frame^\er_\tau$ such that, for every $v\in\domain(\tau)$,
\[
  \vartheta_\tau(s)(v)=\begin{cases}
    * & \text{if $\tau(v)=\integer$}\\
    \{\pi\in s\mid k(\pi)\le\tau(v)\} & \text{if $\tau(v)\in\mathcal{K}$.}
  \end{cases}
\]
The \emph{embedding} $\theta_\tau(\ee)\in\er_\tau$ of
$\ee\in\e_\tau$ is $\theta_\tau(\ee)=\xi_\tau(\vartheta_\tau(\ee)\sep
\vartheta_{\widetilde{\tau}}(\ee))$.
\end{definition}
\begin{example}\label{ex:implementation}
\ifcorr{%
Let $\tau_{w_1}$ be as given in Figure~\ref{fig:state} and
$\ee=\{\overline{\pi},\pi_1,\pi_2,\pi_3\}\in\e_{\tau_{w_1}}$
(Example~\ref{ex:ew}). Then
\[
  \theta_{\tau_{w_1}}(\ee)=\left[\begin{array}{l}
      \mathtt{f}\mapsto\{\pi_2,\pi_3\},\mathtt{n}\mapsto\{\pi_2,\pi_3\}\\
      \Out\mapsto *,\mathtt{this}\mapsto\{\overline{\pi}\}
    \end{array}\right]\sep
    \left[\begin{array}{l}
      \mathtt{next}\mapsto\{\pi_2,\pi_3\}\\
      \mathtt{rotation}\mapsto\{\pi_1\}
    \end{array}\right]
\]
where the missing fields are implicitly bound
to $*$ since they have $\mathit{int}$ type.
\qed
}
\end{example}

Proposition~\ref{prop:inclusion} states that the embedding
of Definition~\ref{def:implementation} is correct.
The proof proceeds by showing
that $\theta_\tau(e)$ is an element of $\er_\tau$ and
approximates exactly the same concrete states as $e$, that is,
for every element of $\e$ there is an element
of $\er$ which represents exactly the same set of concrete states.
\begin{proposition}\label{prop:inclusion}
Let $\gamma_\tau^\e$ and $\gamma_\tau^\er$ be the
concretisation maps induced by the abstraction maps of
Definitions~\ref{def:alpha} and~\ref{def:er_abstraction}, respectively.
Then $\gamma_\tau^\e(\e_\tau)\subseteq\gamma_\tau^\er(\er_\tau)$.
\end{proposition}
The following example shows that the inclusion relation
in Proposition~\ref{prop:inclusion} must be strict.
\begin{example}\label{ex:strict_inclusion}
\ifcorr{%
Let $\tau_{w_1}$ be as given in Figure~\ref{fig:state}.
By Example~\ref{ex:ew} we know that $\emptyset\in\e_{\tau_{w_1}}$ and that
every $\ee\in\e_{\tau_{w_1}}\!\setminus\{\emptyset\}$
must contain $\overline{\pi}$. Moreover, we know that
$\{\pi_1\}\not\in\e_{\tau_{w_1}}$ and
$\{\pi_4\}\not\in\e_{\tau_{w_1}}$. Hence
\[
  \#\e_{\tau_{w_1}}\le 1+(\#\wp(\{\pi_1,\pi_2,\pi_3,\pi_4\})-2)<2^4.
\]
For what concerns $\er_{\tau_{w_1}}$, note that for every
$\ee_1,\ee_2\in\wp(\{\pi_2,\pi_3\})$ the element
\[
  [\mathtt{f}\mapsto\ee_1,\mathtt{n}\mapsto\ee_2,\Out\mapsto *,
    \mathtt{this}\mapsto\{\overline{\pi}\}]\sep[\mathtt{next}
    \mapsto\emptyset,\mathtt{rotation}\mapsto\emptyset,\ldots]
\]
is a fixpoint of $\xi_\tau$ and hence an element of $\er_{\tau_{w_1}}$
(Definition~\ref{def:er_domain}). Hence
\[
  \#\er_{\tau_{w_1}}\ge(\#\wp(\{\pi_2,\pi_3\}))^2=2^4>\#\e_{\tau_{w_1}}.
\]
\qed
}
\end{example}
\section{Implementation}\label{sec:implementation}
\ifcorr{}
\ifthenelse{\boolean{PROOFSONLY}}{}
{
In this section, we present our practical evaluation of the
abstract domain $\er$.
In Subsection~\ref{subsec:julia}, we describe the implementation of
$\er$ used to do the experiments and, in
Subsection~\ref{subsec:tests}, we present the experimental results.

%
%
\subsection{Analysing Java Bytecode}\label{subsec:julia}
We implemented the abstract domain $\er$ inside \textsc{Julia}~\cite{julia}.
This is a generic static analyser written in Java
that is designed for analysing full Java bytecode.
Generic means that \textsc{Julia} does not
embed any abstract domain but, instead, can be instantiated
for a specific static analysis once
an appropriate abstract domain and the attached abstract
operations are provided.
For instance, \textsc{Julia} can perform \emph{rapid type
analysis} (a kind of \emph{class analysis}~\cite{BaconS96})
or instead escape analysis through $\er$
by simply swapping these abstract domains.

In order to target the escape analysis of real Java bytecode programs,
the implementation had to address a number of problems due to
features of the Java bytecode itself.
We describe the main problems and how we addressed them.
These problems were:
\begin{enumerate}
\item
\label{prob:local-stack}
      the Java Virtual Machine frame contains
      both local variables and an operand stack and the number of
      elements in the operand stack can change within the same method,
      although its size at a given program point is fixed and statically known;
\item
\label{prob:library-classes}
      a very large number of library classes are likely to be called and,
      hence, would need to be analysed;
\item
\label{prob:gotos}
      Java bytecode is unstructured \ie lacking any explicit
      scope structure and code is weaved through an extensive use
      of explicit \texttt{goto} jumps;
\item
\label{prob:exceptions}
      since the Java bytecode makes extensive use of exceptions, the
      control flow for exceptions must also be considered.
\item
\label{prob:static}
      Java bytecode has static fields, which are like global variables
      of traditional imperative languages, and are always in scope,
      so that the objects bound to static fields cannot be garbage-collected.
\end{enumerate}

We solved Problem (\ref{prob:local-stack}) by rewording our notion of frame
(Definition~\ref{def:frames1}) into a set of local variables and a stack of
variables. The number of stack variables (elements) in a given program point is
statically determined since \texttt{.class} files must be
verifiable~\cite{LindholmY99}.
Since Java bytecode holds intermediate results in the operand stack,
the latter plays the role of our $\rs$ variable.

%

We dealt with Problem (\ref{prob:library-classes})
by analysing some library classes only, and making
\emph{worst-case assumptions}~\cite{CousotC02}
about the behaviour of calls to methods
of other classes. This means that we assume that such calls can potentially
do everything, such as storing the parameters into (instance or class)
fields or returning
objects created in every creation point $\pi$ (with the restriction, however,
that $\pi$ creates objects
of class compatible with the return type of the method). It is easy to see
how an extensive use of this policy quickly leads to imprecision.
The situation
is made worse in Java (bytecode) because of constructor chaining, stating that
every call to a constructor eventually leads to a constructor in
the \emph{library} class \texttt{java.lang.Object}~\cite{ArnoldGH00}.
To cope with these problems, we allowed the analyser to access at least
the code of \texttt{java.lang.Object}.
We also allowed the analyser access to yet more library classes, leading
to more precise but also more costly analyses.
For many native methods, whose Java bytecode is not available, we have provided
hand-made bytecode stubs which agree with the declared abstract behaviour
of the methods.

Problem (\ref{prob:gotos})
was solved by building a graph of blocks of codes, each
bound to all its possible successors in the control-flow.
We use
class hierarchy analysis to deal with virtual calls whose target is
not explicitly embedded in the code~\cite{DeanGC95}.
Java bytecode subroutines (\ie the \texttt{jsr}/\texttt{ret} mechanism)
are handled by linking each block ending with a \texttt{jsr} to the block
starting with its target. The block ending with \texttt{ret} is then
conservatively linked with all blocks starting
with an instruction immediately following a \texttt{jsr} bytecode
\emph{in the same method} of \texttt{ret}.
The restriction to the
same method is correct because of a constraint imposed on
valid Java bytecode by the verification algorithm~\cite{LindholmY99}.
The resulting graph is the same as that of \emph{dominators}
that are defined in~\cite{ASU86} for much simpler languages.
The graph is then used for a fixpoint computation
by following the structure of its strongly connected components.

We solved Problem (\ref{prob:exceptions})
by using the technique pioneered in~\cite{JacobsP03}.
It consists in denoting a piece of code $c$ through a map from the input state
to the \emph{normal} output state
\emph{and} an \emph{exceptional} output state, representing the
state of the Java virtual machine if an exception has been thrown inside $c$.
Composition of commands uses the normal output state~\cite{SpotoJ03},
but composition with exception handlers uses the exceptional final state.

We dealt with Problem (\ref{prob:static}) by modifying the abstract garbage
collector of Definition~\ref{def:xi} so that it does not
garbage collect the creation points reachable from static fields.
Technically, this amounts to adding to the map $\rho_\tau$ of
Definition~\ref{def:xi} the creation points bound to the approximation
of the static fields of the classes of the program.

The abstract domain $\er$ is implemented inside
\textsc{Julia} as a Java class extending
an abstract (in the sense of Java~\cite{ArnoldGH00})
class standing for a generic abstract domain.
This class contains methods that compute the denotation of
every single Java bytecode
(\ie denotations similar to those given in Figure~\ref{fig:operations_er}
for our simplified bytecode).

The choice of representation for a denotation affects the speed of the
analysis. The representation we chose was a set-constraint~\cite{DovierPPR00}
between its input and output variables.
For instance, the denotation for the $\mathsf{new}^\pi$ operation in
Figure~\ref{fig:operations_er} is implemented as a constraint
$\{\pi\}\subseteq S$ over the unknown $S$. It states that the creation
point $\pi$ must belong to the set $S$ of the creation points for $\rs$ in the
output of the operations (\ie for the top of the operand stack when considering
the real Java bytecode).
We use $\subseteq$ instead of $=$ since there might be many possible
ways of reaching the program point that follows $\mathsf{new}^\pi$.
We use default reasoning to state that the other variables are unchanged.
This is a generalisation of the technique we introduced in~\cite{HillS03}.
As another example, the denotation for the $\mathsf{get\_var}^v$
operation in Figure~\ref{fig:operations_er}
is implemented as a constraint $S_1\subseteq S_2$ over the unknowns
$S_1$ and $S_2$. It states that the set $S_2$
of creation points for $\rs$ in the output must contain
the set $S_1$ of creation points for $v$ in the input.
We solve the set-constraints constructed from a program by propagation of
creation points. Namely, a constraint such as $\{\pi\}\subseteq S$ propagates
$\pi$ into $S$. A constraint such as
$S_1\subseteq S_2$ propagates the creation points inside $S_1$
into $S_2$. Propagation starts with empty
approximations for the unknowns and
continues until there is no further growth in these approximations.
We have implemented this propagation by exploiting a preliminary
topological sort of the unknowns of the constraints. Namely,
a constraint $S_1\subseteq S_2$ induces a pre-order (a reflexive and
transitive relation) $S_1\le S_2$. A topological sort \wrt this pre-order
builds a tree of strongly-connected
components of unknowns. A strongly-connected component represents
a set of mutually dependent unknowns. We propagate the
creation points by following the topological ordering backwards, so that
we can consider one strongly-connected component at a time. This technique
significantly speeds up the propagation.

There were two alternative choices we might have taken for the representation.
The simplest would have been an extensional definition,
in the form of an exhaustive
input/output tabling; but that would have been far too slow.
Alternatively, we could have used binary decision diagrams~\cite{Bryant86}
to represent the denotations; this traditional approach can
represent the denotations in a compact and efficient way.
This technique is certainly possible,
but it requires more technical work since we
have to code maps over sets of creation points through Boolean functions.
Moreover, since bytecodes usually apply local modifications to the
state (for instance, the $\mathsf{put\_var}^v$ bytecode in
Figure~\ref{fig:operations_er} leaves all variables other than $v$
untouched), they would be coded into binary decision diagrams
which mainly assert that the output variables are a copy of their input
counterparts. In terms of Boolean functions, this means that
such functions would contain a lot of \emph{if and only if} constraints,
which significantly increases the size of the diagram. By using
set-constraints, we solve this problem through default reasoning.

The use of set-constraints is appealing since we can easily use the
same unknown to represent two or more distinct approximations.
For instance,
different unknowns might represent the approximation of a field at
different program points, or rather the same unknown might represent
all those approximations. The second choice leads to a less precise
analysis but also to fewer unknowns and constraints than
the first choice. Thus, the second choice should lead to faster analyses.
In the first case we say that the approximation of the fields
is \emph{flow-sensitive}, while in the second case we say that it is
\emph{flow-insensitive}.
The same idea can be used for the approximations of local variables or
even operand stack elements. In Section~\ref{subsec:tests} we evaluate
the practical consequences of merging different approximations into
the same unknown.

The use of set-constraints for the representation
has, however, also a few negative consequences.
To keep the implementation simple and fast, our set-constraints
are built from equality, union and intersection only. But these
operations do not allow us
to represent
tests on the input variables (such as in \textsf{put\_field}, see
Figure~\ref{fig:operations_er}). Hence conservative approximations
must be made. For instance, the third case of the definition of
\textsf{put\_field} is \emph{always} used.
This is correct since it is a conservative approximation of all three cases.
Moreover, it does not introduce a significant precision loss since the first
alternative of the definition of \textsf{put\_field} deals with the
pathological case when a given \textsf{put\_field} is \emph{always} applied
to a $\nil$ receiver; and the second alternative deals with
the case when a given
\textsf{put\_field} is \emph{always} applied to a receiver
which is made unreachable by the evaluation of the value which must be
put inside the field. Both the first two alternatives correspond to legal
but quite unusual ways of using \textsf{put\_field}
and are almost never applicable.
For efficiency reasons, the $\xi$ garbage collector is only used
at the end of a method.
This is a correct approximation since $\xi$ is an
lco (Proposition~\ref{prop:xi_lco})
although \emph{forgetting} some of its applications can lose precision.
Also for efficiency reasons, we have used
memoisation to cache repeated calls to the abstract garbage collector.

We have described how we map the input abstract
state to the output (and exceptional) abstract state.
However, the information needed for stack allocation is related to some
internal points of the program.
For instance, in the program in Figure~\ref{fig:program},
we would like to know if the creation point $\pi_4$ inside
\texttt{rotate} could be stack allocated.
For this, we need to know the set of the
creation points of objects that are reachable at the end of \texttt{rotate}
from each return value,
from the (possible) objects thrown as an exception,
or from the fields of the objects bound to its parameter
(\ie the set $E$ of Subsection~\ref{subsec:e}).
Therefore, we need information related to some
internal program points.
This can be obtained by placing a \emph{watchpoint}
at every exit point of a method such as \texttt{rotate}.
Note that, with the analyser \textsc{Julia}, we can do this automatically
and obtain the set of creation points that can be stack allocated.
\subsection{Experimental Evaluation}\label{subsec:tests}
We report our experiments with the escape analysis through $\er$ of
some Java applications:
\texttt{Figures} is the program in Figure~\ref{fig:program}
fed with a list of \texttt{Circle}s;
\texttt{LimVect} is a small Java program used in~\cite{Blanchet03};
\texttt{Dhrystone} version $2.1$ is a testbench for numerical computations
(most of the arrays it creates can be stack allocated);
\texttt{ImageVwr} is an image visualisation applet;
\texttt{Morph} is an image morphing program;
\texttt{JLex} version
$1.2.6$ is the Java version of the well-known \texttt{lex}
lexical analysers generator;
\texttt{JavaCup} version $0.10j$ and \texttt{Javacc} version $3.2$
are compilers' compilers;
\texttt{Julia} version $0.39$ is our \textsc{Julia} analyser itself;
\texttt{Jess} version $6.1p7$
is a rule engine and scripting language for developing
rule-based expert systems.
All these benchmarks are free software
except \texttt{Jess} which is copyrighted.
Some of these programs were
analysed in~\cite{Blanchet03}; these are \texttt{LimVect}, \texttt{Dhrystone},
\texttt{JLex}, an older version of \texttt{Javacc} and an older version
of \texttt{Jess}.
Note that the newer versions of \texttt{Javacc} and \texttt{Jess}
considered here are bigger
than those used in~\cite{Blanchet03}.

Our experiments have been performed on a \textsf{Pentium}
$2.1$ Ghz machine with $1024$ megabytes of RAM, running
Linux $2.6$ and Sun Java Development Kit version
$1.5.0$ with HotSpot just-in-time compiler.

For each experiment, we report how many
Java classes, methods and bytecodes are analysed, as well as
the time taken by the analysis (in seconds)
to build and solve the set of constraints
generated for our escape analysis.
We first show the \emph{static} precision of the analyses
(Subsection~\ref{subsub:static}). Namely, we report the number of
creation points which can be stack allocated. We study how the
precision of the analyses is affected by flow sensitivity and by the ability
to approximate precisely each field.
Later (Subsection~\ref{subsub:dynamic}),
we report the \emph{dynamic} precision of the analyses \ie
the number of creation operations which are stack allocated \emph{at run-time}
and their relative ratio \wrt those which are heap allocated. We also
provide information on the amount of memory which is stack allocated rather
than heap allocated at run-time.
Finally (Subsection~\ref{subsub:cost}) we briefly discuss the cost of the
analyses.
\subsubsection{Static Tests}\label{subsub:static}
\begin{figure}[t]
\[\begin{array}{|r|r|r|r||r|r|r|r|r|r|}\hline
  \multicolumn{1}{|c}{\mbox{benchmark}} &
  \multicolumn{1}{|c}{\mbox{clss}} &
  \multicolumn{1}{|c}{\mbox{meth}} &
  \multicolumn{1}{|c||}{\mbox{bytec}} &
  \multicolumn{1}{|c}{\mbox{time}} &
  \multicolumn{2}{|c}{\textit{SA}
  } &
  \multicolumn{1}{|c|}{\textit{TT}
  } &
  \multicolumn{1}{|c|}{\textit{NC}
  } &
  \multicolumn{1}{|c|}{\textit{LIN}
  } \\\hline
  \mathtt{Figures} & 6 & 17 & 146
    & 0.06 & 1 & (20\%) & 0.41 & 109 & 2.067\\\hline
  \mathtt{LimVect} & 2 & 8 & 48
    & 0.02 & 1 & (0\%) & 0.42 & 46 & 1.631\\\hline
  \mathtt{Dhrystone} & 7 & 24 & 610
    & 0.17 & 4 & (25\%) & 0.28 & 253 & 1.640\\\hline
  \mathtt{ImageVwr} & 3 & 23 & 1238
    & 0.35 & 0 & (0\%) & 0.28 & 412 & 1.669\\\hline
  \mathtt{Morph} & 1 & 14 & 1367
    & 0.30 & 0 & (0\%) & 0.22 & 253 & 1.355\\\hline
  \mathtt{JLex} & 26 & 138 & 12520
    & 0.76 & 2 & (0\%) & 0.06 & 2904 & 1.974\\\hline
  \mathtt{JavaCup} & 37 & 317 & 14390
    & 1.76 & 1 & (0\%) & 0.12 & 5577 & 2.206\\\hline
  \mathtt{Julia} & 164 & 821 & 28507
    & 10.45 & 6 & (0\%) & 0.37 & 17653 & 2.672\\\hline
  \mathtt{Jess} & 268 & 1543 & 51663
    & 31.04 & 2 & (0\%) & 0.60 & 45614 & 2.914\\\hline
  \mathtt{Javacc} & 65 & 953 & 79325
    & 15.32 & 0 & (0\%) & 0.19 & 17759 & 2.162\\\hline
\end{array}\]
\caption{Flow insensitive escape analyses with $\er$. Fields are merged. Only
\texttt{java.lang.Object} is included in the analysis.
$\mathit{SA}$ is the number of creation points which are stack allocated;
$\mathit{TT}$ is the time per one thousand bytecodes;
$\mathit{NC}$ is the number of constraints generated;
$\mathit{LIN}$ is the linearity of the set of constraints.
}
  \label{fig:analyses_er}
\end{figure}
We start from the fastest but also less precise way of using our escape
analysis. Namely, the analyses are completely flow insensitive
and field insensitive, in the sense that
the fields are approximated into one variable
and, except for \texttt{java.lang.Object}, library classes are not included.
As we said in Subsection~\ref{subsec:julia}, calls to methods
of other library classes are approximated through a worst-case assumption.
In particular, this assumption states that the parameters passed to the call
escape, since they might be stored into a static field, and hence be accessible
after the call has returned. Because of constructor chaining, all
object creations result in a call to the constructor of
\texttt{java.lang.Object}. This is why the inclusion of that class is
a minimum requirement to the precision of the analysis. Otherwise, every
newly created object would escape as soon as it is initialised.
The results are shown in Figure~\ref{fig:analyses_er}, where for each
benchmark we report the number of classes, methods and bytecodes
analysed and the time of the analysis (in seconds).

Figure~\ref{fig:analyses_er} also reports the number of set-constraints
generated for the analysis.
These constraints are organised into a graph.
Each variable in the constraints is a node in the graph; nodes
are connected if they are related by some constraint.
The \emph{linearity} column reports
the average size of a strongly-connected component.
Linearity is equal to $1.000$ for fully non-recursive programs without cycles.
Higher values of linearity represent programs which use recursion and cycles
extensively. For a given number of constraints, their solution is computed
more efficiently if linearity is low.

Although the analyses in Figure~\ref{fig:analyses_er} are relatively fast,
it can be seen that almost no creation points
are found to be stack allocatable.
The analyses can be made more precise if flow sensitivity is used, at least
for the operand stack.
Results using this level of flow sensitivity are shown in
Figure~\ref{fig:analyses_er2}.
They are more precise than those in Figure~\ref{fig:analyses_er},
but the analyses are also more expensive.
If we also analyse some library classes, the precision of the analyses
improves further.
Namely, we decided to add part of the \texttt{java.lang} and
\texttt{java.util} standard Java packages. Such classes are chosen
in such a way to include typical candidates for stack allocation and to
form an upward closed set, so that constructor chaining
for those classes never goes out of the set of analysed classes.
The results with these additions are shown
in Figure~\ref{fig:analyses_er3}.
We only count the creation points inside the classes of the application,
so that numbers are comparable with those in Figures~\ref{fig:analyses_er}
and~\ref{fig:analyses_er2}.
In comparison with Figure~\ref{fig:analyses_er2}, we manage to stack allocate
many more creation points, but with a further increase in
the cost of the analyses.
\begin{figure}
\[\begin{array}{|r|r|r|r||r|r|r|r|r|r|}\hline
  \multicolumn{1}{|c}{\mbox{benchmark}} &
  \multicolumn{1}{|c}{\mbox{clss}} &
  \multicolumn{1}{|c}{\mbox{meth}} &
  \multicolumn{1}{|c||}{\mbox{bytec}} &
  \multicolumn{1}{|c}{\mbox{time}} &
  \multicolumn{2}{|c}{\textit{SA}
  } &
  \multicolumn{1}{|c|}{\textit{TT}
  } &
  \multicolumn{1}{|c|}{\textit{NC}
  } &
  \multicolumn{1}{|c|}{\textit{LIN}
  } \\\hline
  \mathtt{Figures} & 6 & 17 & 146
    & 0.08 & 3 & (60\%) & 0.54 & 454 & 1.003\\\hline
  \mathtt{LimVect} & 2 & 8 & 48
    & 0.06 & 1 & (33\%) & 1.25 & 214 & 1.007\\\hline
  \mathtt{Dhrystone} & 7 & 24 & 610
    & 0.20 & 8 & (50\%) & 0.33 & 2784 & 1.174\\\hline
  \mathtt{ImageVwr} & 3 & 23 & 1238
    & 0.50 & 0 & (0\%) & 0.40 & 7831 & 1.377\\\hline
  \mathtt{Morph} & 1 & 14 & 1367
    & 0.35 & 0 & (0\%) & 0.26 & 6483 & 1.110\\\hline
  \mathtt{JLex} & 26 & 138 & 12520
    & 2.40 & 11 & (5\%) & 0.19 & 60379 & 1.298\\\hline
  \mathtt{JavaCup} & 37 & 317 & 14390
    & 14.98 & 15 & (3\%) & 1.04 & 124097 & 1.578\\\hline
  \mathtt{Julia} & 164 & 821 & 28472
    & 33.30 & 30 & (3\%) & 1.17 & 217874 & 1.788\\\hline
  \mathtt{Jess} & 268 & 1543 & 51663
    & 51.77 & 51 & (3\%) & 1.00 & 335010 & 1.645\\\hline
  \mathtt{Javacc} & 65 & 953 & 79325
    & 239.22 & 45 & (3\%) & 3.01 & 382862 & 1.629\\\hline
\end{array}\]
\caption{Flow sensitive (on the operand stack only)
escape analyses with $\er$. Fields are merged. Only
\texttt{java.lang.Object} is included in the analysis.}
  \label{fig:analyses_er2}
\end{figure}

\begin{figure}
\[\begin{array}{|r|r|r|r||r|r|r|r|r|r|}\hline
  \multicolumn{1}{|c}{\mbox{benchmark}} &
  \multicolumn{1}{|c}{\mbox{clss}} &
  \multicolumn{1}{|c}{\mbox{meth}} &
  \multicolumn{1}{|c||}{\mbox{bytec}} &
  \multicolumn{1}{|c}{\mbox{time}} &
  \multicolumn{2}{|c}{\textit{SA}
  } &
  \multicolumn{1}{|c|}{\textit{TT}
  } &
  \multicolumn{1}{|c|}{\textit{NC}
  } &
  \multicolumn{1}{|c|}{\textit{LIN}
  } \\\hline
  \mathtt{Figures} & 6 & 17 & 146
    & 0.10 & 3 & (60\%) & 0.68 & 454 & 1.003\\\hline
  \mathtt{LimVect} & 4 & 11 & 1057
    & 0.18 & 1 & (33\%) & 0.17 & 2743 & 1.007\\\hline
  \mathtt{Dhrystone} & 14 & 54 & 2240
    & 0.35 & 12 & (75\%) & 0.16 & 7157 & 1.116\\\hline
  \mathtt{ImageVwr} & 47 & 209 & 8557
    & 1.64 & 1 & (8\%) & 0.19 & 43152 & 1.242\\\hline
  \mathtt{Morph} & 18 & 93 & 7302
    & 1.20 & 1 & (5\%) & 0.16 & 33512 & 1.168\\\hline
  \mathtt{JLex} & 72 & 396 & 21025
    & 3.71 & 48 & (22\%) & 0.18 & 94243 & 1.248\\\hline
  \mathtt{JavaCup} & 84 & 569 & 23196
    & 11.98 & 213 & (39\%) & 0.52 & 147139 & 1.434\\\hline
  \mathtt{Julia} & 530 & 2007 & 60846
    & 50.47 & 331 & (41\%) & 0.83 & 348621 & 1.551\\\hline
  \mathtt{Jess} & 363 & 2151 & 71329
    & 53.71 & 289 & (22\%) & 0.75 & 399188 & 1.539\\\hline
  \mathtt{Javacc} & 160 & 1489 & 97981
    & 65.75 & 878 & (61\%) & 0.67 & 418552 & 1.436\\\hline
\end{array}\]
\caption{Flow sensitive (on the operand stack only) escape analyses with the
abstract domain $\er$.
Fields are merged. The standard library classes
\texttt{java.lang.}$\{$\texttt{Object},
\texttt{CharSequence},
\texttt{String*},
\texttt{AbstractStringBuilder},
\texttt{Integer},\texttt{Number},
\texttt{Character}$\}$ and
\texttt{java.util.}$\{$\texttt{AbstractList},
\texttt{AbstractCollection},
\texttt{Vector},
\texttt{HashMap},\texttt{Hashtable},
\texttt{AbstractMap}$\}$ are included in the analysis.
For $\mathtt{Julia}$,
we also included the \textsf{bcel} libraries for bytecode manipulation.}
  \label{fig:analyses_er3}
\end{figure}
\begin{figure}
\[\begin{array}{|r|r|r|r||r|r|r|r|r|r|}\hline
  \multicolumn{1}{|c}{\mbox{benchmark}} &
  \multicolumn{1}{|c}{\mbox{clss}} &
  \multicolumn{1}{|c}{\mbox{meth}} &
  \multicolumn{1}{|c||}{\mbox{bytec}} &
  \multicolumn{1}{|c}{\mbox{time}} &
  \multicolumn{2}{|c}{\textit{SA}
  } &
  \multicolumn{1}{|c|}{\textit{TT}
  } &
  \multicolumn{1}{|c|}{\textit{NC}
  } &
  \multicolumn{1}{|c|}{\textit{LIN}
  } \\\hline
  \mathtt{Figures} & 6 & 17 & 146
    & 0.13 & 3 & (60\%) & 0.89 & 454 & 1.003\\\hline
  \mathtt{LimVect} & 4 & 11 & 1057
    & 0.18 & 1 & (33\%) & 0.17 & 2743 & 1.001\\\hline
  \mathtt{Dhrystone} & 14 & 54 & 2240
    & 0.34 & 12 & (75\%) & 0.15 & 7157 & 1.080\\\hline
  \mathtt{ImageVwr} & 47 & 209 & 8557
    & 1.76 & 1 & (8\%) & 0.20 & 43174 & 1.218\\\hline
  \mathtt{Morph} & 18 & 93 & 7302
    & 1.14 & 1 & (5\%) & 0.16 & 33526 & 1.158\\\hline
  \mathtt{JLex} & 72 & 396 & 21025
    & 3.11 & 49 & (23\%) & 0.15 & 94383 & 1.153\\\hline
  \mathtt{JavaCup} & 84 & 569 & 23196
    & 9.24 & 215 & (39\%) & 0.40 & 147159 & 1.319\\\hline
  \mathtt{Julia} & 530 & 2007 & 60846
    & 38.61 & 336 & (41\%) & 0.63 & 348799 & 1.428\\\hline
  \mathtt{Jess} & 363 & 2151 & 71329
    & 56.71 & 289 & (22\%) & 0.79 & 399289 & 1.488\\\hline
  \mathtt{Javacc} & 160 & 1489 & 97981
    & 59.45 & 895 & (62\%) & 0.61 & 419067 & 1.378\\\hline
\end{array}\]
\caption{Flow sensitive (on the operand stack only) escape analyses with the
abstract domain $\er$.
A specific approximation is used for each field.
The same library classes as in Figure~\ref{fig:analyses_er3}
are included in the analysis.
For $\mathtt{Julia}$,
we also included the \textsf{bcel} libraries for bytecode manipulation.}
  \label{fig:analyses_er4}
\end{figure}

The final experiments used the full power of
the $\er$ abstract domain by providing a (flow insensitive)
specific approximation for each field. The results are shown in
Figure~\ref{fig:analyses_er4}. The precision is just slightly better
than in Figure~\ref{fig:analyses_er3}, and the analyses
require less time. They are sometimes even faster than those
in Figure~\ref{fig:analyses_er2}.
This reduction in time might seem surprising. However, this is a consequence
of the fact that a field-specific approximation slightly increases
the number of constraints, but reduces
linearity and the average size of creation point sets;
hence, less time is needed to solve the constraints
(compare the linearity columns in Figures~\ref{fig:analyses_er3}
and~\ref{fig:analyses_er4}). A similar behaviour has been experienced
in~\cite{RountevMR01}.
More generally, it has been witnessed in different contexts that increasing the
precision of a static analysis may yield faster computations, since
an imprecise analysis yields spurious execution paths which
slow down the analysis itself.

Beyond these experiments, we also tried to include more library classes in the
analysis (such as all \texttt{java.lang} and \texttt{java.util} packages),
but the results were very similar to those in Figure~\ref{fig:analyses_er4},
confirming the claim in~\cite{Blanchet03} that most of the
stack allocatable objects are arrays,
\texttt{java.lang.StringBuffer}s and a few objects of the collection
classes (vectors and sets).
We also tried to use flow sensitivity for the local variables and the
field approximations but this did not improve the results.
This behaviour can be
explained, for local variables, by observing that typical Java compilers
do not try to recycle local variables if they can be used for different tasks
in different parts of a method. Hence one approximation per method is
enough.
Note that both conclusions agree with the results provided
in~\cite{ChoiGSSM03} where flow sensitivity looks
useless for stack allocation and  a bounded field approximation
reduces the precision of stack allocation in at most one case in ten.
The analysis in~\cite{ChoiGSSM03} works for Java instead of Java
bytecode, so flow sensitivity for the operand stack is meaningless in their
case.

The precision of the analyses seems similar to that of the experiments
reported in~\cite{Blanchet03}. The results reported in
Figure~\ref{fig:analyses_er4} for the first five benchmarks are actually
optimal, in the sense that no other creation point can ever be stack allocated.
For the other five benchmarks, the exact comparison is hard since the
older versions of the benchmarks, as analysed in~\cite{Blanchet03}, are
not available anymore. See, however, Section~\ref{sec:discussion}
for a theoretical comparison.

The overall conclusion we draw from our experiments is that flow sensitivity
is important, but only for the stack variables. The inclusion of library
classes is also essential for the precision although,
in practice, only very few
classes are needed. The ability to approximate each field
individually does not contribute significantly to the precision of the
analyses, but improves their efficiency.
\subsubsection{Dynamic Tests}\label{subsub:dynamic}
The static measurements in Subsection~\ref{subsub:static} have been
useful to compare the relative precision and cost of different
implementations of our escape analysis.
However, another piece of information, important
for an escape analysis, is the number of creation operations that are actually
avoided \emph{at run-time} because their creation point has been
stack allocated.
As well as the size of the objects which are stack allocated
\wrt the size of the objects which are heap allocated.
We computed these measurements for
the analyses reported in Figure~\ref{fig:analyses_er4} only, which are
the most precise escape analyses which we managed to implement with
$\er$. Some results are shown in Figure~\ref{fig:dynamic}. For each
benchmark, we report the number of objects and the amount of memory allocated
in the stack or in the heap.
In Section~\ref{sec:discussion}, these results are compared with
results reported for other escape analysers. Here we just note that
the poor result for the escape analysis of \texttt{Julia} is a consequence
of the fact that \texttt{Julia} mainly computes a large set of constraints
which escape from their creating methods to flow into the methods that
solve them. We note that escape analysis is of little use for this
type of program.
%
\begin{figure}
\[
\begin{array}{c}
\begin{array}{|r||r|r|r|r|}\hline
  \multicolumn{1}{|c||}{\mbox{benchmark}} &
  \multicolumn{2}{|c}{\begin{array}{c}
    \mbox{objects in}\\
    \mbox{the stack}
  \end{array}} &
  \multicolumn{2}{|c|}{\begin{array}{c}
    \mbox{objects in}\\
    \mbox{the heap}
  \end{array}} \\\hline
  \mathtt{Figures} & 101 & 97.11\% & 3 & 2.89\% \\\hline
  \mathtt{LimVect} & 1 & 33.33\% & 2 & 66.66\% \\\hline
  \mathtt{Dhrystone} & 200009 & 99.99\% & 4 & 0.01\% \\\hline
  \mathtt{JLex} & 672 & 8.92\% & 6854 & 91.08\% \\\hline
  \mathtt{JavaCup} & 141875 & 62.37\% & 85564 & 37.63\% \\\hline
  \mathtt{Julia} & 5534 & 6.56\% & 78748 & 93.44\% \\\hline
  \mathtt{Jess} & 924 & 7.81\% & 10897 & 92.19\% \\\hline
  \mathtt{Javacc} & 26461 & 43.72\% & 34050 & 56.28\% \\\hline
\end{array}\\
\mbox{}\\
\begin{array}{|r||r|r|r|r|}\hline
  \multicolumn{1}{|c||}{\mbox{benchmark}} &
  \multicolumn{2}{|c}{\begin{array}{c}
    \mbox{memory in}\\
    \mbox{the stack}
  \end{array}} &
  \multicolumn{2}{|c|}{\begin{array}{c}
    \mbox{memory in}\\
    \mbox{the heap}
  \end{array}} \\\hline
  \mathtt{Figures} & 2024 & 98.82\% & 24 & 1.18\%\\\hline
  \mathtt{LimVect} & 12 & 30.00\% & 28 & 70.00\%\\\hline
  \mathtt{Dhrystone} & 1600140 & 96.03\% & 66144 & 3.97\%\\\hline
  \mathtt{JLex} & 147060 & 47.77\% & 160772 & 52.23\% \\\hline
  \mathtt{JavaCup} & 1580336 & 48.76\% & 1660516 & 51.24\% \\\hline
  \mathtt{Julia} & 70764 & 5.57\% & 1198120 & 94.47\%\\\hline
  \mathtt{Jess} & 13328 & 4.90\% & 258128 & 95.10\% \\\hline
  \mathtt{Javacc} & 604244 & 40.72\% & 883224 & 59.28\% \\\hline
\end{array}
\end{array}
\]
\caption{The dynamic statistics for our benchmarks. Memory is
expressed in bytes. \texttt{Dhrystone} performs its numerical benchmark
$100000$ times; \texttt{JLex} is applied to \texttt{sample.tex}.
\texttt{JavaCup} is
applied to the grammar \texttt{tiger.cup} (included in the distribution
of \textsc{Julia}) with the \texttt{-dump\_states} option on.
\texttt{Julia} is applied to the escape analysis
of \texttt{Dhrystone} performed as in Figure~\ref{fig:analyses_er4}.
\texttt{Javacc} is applied to the grammar \texttt{Java1.1.jj}.
\texttt{Jess} is applied to the solution of \texttt{fullmab.clp}.}
  \label{fig:dynamic}
\end{figure}
\subsubsection{Cost of the Analysis}\label{subsub:cost}
Figure~\ref{fig:analyses_er4} shows that one minute is more than enough to
analyse each of the benchmarks, some of them featuring more than
$2000$ methods.
The \textsc{Julia} analyser is under development,
so that analysis times can only improve.
In theory, as a consequence of using sets of creation
points as variable approximations (Section~\ref{subsec:julia}),
their computation as a
fixpoint of a system of set constraints might require an exponential
number of iterations.
Thus, an important observation from
the same Figure~\ref{fig:analyses_er4} is that there is no
exponential blow-up of the analysis time with the size of the benchmarks
(see the \emph{time per $1000$ bytecodes} column $\mathit{TT}$);
hence, it appears from these results that the worst-case scenario,
leading to an exponential blow-up, is not frequent in practice.
Moreover, from Figure~\ref{fig:analyses_er4} it seems
that the cost of the analysis is not only related to the size of the
benchmark, but also to its linearity. See for instance the case of
\texttt{JLex} and \texttt{JavaCup} in Figure~\ref{fig:analyses_er4}, which have
comparable size (\ie number of bytecodes) but quite different linearity.
}
\section{Discussion}\label{sec:discussion}
\ifthenelse{\boolean{PROOFSONLY}}{}{
The first escape
analysis~\cite{ParkG92} was designed for functional languages
where lists are the representative data structure.
Hence the analysis
was meant to stack allocate portions of lists which do not escape
from the function which creates them. That analysis was
later made more efficient in~\cite{Deutsch97} and finally extended to some
imperative constructs and applied to very large programs~\cite{Blanchet98}.

More recently, escape analysis has been studied for object-oriented
languages. In this context, there are actually different formalisations
of the meaning of \emph{escaping}.
While we assume that an object $o$ escapes from
its creating method if it is still reachable after this method terminates,
others (such as~\cite{Ruf00,SalcianuR01,WhaleyL04})
further require that $o$ is actually used after the method terminates.
This results in less conservative and hence
potentially more precise analyses than ours. Note, however,
that our notion of \emph{escaping} allows us to analyse libraries whose calling
contexts are not known at the time of the analysis.

The first work which can be related to escape analysis seems to us
to be~\cite{RuggieriM88} where
a \emph{lifetime analysis} propagates the \emph{sources}
of data structures. This same idea has been used in~\cite{Agrawal99}
where the traditional reaching definition analysis is extended to
object-oriented programs.
Note that, to improve efficiency,~\cite{Agrawal99}
made the escape analysis demand-driven.

Many escape analyses use however some graph-based
representation of the memory of the system, typically derived from
previous work on points-to analysis. Escaping objects are then identified
by applying some form of reachability on this graph. Examples
are~\cite{WhaleyR99,SalcianuR01,VivienR01,ChoiGSSM03,WhaleyL04}.
Points-to information lets such analyses select the target of virtual calls
on the basis of the class of the objects pointed to by the
receiver of the virtual call. Although these works abstract
\emph{the memory graph} into a graph while we abstract \emph{the state}
into $\e$ or $\er$, we think that the information
expressed by $\e$ or $\er$ can be derived by
abstracting such graphs. Hence our analyses should be
less precise but more efficient
than~\cite{WhaleyR99,SalcianuR01,VivienR01,ChoiGSSM03,WhaleyL04}.
Namely, we miss the sharing information contained in a graph-based
representation of the memory of the system.
Note that, to improve efficiency, some escape analyses such as~\cite{VivienR01}
have been made incremental.

A large group of escape analyses use constraints (typically,
set-constraints) for expressing escape analyses. These
include~\cite{BogdaH99,GayS00,Ruf00,StreckenbachS00,RountevMR01},
although~\cite{StreckenbachS00} assumes that points-to information is
available at the time of the analysis.
The escape analysis in~\cite{Blanchet03} is unique
in that it uses integer \emph{contexts} to
specify the part of a data structure which can escape.
However, integer contexts are
an approximation of the full typing information used in
our abstract domain $\er$ as well as in most of the previously
cited escape analyses. This possible imprecision has been observed also
by~\cite{ChoiGSSM03} where it is said (without formal proof)
that their analysis is \emph{inherently more precise} than that
defined in~\cite{Blanchet03}.
The simplicity of Blanchet's escape analysis is however appealing,
and the experimental times he reports in~\cite{Blanchet03}
currently score as the fastest of all the cited analyses that provide timings.

The way we deal with exceptions (Subsection~\ref{subsec:julia})
is largely inspired by~\cite{JacobsP03}.
It is also similar to the technique in~\cite{ChoiGHS99}, although their
optimised factorisation is not currently implemented
inside our \textsc{Julia} analyser. Once implemented, we expect it to improve
the overall efficiency of the analyses shown in Figure~\ref{fig:analyses_er4}.

Some escape analyses have been formally proved correct.
Namely,~\cite{WhaleyR99} has been proved correct in~\cite{Salcianu01t},
and~\cite{ChoiGSSM03} in~\cite{ChoiGSSM02}. Neither proof is based
on abstract interpretation, and no optimality result is proved.
The proof in~\cite{Blanchet03} is closer to ours, since it
is based on abstract interpretation. However, a Galois connection is proved
to exist between the concrete and the abstract domain, rather than
a Galois insertion. Moreover, no optimality result for the abstract operations
is provided.

To the best of our knowledge,
our notion of abstract garbage collector (Definitions~\ref{def:delta}
and~\ref{def:xi}) and its use for deriving Galois \emph{insertions}
rather than \emph{connections} (Propositions~\ref{prop:e_insertion}
and~\ref{prop:er_insertion}) is new.
A similar idea has been used in~\cite{ChoiGSSM03}, which
removes from the connection graphs (their abstraction)
that part that consists only of captured nodes (unreachable from
parameters, result and static variables). However, this was not
related to the Galois insertion property.
Static types of variables are used in~\cite{LhotakH03} to improve the
precision of a points-to analysis by removing spurious points-to
sets which are not allowed by the static typing of the variables.
This idea is somehow similar to our garbage-collectors, but
no connection is shown to the Galois insertion property.

It is quite hard to compare the available escape analyses \wrt precision.
From a theoretical point of view, their definitions are sometimes
so different that a formal comparison inside the same framework is
impossible. However, below we make some informal comparisons
and discuss some of the issues that affect the precision.
\begin{itemize}
\item Escape analyses using graph-based representations of the heap should
be the most precise overall since they express points-to and sharing
information~\cite{WhaleyR99,SalcianuR01,VivienR01,ChoiGSSM03,WhaleyL04}.
\item Precision is also significantly affected by the call-graph construction
algorithm used for the analysis which is largely independent from the
analysis itself~\cite{TipP00,GroveC01}.
As $\er$ couples each variable with the set of its creation points,
class information can be recovered and the call-graph constructed.
This is also typical of those analyses that compute some form
of points-to information.
\item Another issue related to precision is the
level of flow and context sensitivity of the analysis. Our experiments
(Section~\ref{subsec:tests})
have shown that flow sensitivity is important
for the operand stack only. This agrees with the experiments reported
in~\cite{ChoiGSSM03}, since the operand stack is
a feature of Java bytecode not present in Java
(the target language of~\cite{ChoiGSSM03}).
Context sensitivity \ie the
ability to name a given creation point in method $m$
with different names, depending
on the call site of $m$, is advocated as an important feature
of escape analyses~\cite{WhaleyR99,VivienR01,Blanchet03}. This idea
is taken further in~\cite{WhaleyL04}, where method bodies are
\emph{cloned} for each different calling context in order to improve
the precision of the analysis. Our analysis decorates each given program
point with a unique name, and hence misses this extra axis of
precision. Note, however, that method cloning can safely be
applied before many escape analyses, including ours.
\item The optimality of the abstract operations or algorithm used
for the analysis also affects its precision, since optimality entails that the
analysis uses the abstract information in the most precise possible way.
To the best of our knowledge, we are the first to prove an optimality
result for the abstract operations of an escape analysis, which is a
significant step forward although we are aware that the composition
of optimal operations to build an abstract semantics is not necessarily
optimal.
\item A final source of precision comes from the
preliminary inlining of small methods before the actual analysis is applied.
Inlining can only enlarge the possibilities for stack allocation,
although care must be taken to avoid any exponential code explosion.
As far as we can see, only~\cite{Blanchet03} applies method
inlining before the escape analysis. This technique is largely independent
from the subsequent escape analysis, so it could be applied with
other escape analyses, including our own.
\item Some specially designed analyses should perform better on some
specific applications. For instance, the analysis in~\cite{Ruf00}
is probably the most precise one
\wrt synchronisation elimination. This is because it allows one
to remove unnecessary synchronisation (locking)
on objects which do escape from their creating
thread provided that, at run-time, they are locked by at most one thread.
As another example the analysis in~\cite{SalcianuR01} is expected to
perform better on multithreaded applications, since it models precisely
inter-thread interactions. All other escape analyses (including ours)
conservatively assume instead that everything stored inside a thread object
(and the thread object itself) escapes.
\end{itemize}

From an experimental point of view, a comparison of different escape analyses
is possible although hard and sometimes contradictory. This difficulty
is due to the fact that some analyses have been evaluated
\wrt stack allocation, others \wrt
synchonization elimination, and others \wrt both. Moreover, sometimes
only compile-time (\emph{static}) statistics are provided
(such as~\cite{VivienR01}), sometimes
only run-time (\emph{dynamic}) statistics (such as~\cite{Blanchet03}),
sometimes both (such as~\cite{ChoiGSSM03} and ourselves).
Still, some statistics include the library classes (such
as~\cite{BogdaH99,Ruf00,Blanchet03}), others only report numbers for
the user classes (such as~\cite{ChoiGSSM03}, which analyses library
code during the analysis but does not transform it for stack allocation,
exactly as we do in Subsection~\ref{subsec:tests}).
Furthermore, there is no standard set of benchmarks evaluated across
all different escape analyses.
If some benchmark is shared by different
analyses, their version number is not necessarily the same.
For what concerns the dynamic statistics, the input provided to the
benchmark is important. Hence we provided this information
in Figure~\ref{fig:dynamic}, trying to make it as similar
to that in~\cite{Blanchet03} as possible. However, others do not
specify this information (such as~\cite{ChoiGSSM03}). Finally,
analysis times are not always disclosed, making a fair comparison harder.
Let us anyway compare the
benchmarks that we share with~\cite{Blanchet03}, \cite{ChoiGSSM03}
and~\cite{RountevMR01}. From Figure~\ref{fig:dynamic}, we see
that we perform equally on \texttt{Dhrystone}
\wrt~\cite{Blanchet03} (which, however, does not specify the
parameter passed to \texttt{Dhrystone}, which completely modifies
the run-time behaviour of \texttt{Dhrystone}).
For \texttt{JLex}, we stack allocate $47.77\%$ of the memory
while~\cite{Blanchet03} stack allocates $26\%$;
we stack allocate $8.92\%$ of the run-time objects,
while~\cite{RountevMR01} stack allocates $31.6\%$;
we found $23\%$ of the allocation sites to be stack allocatable
(Figure~\ref{fig:analyses_er4}) while~\cite{RountevMR01} found $27\%$.
For \texttt{JavaCup}, we stack allocate $48.76\%$ of the memory,
while~\cite{ChoiGSSM03} stack allocates $17\%$. We stack allocate $39\%$ of
the allocation sites, while~\cite{RountevMR01} stack allocates $30\%$.
For \texttt{Jess}, we only stack allocate $4.90\%$ of the memory,
while~\cite{Blanchet03} manages to stack allocate $27\%$
and~\cite{RountevMR01} $17.9\%$.
This case may be a consequence of
a preliminary lack of methods inlining,
since the \texttt{Jess} program actually
contains a large set of very small methods.
For \texttt{Javacc}, we stack allocate $40.72\%$ of the memory,
while~\cite{Blanchet03} stack allocates $43\%$; we stack allocate
$43.72\%$ of the run-time objects, while~\cite{RountevMR01}
stack allocates $45.8\%$; this is contradictory with the fact
that we found $62\%$ of the allocation sites to be
stack allocatable (Figure~\ref{fig:analyses_er4}) while~\cite{RountevMR01}
found only $29\%$.
}
\section{Conclusion}\label{sec:conclusion}
We have presented a formal development of an escape analysis by
abstract interpretation, providing optimality results in the form
of a Galois insertion from the concrete to the abstract domain
and of the definition of optimal abstract operations. This escape
analysis has been implemented and applied to full Java (bytecode).
This results in an escape analyser which is probably less precise
than others already developed, but still performs well in practice from the
points of view of its cost and precision
\ifthenelse{\boolean{PROOFSONLY}}{}{(Sections~\ref{subsec:tests}
and~\ref{sec:discussion})}.

A first, basic escape domain $\e$ is defined
as a property of concrete states (Definition~\ref{def:property_e}).
This domain is simple but non-trivial since
\begin{itemize}
\item The set of the creation points of the objects reachable from the current
  state can both grow ($\mathsf{new}$) and shrink ($\delta$);
  \ie \emph{static type information contains escape information}
  (Examples~\ref{ex:rho_not_onto} and~\ref{ex:abstract_operations_e});
\item That set is useful, sometimes, to restrict the
  possible targets of a virtual call \ie
  \emph{escape information contains class information}
  (Example~\ref{ex:abstract_operations_e}).
\end{itemize}
However, the escape analysis induced by our domain $\e$ is
not precise enough from a computational point of view,
since it induces rather imprecise abstract operations.
We have therefore defined a refinement $\er$ of $\e$, on the basis of the
information that $\e$ lacks, in order to attain better precision.
The relation between $\er$ and $\e$ is similar to that between Palsberg and
Schwartzbach's class analysis~\cite{PalsbergS91,SpotoJ03}
and \emph{rapid type analysis}~\cite{BaconS96}
although, while all objects stored in memory are
considered in~\cite{BaconS96,SpotoJ03,PalsbergS91},
only those actually reachable from the variables in scope
are considered by the domains $\e$
and $\er$ (Definitions~\ref{def:alpha} and~\ref{def:er_abstraction}).
The ability to describe only the reachable objects, through the use
of an abstract garbage collector ($\delta$ in Figure~\ref{fig:operations_e}
and $\xi$ in Figure~\ref{fig:operations_er}), improves the
precision of the analysis, since it becomes focused on only those objects
that can actually affect the concrete execution of the program.

It is interesting to consider if this notion of reachability and the use
of an abstract garbage collector can be applied to other static analyses
of the run-time heap as well. Namely, class, shape, sharing and cyclicity
analyses might benefit from them.
%
\begin{acknowledgements}
This work has been funded by the Italian
MURST grant \emph{Abstract Interpretation,
Type Systems and Control-Flow Analysis} and by the British
EPSRC grant GR/R53401.
\end{acknowledgements}
%
\bibliographystyle{plain}
\ifthenelse{\boolean{PROOFSONLY}}{

}{
\bibliography{escape01}

\begin{thebibliography}{10}

\bibitem{Agrawal99}
G.~Agrawal.
\newblock Simultaneous {D}emand-{D}riven {D}ata-flow and {C}all {G}raph
  {A}nalysis.
\newblock In {\em Proc.\ of the International Conference on Software
  Maintenance (ICSM'99)}, pages 453--462, Oxford, UK, September 1999. IEEE
  Computer Society.

\bibitem{ArnoldGH00}
K.~Arnold, J.~Gosling, and D.~Holmes.
\newblock {\em The {J}ava$^{TM}$ {P}rogramming {L}anguage}.
\newblock Addison-Wesley, third edition, 2000.

\bibitem{BaconS96}
D.~F. Bacon and P.~F. Sweeney.
\newblock Fast {S}tatic {A}nalysis of {C}++ {V}irtual {F}unction {C}alls.
\newblock In {\em Proc.\ of OOPSLA'96}, volume 31(10) of {\em ACM SIGPLAN
  Notices}, pages 324--341, New York, 1996. ACM Press.

\bibitem{Blanchet98}
B.~Blanchet.
\newblock Escape {A}nalysis: {C}orrectness {P}roof, {I}mplementation and
  {E}xperimental {R}esults.
\newblock In {\em 25th ACM SIGPLAN-SIGACT Symposium of Principles of
  Programming Languages (POPL'98)}, pages 25--37, San Diego, CA, USA, January
  1998. ACM Press.

\bibitem{Blanchet03}
B.~Blanchet.
\newblock Escape {A}nalysis for {J}ava: {T}heory and {P}ractice.
\newblock {\em ACM TOPLAS}, 25(6):713--775, November 2003.

\bibitem{BogdaH99}
J.~Bogda and U.~H\"olzle.
\newblock Removing {U}nnecessary {S}ynchronization in {J}ava.
\newblock In {\em Proc.\ of OOPSLA'99}, volume 34(10) of {\em SIGPLAN Notices},
  pages 35--46, Denver, Colorado, USA, November 1999.

\bibitem{BossiGLM94}
A.~Bossi, M.~Gabbrielli, G.~Levi, and M.~Martelli.
\newblock The s-{S}emantics {A}pproach: {T}heory and {A}pplications.
\newblock {\em Journal of Logic Programming}, 19/20:149--197, 1994.

\bibitem{ChoiGSSM03}
J.-D. Choi, M.~Gupta, M.~J. Serrano, V.~C. Sreedhar, and S.~P. Midkiff.
\newblock Stack {A}llocation and {S}ynchronization {O}ptimizations for {J}ava
  {U}sing {E}scape {A}nalysis.
\newblock {\em ACM TOPLAS}, 25(6):876--910, November 2003.

\bibitem{CortesiFW98}
A.~Cortesi, G.~Fil\'e, and W.~Winsborough.
\newblock The {Q}uotient of an {A}bstract {I}nterpretation.
\newblock {\em Theoretical Computer Science}, 202(1-2):163--192, 1998.

\bibitem{CousotC77}
P.~Cousot and R.~Cousot.
\newblock Abstract {I}nterpretation: {A} {U}nified {L}attice {M}odel for
  {S}tatic {A}nalysis of {P}rograms by {C}onstruction or {A}pproximation of
  {F}ixpoints.
\newblock In {\em Proc.\ of POPL'77}, pages 238--252, 1977.

\bibitem{CousotC92}
P.~Cousot and R.~Cousot.
\newblock Abstract {I}nterpretation and {A}pplications to {L}ogic {P}rograms.
\newblock {\em Journal of Logic Programming}, 13(2 \& 3):103--179, 1992.

\bibitem{Dams96}
D.~R. Dams.
\newblock {\em Abstract {I}nterpretation and {P}artition {R}efinement for
  {M}odel {C}hecking}.
\newblock PhD thesis, Eindhoven University of Technology, The Netherlands, July
  1996.

\bibitem{Deutsch97}
A.~Deutsch.
\newblock On the {C}omplexity of {E}scape {A}nalysis.
\newblock In {\em 24th ACM SIGPLAN-SIGACT Symposium on Principles of
  Programming Languages (POPL'97)}, pages 358--371, Paris, France, January
  1997. ACM Press.

\bibitem{GayS00}
D.~Gay and B.~Steensgaard.
\newblock Fast {E}scape {A}nalysis and {S}tack {A}llocation for
  {O}bject-{B}ased {P}rograms.
\newblock In D.~A. Watt, editor, {\em Compiler Construction, 9th International
  Conference (CC'00)}, volume 1781 of {\em Lecture Notes in Computer Science},
  pages 82--93. Springer-Verlag, Berlin, March 2000.

\bibitem{GiacobazziR97}
R.~Giacobazzi and F.~Ranzato.
\newblock Refining and {C}ompressing {A}bstract {D}omains.
\newblock In {\em Proc.\ of the 24th International Colloquium on Automata,
  Languages and Programming (ICALP'97)}, volume 1256 of {\em LNCS}, pages
  771--781. Springer-Verlag, 1997.

\bibitem{GiacobazziS98}
R.~Giacobazzi and F.~Scozzari.
\newblock A {L}ogical {M}odel for {R}elational {A}bstract {D}omains.
\newblock {\em ACM Transactions on Programming Languages and Systems},
  20(5):1067--1109, 1998.

\bibitem{HillS02b}
P.~M. Hill and F.~Spoto.
\newblock A {F}oundation of {E}scape {A}nalysis.
\newblock In H.~Kirchner and C.~Ringeissen, editors, {\em Proc.\ of AMAST'02},
  volume 2422 of {\em LNCS}, pages 380--395, St.\ Gilles les Bains, La
  R\'eunion island, France, September 2002. Springer-Verlag.

\bibitem{HillS02}
P.~M. Hill and F.~Spoto.
\newblock A {R}efinement of the {E}scape {P}roperty.
\newblock In A.~Cortesi, editor, {\em Proc.\ of the VMCAI'02 workshop on
  Verification, Model-Checking and Abstract Interpretation}, volume 2294 of
  {\em Lecture Notes in Computer Science}, pages 154--166, Venice, Italy,
  January 2002. Springer-Verlag.

\bibitem{Jensen97}
T.~Jensen.
\newblock Disjunctive {P}rogram {A}nalysis for {A}lgebraic {D}ata {T}ypes.
\newblock {\em {ACM} Transactions on Programming Languages and Systems},
  19(5):752--804, 1997.

\bibitem{JS87}
N.~D. Jones and H.~S{\o}ndergaard.
\newblock A {S}emantics-based {F}ramework for the {A}bstract {I}nterpretation
  of \textsf{Prolog}.
\newblock In S.~Abramsky and C.~Hankin, editors, {\em Abstract Interpretation
  of Declarative Languages}, pages 123--142. Ellis Horwood Ltd, 1987.

\bibitem{LindholmY99}
T.~Lindholm and F.~Yellin.
\newblock {\em The {J}ava$^{\mathit{TM}}$ {V}irtual {M}achine {S}pecification}.
\newblock Addison-Wesley, second edition, 1999.

\bibitem{CousotC79}
Cousot. P. and R.~Cousot.
\newblock Systematic {D}esign of {P}rogram {A}nalysis {F}rameworks.
\newblock In {\em Proc.\ of the Sixth Annual ACM Symposium on Principles of
  Programming Languages (POPL'79)}, pages 269--282, San Antonio, Texas, 1979.
  ACM.

\bibitem{PalsbergS91}
J.~Palsberg and M.~I. Schwartzbach.
\newblock Object-{O}riented {T}ype {I}nference.
\newblock In {\em Proc.\ of OOPSLA'91}, volume 26(11) of {\em ACM SIGPLAN
  Notices}, pages 146--161. ACM Press, November 1991.

\bibitem{ParkG92}
Y.~G. Park and B.~Goldberg.
\newblock Escape {A}nalysis on {L}ists.
\newblock In {\em ACM SIGPLAN'92 Conference on Programming Language Design and
  Implementation (PLDI'92)}, volume 27(7) of {\em SIGPLAN Notices}, pages
  116--127, San Francisco, California, USA, June 1992.

\bibitem{RountevMR01}
A.~Rountev, A.~Milanova, and B.~G. Ryder.
\newblock Points-to {A}nalysis for {J}ava {U}sing {A}nnotated {C}onstraints.
\newblock In {\em Proc.\ of ACM SIGPLAN Conference on Object-Oriented
  Programming Systems, Languages and Applications (OOPSLA'01)}, volume 36(11)
  of {\em ACM SIGPLAN}, pages 43--55, Tampa, Florida, USA, October 2001.

\bibitem{Ruf00}
E.~Ruf.
\newblock Effective {S}ynchronization {R}emoval for {J}ava.
\newblock In {\em ACM SIGPLAN Conference on Programming Language Design and
  Implementation (PLDI'00)}, volume 35(5) of {\em SIGPLAN Notices}, pages
  208--218, Vancouver, British Columbia, Canada, June 2000.

\bibitem{RuggieriM88}
C.~Ruggieri and T.~P. Murtagh.
\newblock Lifetime {A}nalysis of {D}ynamically {A}llocated {O}bjects.
\newblock In {\em 15th ACM Symposium on Principles of Programming Languages
  (POPL'88)}, pages 285--293, San Diego, California, USA, January 1988.

\bibitem{SalcianuR01}
A.~Salcianu and M.~Rinard.
\newblock Pointer and {E}scape {A}nalysis for {M}ultithreaded {P}rograms.
\newblock In {\em Proc.\ of ACM SIGPLAN Symposium on Principles and Practice of
  Parallel Programming (PPoPP'01)}, volume 36(7) of {\em SIGPLAN Notices},
  pages 12--23, Snowbird, Utah, USA, July 2001.

\bibitem{Scozzari00}
F.~Scozzari.
\newblock Logical {O}ptimality of {G}roundness {A}nalysis.
\newblock {\em Theoretical Computer Science}, 277(1-2):149--184, 2002.

\bibitem{Sondergaard86}
H.~S{\o}ndergaard.
\newblock An {A}pplication of {A}bstract {I}nterpretation of {L}ogic
  {P}rograms: {O}ccur {C}heck {R}eduction.
\newblock In B.~Robinet and R.~Wilhelm, editors, {\em Proc.\ of the European
  Symposium on Programming (ESOP)}, volume 213 of {\em Lecture Notes in
  Computer Science}, pages 327--338, Saarbr{\"u}cken, Federal Republic of
  Germany, March 1986. Springer.

\bibitem{SpotoJ03}
F.~Spoto and T.~Jensen.
\newblock {C}lass {A}nalyses as {A}bstract {I}nterpretations of {T}race
  {S}emantics.
\newblock {\em ACM Transactions on Programming Languages and Systems (TOPLAS)},
  25(5):578--630, September 2003.

\bibitem{StreckenbachS00}
M.~Streckenbach and G.~Snelting.
\newblock Points-to for {J}ava: {A} {G}eneral {F}ramework and an {E}mpirical
  {C}omparison.
\newblock Technical report, Universit\"at Passau, Germany, November 2000.

\bibitem{Ta55}
A.~Tarski.
\newblock A {L}attice-theoretical {F}ixpoint {T}heorem and its {A}pplications.
\newblock {\em Pacific J. Math.}, 5:285--309, 1955.

\bibitem{VivienR01}
F.~Vivien and M.~Rinard.
\newblock Incrementalized {P}ointer and {E}scape {A}nalysis.
\newblock In {\em Proc.\ of ACM SIGPLAN Conference on Programming Language
  Design and Implementation (PLDI'01)}, volume 36(5) of {\em SIGPLAN Notices},
  pages 35--46, Snowbird, Utah, USA, June 2001.

\bibitem{WhaleyL04}
J.~Whaley and M.~S. Lam.
\newblock Cloning-{B}ased {C}ontext-{S}ensitive {P}ointer {A}lias {A}nalysis
  {U}sing {B}inary {D}ecision {D}iagrams.
\newblock In W.~Pugh and C.~Chambers, editors, {\em Proc.\ of ACM SIGPLAN 2004
  Conference on Programming Language Design and Implementation (PLDI'04)},
  pages 131--144, Washington, DC, USA, June 2004. ACM.

\bibitem{WhaleyR99}
J.~Whaley and M.~C. Rinard.
\newblock Compositional {P}ointer and {E}scape {A}nalysis for {J}ava
  {P}rograms.
\newblock In {\em 1999 ACM SIGPLAN Conference on Object-Oriented Programming
  Systems, Languages and Applications (OOPSLA'99)}, volume 34(1) of {\em
  SIGPLAN Notices}, pages 187--206, Denver, Colorado, USA, November 1999.

\bibitem{Winskel93}
G.~Winskel.
\newblock {\em The {F}ormal {S}emantics of {P}rogramming {L}anguages}.
\newblock The MIT Press, 1993.

\end{thebibliography}
}
%

\ifthenelse{\boolean{WITHAPPENDIX}}{
%
\appendix
\label{sec:proofs}
%
\section{Proofs of Propositions~\ref{prop:delta_lco},
\ref{prop:delta_fixpoints} and~\ref{prop:operations_e}
in Section~\ref{sec:edomain}.}

\summary{Proposition \ref{prop:delta_lco}.}
{Let $i\in\nat$.
The abstract garbage collectors $\delta_\tau^i$ and $\delta_\tau$ are lco's.}
\myproofbis{
Since $\delta_\tau=\delta_\tau^{\#\Pi}$, it is enough to prove the result for
$\delta_\tau^i$ only.
By Definition~\ref{def:delta}, the maps $\delta_\tau^i$ for $i\in\nat$
are reductive and monotonic.
We prove idempotency by induction over $i\in\nat$.
Let $\ee\subseteq\Pi$.
We have $\delta_\tau^0\delta_\tau^0(\ee)=
\delta_\tau^0(\emptyset)=\emptyset=\delta_\tau^0(\ee)$.
Assume that the result holds for a given $i\in\nat$.
If $\mathtt{this}\in\domain(\tau)$ and there
is no $\pi\in \ee$ such that $k(\pi)\le\tau(\mathtt{this})$, then
$\delta_\tau^i\delta_\tau^i(\ee)=\delta_\tau^i(\emptyset)=\emptyset=
\delta_\tau^i(\ee)$.
Suppose now that, if $\mathtt{this}\in\domain(\tau)$, then
there exists  $\pi\in \ee$ such that $k(\pi)\le\tau(\mathtt{this})$.
By reductivity,
$\delta_\tau^{i+1}\delta_\tau^{i+1}(\ee)\subseteq\delta_\tau^{i+1}(\ee)$.
We prove that the converse inclusion holds.
We have
\begin{equation}\elabel{eq:idempotency}
  \delta_\tau^{i+1}\delta_\tau^{i+1}(\ee)=\cup\left\{
    \{\pi\}\cup\delta^i_{F(k(\pi))}\delta_\tau^{i+1}(\ee)\left|
    \begin{array}{l}
      \kappa\in\codom(\tau)\cap\mathcal{K}\\
      \pi\in\delta_\tau^{i+1}(\ee),\ k(\pi)\le\kappa
    \end{array}\right.\right\}.
\end{equation}
Let $\kappa\in\codom(\tau)\cap\mathcal{K}$ and $\pi\in\Pi$ be such that
$k(\pi)\le\kappa$.
If $\pi\in\delta_\tau^{i+1}(\ee)$ then, by reductivity, we have $\pi\in \ee$.
Conversely, if $\pi\in \ee$ then, by Definition~\ref{def:delta},
$\pi\in\delta_\tau^{i+1}(\ee)$.
We conclude from~\eref{eq:idempotency} that
\begin{align*}
  \delta_\tau^{i+1}\delta_\tau^{i+1}(\ee)
    &= \cup \left\{\{\pi\}\cup\delta^i_{F(k(\pi))}\delta_\tau^{i+1}(\ee)\left|
    \begin{array}{l}
      \kappa\in\codom(\tau)\cap\mathcal{K}\\
      \pi\in \ee,\ k(\pi)\le\kappa
    \end{array}
    \right.\right\}\\
  \text{(monotonicity)}
    &\supseteq
    \cup\left\{\{\pi\}\cup\delta^i_{F(k(\pi))}\delta_{F(\pi)}^i(\ee)\left|
      \begin{array}{l}
      \kappa\in\codom(\tau)\cap\mathcal{K}\\
      \pi\in \ee,\ k(\pi)\le\kappa
    \end{array}
    \right.\right\}\\
  \text{(ind. hypothesis)}
    &= \cup \left\{\{\pi\}\cup\delta_{F(k(\pi))}^i(\ee)\left|
    \begin{array}{l}
      \kappa\in\codom(\tau)\cap\mathcal{K}\\
      \pi\in \ee,\ k(\pi)\le\kappa
    \end{array}
    \right.\right\}\\
    &= \delta_\tau^{i+1}(\ee).
\end{align*}
}

To prove Proposition~\ref{prop:delta_fixpoints},
we need some preliminary definitions and results. We start by defining,
for every $i\in\nat$, a map $\alpha_\tau^i$ which, for sufficiently large $i$,
coincides with $\alpha_\tau^\e$ (Definition~\ref{def:alpha}).
\begin{definition}\label{def:pre_small_step}
Let $i\in\nat$.
We define the map $\alpha_\tau^i:\wp(\Sigma_\tau)
\mapsto\Pi$ as $\alpha_\tau^i(S)=\{o.\pi\mid\sigma\in S\text{ and }o\in
O_\tau^i(\sigma)\}$ (see Definition \ref{def:reachability} for
$O_\tau^i$).
\end{definition}
\begin{corollary}\label{cor:alternative}
Let $S\subseteq\Sigma_\tau$ and $i\ge 0$. We have
\[
  \alpha_\tau^{i+1}(S)=\bigcup\left\{\{o.\pi\}\cup
    \alpha^i_{F(k(o.\pi))}(o.\phi\sep\mu)
    \left|\begin{array}{l}
      \phi\sep\mu\in\Sigma_\tau,\ v\in\dom(\tau)\\
      \phi(v)\in\Loc,\ o=\mu\phi(v)
    \end{array}\right.\right\}~.
\]
\end{corollary}
\myproofbis{
By Definitions~\ref{def:pre_small_step} and~\ref{def:reachability}.}
Lemma~\ref{lem:small_step} states that $\alpha_\tau^i$ (and
hence also $\alpha_\tau^\e$ itself) yields sets of creation
points that do not contain garbage.
\begin{lemma}\label{lem:small_step}
Let $\sigma\in\Sigma_\tau$ and $i\in\nat$.
Then $\alpha_\tau^i(\sigma)=\delta_\tau^i\alpha_\tau^i(\sigma)$.
\end{lemma}
\myproofbis{
By reductivity (Proposition \ref{prop:delta_lco}), we have
$\alpha_\tau^i(\sigma)\supseteq\delta_\tau^i\alpha_\tau^i(\sigma)$.
It remains to prove
$\alpha_\tau^i(\sigma)\subseteq\delta_\tau^i\alpha_\tau^i(\sigma)$.
Let $\sigma=\phi\sep\mu$. We proceed by induction on $i$. We have
$\alpha_\tau^0(\sigma)=\emptyset=\delta_\tau^0\alpha_\tau^0(\sigma)$.
Assume that the property holds for a given $i\in\nat$.
Let $\tau'=F(k(o.\pi))$ and
$X=\{\mu\phi(v)\mid v\in\dom(\phi)\text{ and }\phi(v)\in Loc\}$.
By Corollary~\ref{cor:alternative},
\begin{align}
  \alpha_\tau^{i+1}(\sigma)&=\cup\{\{o.\pi\}\cup\alpha^i_{\tau'}
    (o.\phi\sep\mu)\mid o\in X\}\notag\\
  \text{(inductive hypothesis)}
    &=\cup\{\{o.\pi\}\cup\delta^i_{\tau'}\alpha^i_{\tau'}
    (o.\phi\sep\mu)\mid o\in X\}~.\elabel{eq:alphadelta1}
\end{align}
By Corollary~\ref{cor:alternative}, we have $\alpha_{\tau'}^i(o.\phi\sep\mu)
\subseteq\alpha_\tau^{i+1}(\sigma)$ and,
by Proposition~\ref{prop:delta_lco}, \eref{eq:alphadelta1} is contained in
\begin{equation}\elabel{eq:alphadelta}
  \cup\{\{o.\pi\}\cup\delta^i_{\tau'}\alpha^{i+1}_\tau(\sigma)\mid o\in X\}~.
\end{equation}
Note that, given $o\in X$, we can always find $\kappa\in\codom(\tau)\cap
\mathcal{K}$ such that $k(o.\pi)\le\kappa$. Indeed, for the definition of
$X$, there exists $v\in\dom(\phi)=\dom(\tau)$ such that $\phi(v)\in Loc$ and
$o=\mu\phi(v)$.
By Definition~\ref{def:domains2}, we have $\tau(v)\in\mathcal{K}$.
By Definition~\ref{def:proptotau}, we have $k(o.\pi)=k((\mu\phi(v)).\pi)
\le\tau(v)$. Hence letting $\kappa=\tau(v)$, \eref{eq:alphadelta} is
\begin{align*}
  &\cup\left\{\{o.\pi\}\cup\delta^i_{\tau'}\alpha^{i+1}_\tau(\sigma)\left|
   \begin{array}{l}
     o\in X,\ \kappa\in\codom(\tau)\cap\mathcal{K}\\
     k(o.\pi)\le\kappa
   \end{array}\right.\right\}\\
  \text{(Corollary~\ref{cor:alternative})}\subseteq&
    \cup\left\{\{\pi\}\cup\delta^i_{\tau'}\alpha^{i+1}_\tau(\sigma)\left|
    \begin{array}{l}
      \pi\in\alpha_\tau^{i+1}(\sigma),\ \kappa\in\codom(\tau)\cap\mathcal{K}\\
      k(\pi)\le\kappa
    \end{array}\right.\right\}\\
  \text{(Definition~\ref{def:delta})}
    =&\ \delta_\tau^{i+1}\alpha_\tau^{i+1}(\sigma)~.
\end{align*}
Note that the last step is correct since
if $\mathtt{this}\in\dom(\tau)$ we have $\phi(\mathtt{this})\not=\nil$
(Definition~\ref{def:concrete_states}). Hence $(\mu\phi(\mathtt{this})).\pi
\in\alpha_\tau^{i+1}(\sigma)$ and $k((\mu\phi(\mathtt{this})).\pi)\le
\tau(\mathtt{this})$ (Definition~\ref{def:weak_correctness}). We conclude
that, if $\mathtt{this}\in\dom(\tau)$, then there exists
$\pi\in\alpha_\tau^{i+1}(\sigma)$ such that
$k(\pi)\le\tau(\mathtt{this})$.}

Let $\ee$ be a set of creation points.
We now define frames and memories which use all possible creation points
in $\ee$ allowed by the type environment of the variables. In this sense,
they are the \emph{richest} frames and memories containing creation points
from $\ee$ only.
\begin{definition}\label{def:worst}
Let $\{\pi_1,\ldots,\pi_n\}$ be an enumeration without repetitions
of $\Pi$. Let $l_1,\ldots,l_n$ be distinct locations.
Let $\ee\subseteq\Pi$ and $w\in\dom(\tau)$ such that
$\tau(w)\in\mathcal{K}$.
We define
\begin{align*}
  L_\tau(\ee,w)&=\{l_i\mid 1\le i\le n,\ \pi_i\in \ee\text{ and }k(\pi_i)
    \le\tau(w)\}~,\\
  \overline{\phi}_\tau(\ee)&=\left\{\phi\in\Frame_\tau\left|
  \begin{array}{l}
    \text{for every $v\in\dom(\tau)$}\\
    \text{ $\tau(v)=\integer\Rightarrow\phi(v)=0$}\\
    \text{ $\tau(v)\in\mathcal{K}$, $L_\tau(\ee,v)=\emptyset\Rightarrow
      \phi(v)=\nil$}\\
    \text{ $\tau(v)\in\mathcal{K}$, $L_\tau(\ee,v)\not=\emptyset\Rightarrow
      \phi(v)\in L_\tau(\ee,v)$}
  \end{array}\right.\right\},\\
  \overline{\mu}(\ee)&=\left\{\mu\in\Memory\left|\begin{array}{l}
    \mu=[l_1\mapsto\pi_1\sep\phi_1,\ldots,l_n\mapsto\pi_n\sep\phi_n]\\
    \text{and }\phi_i\in\overline{\phi}_{F(k(\pi_i))}(\ee)
    \text{ for $i=1,\ldots,n$}
  \end{array}\right.\right\}~.
\end{align*}
\end{definition}

We prove now some properties of the frames and memories of Definition
\ref{def:worst}.
\begin{lemma}\label{lem:compatibility}
Let $\ee_1,\ee_2\subseteq\Pi$,
$\phi\in\overline{\phi}_\tau(\ee_1)$
and $\mu\in\overline{\mu}(\ee_2)$.
Then
\begin{romanenumerate}
\item $\phi\sep\mu:\tau$;
\item $\phi\sep\mu\in\Sigma_\tau$ iff
$\mathtt{this}\not\in\dom(\tau)$ or there
exists $\pi\in \ee_1$ s.t.\ $k(\pi)\le\tau(\mathtt{this})$;
\item If $\phi\sep\mu\in\Sigma_\tau$ then
$\alpha_\tau(\phi\sep\mu)\subseteq \ee_1\cup \ee_2$.
\end{romanenumerate}
\end{lemma}
\myproofbis{\mbox{}
\begin{romanenumerate}
\item Condition 1 of Definition~\ref{def:proptotau} is satisfied since
we have that
$\codom(\phi)\cap\Loc\subseteq\{l_1,\ldots,l_n\}=\dom(\mu)$.
Moreover, if $v\in\dom(\phi)$ and $\phi(v)\in\Loc$ then
$\phi(v)\in L_\tau(\ee_1,v)$. Thus there exists
$1\le i\le n$ such that $\phi(v)=l_i$, $(\mu\phi(v)).\pi=\pi_i$ and
$k((\mu\phi(v)).\pi)=k(\pi_i)\le
\tau(v)$.
Condition 2 of Definition~\ref{def:proptotau} holds because
if $o\in\codom(\mu)$ then $o.\phi=\phi_i$ for some
$1\le i\le n$.
Since $\phi_i\in\overline{\phi}_{F(k(\pi_i))}(\ee)$, reasoning
as above we conclude that $\phi_i$ is $F(k(\pi_i))$-correct \wrt $\mu$.
Then $\phi\sep\mu:\tau$.
\item
By point i, we know that $\phi\sep\mu:\tau$.
From Definition~\ref{def:concrete_states},
we have $\phi\sep\mu\in\Sigma_\tau$ if
and only if $\mathtt{this}\not\in\dom(\tau)$ or
$\phi(\mathtt{this})\not=\nil$.
By Definition~\ref{def:worst}, the latter case holds if and only if
$L_\tau(\ee_1,\mathtt{this})\not=\emptyset$ \ie if and only if there exists
$\pi\in \ee_1$ such that $k(\pi)\le\tau(\mathtt{this})$.
\item Since $\phi\sep\mu\in\Sigma_\tau$, the $\alpha_\tau$ map
is defined (Definition~\ref{def:alpha}). Let
\[
  L=(\codom(\phi)\cup(\cup\{\codom(o.\phi)\mid o\in\codom(\mu)\}))\cap\Loc~.
\]
Since $\phi\in\overline{\phi}_\tau(\ee_1)$ and
$o.\phi\in\overline{\phi}_{F(k(o.\pi))}(\ee_2)$ for every $o\in\codom(\mu)$,
by Definition~\ref{def:worst}, we have
\[
  \{\mu(l).\pi\mid l\in L\}\subseteq \ee_1\cup \ee_2~.
\]
By Definition~\ref{def:alpha}, we conclude that
\[
  \alpha_\tau(\phi\sep\mu)\subseteq\{\mu(l).\pi\mid l\in L\}
    \subseteq \ee_1\cup \ee_2~.
\]
\end{romanenumerate}%
}

Lemma~\ref{lem:worst} gives an explicit definition of the abstraction
of the set of states
constructed from the frames and memories of Definition~\ref{def:worst}.
\begin{lemma}\label{lem:worst}
Let $\ee_1,\ee_2\subseteq\Pi$, $j\in\nat$ and
\[
  A^j=\alpha_\tau^{j+1}(\{\phi\sep\mu\in\Sigma_\tau\mid
    \phi\in\overline{\phi}_\tau(\ee_1)\text{ and }\mu\in\overline{\mu}(\ee_2)\})~.
\]
Then
\[
  A^j=\begin{cases}
      \emptyset\quad
        \text{if $\mathtt{this}\in\dom(\tau)$ and there is no $\pi\in \ee_1$
          s.t.\ $k(\pi)\le\tau(\mathtt{this})$}\\
      \mbox{}\\
      \cup\left\{
        \{\pi\}\cup\delta^j_{F(k(\pi))}(\ee_2)\left|\begin{array}{l}
          v\in\dom(\tau),\ \tau(v)\in\mathcal{K}\\
          \pi\in \ee_1,\ k(\pi)\le\tau(v)
        \end{array}\right.\right\}\quad\text{otherwise.}
    \end{cases}
\]
\end{lemma}
\myproofbis{
We proceed by induction over $j$.
By Lemma~\ref{lem:compatibility}.ii, if $j=0$ we have
\[
  A^0=\begin{cases}
      \emptyset\quad
        \text{if $\mathtt{this}\in\dom(\tau)$ and there is no $\pi\in \ee_1$
           s.t.\ $k(\pi)\le\tau(\mathtt{this})$}\\
      \mbox{}\\
      \left\{o.\pi\left|\begin{array}{l}
        \phi\in\overline{\phi}_\tau(\ee_1),\ \mu\in\overline{\mu}(\ee_2)
        ,\ v\in\dom(\phi)\\
        \phi(v)\in Loc,\ o=\mu\phi(v)
      \end{array}\right.\right\}\quad\text{otherwise.}
    \end{cases}
\]
By Definition~\ref{def:worst}, the latter case is equal to
\[
  \left\{\pi_i\left|\begin{array}{l}
    v\in\dom(\tau),\ \tau(v)\in\mathcal{K}\\
    1\le i\le n,\ \pi_i\in \ee_1\\
    k(\pi_i)\le\tau(v)
  \end{array}\right.\right\}
    =\cup\left\{\{\pi\}\cup\delta^0_{F(k(\pi))}(\ee_2)\left|\begin{array}{l}
      v\in\dom(\tau)\\
      \tau(v)\in\mathcal{K}\\
      \pi\in \ee_1\\
      k(\pi)\le\tau(v)
    \end{array}\right.\right\}.
\]
Assume now that the result holds for a given $j\in\nat$.
If $\mathtt{this}\in\dom(\tau)$ and there is no $\pi\in \ee_1$ such that
$k(\pi)\le\tau(\mathtt{this})$, by Lemma~\ref{lem:compatibility}.ii,
we have $A^{j+1}=\emptyset$.
Otherwise, by Corollary~\ref{cor:alternative} we have
\begin{equation}\elabel{eq:Ai}
  A^{j+1}\!=\!\cup\left\{\{o.\pi\}\!\cup\!\alpha_{F(k(o.\pi))}^{j+1}
    (o.\phi\sep\mu)\left|\begin{array}{l}
    \phi\!\in\!\overline{\phi}_\tau(\ee_1),\ \mu\!\in\!\overline{\mu}(\ee_2)\\
    v\!\in\!\dom(\phi)\\
    \phi(v)\in\Loc,\ o=\mu\phi(v)\end{array}\right.\right\}.
\end{equation}
As for the base case, we know that $o.\pi$ ranges over
$\{\pi\in \ee_1\mid v\in\dom(\tau),\ \tau(v)\in\mathcal{K},\ k(\pi)\le
\tau(v)\}$. Since $o.\phi\in\overline{\phi}_{F(k(o.\pi))}
(\ee_2)$ is arbitrary (Definition~\ref{def:worst}),
by the inductive hypothesis, \eref{eq:Ai} becomes
\begin{align*}
  &\cup\left\{\{\pi\}\cup
    \alpha_{F(k(\pi))}^{j+1}\left(\left\{\phi\sep\mu\left|
    \begin{array}{l}
      \phi\in\overline{\phi}_{F(k(\pi))}(\ee_2)\\
      \mu\in\overline{\mu}(\ee_2)
    \end{array}\right.\right\}\right)
    \left|\begin{array}{l}
      v\in\dom(\tau)\\
      \tau(v)\in\mathcal{K}\\
      \pi\in \ee_1\\
      k(\pi)\le\tau(v)
    \end{array}\right.\right\}\\
  =&\cup\{\{\pi\}\cup\delta_{F(k(\pi))}^{j+1}(\ee_2)\mid
      v\in\dom(\tau),\ \tau(v)\in\mathcal{K}
      ,\ \pi\in \ee_1,\ k(\pi)\le\tau(v)\}.
\end{align*}
}
\begin{corollary}\label{cor:worst}
Let $\ee_1,\ee_2\subseteq\Pi$. Let
\[
  A_\tau(\ee_1,\ee_2)=\alpha_\tau(\{\phi\sep\mu\in\Sigma_\tau\mid
    \phi\in\overline{\phi}_\tau(\ee_1)\text{ and }\mu\in\overline{\mu}
    (\ee_2)\})~.
\]
Then
\begin{romanenumerate}
\item $
  A_\tau(\ee_1,\ee_2)=\begin{cases}
      \emptyset\quad
        \text{if $\mathtt{this}\in\dom(\tau)$}\\
      \qquad\text{and no $\pi\in \ee_1$ is
          s.t.\ $k(\pi)\le\tau(\mathtt{this})$}\\
      \mbox{}\\
      \cup\left\{
        \{\pi\}\cup\delta_{F(k(\pi))}(\ee_2)\left|\begin{array}{l}
          v\in\dom(\phi),\ \tau(v)\in\mathcal{K}\\
          \pi\in \ee_1,\ k(\pi)\le\tau(v)
        \end{array}\right.\right\}\\
    \quad\text{otherwise,}
    \end{cases}$
\item $A_\tau(\ee_1,\ee_1)=\delta_\tau(\ee_1)$.
\end{romanenumerate}
\end{corollary}
\myproofbis{
Point i follows by Lemma~\ref{lem:worst}
since $j$ is arbitrary. Point ii
follows from point i and Definition~\ref{def:delta}.
}
\begin{corollary}\label{cor:worst2}
Let $\kappa\in\mathcal{K}$, $\tau=[\rs\mapsto\kappa]$, $p$ be a
predicate over $\Pi$ and $\ee\subseteq\Pi$ be such that there exists
$\pi\in \ee$ such that $k(\pi)\le\tau(\rs)$ and $p(\pi)$ holds.
Then
\begin{multline*}
  \alpha_\tau(\{\phi\sep\mu\in\Sigma_\tau\mid\phi\in
    \overline{\phi}_\tau(\ee),\ \mu\in\overline{\mu}(\ee),\ p(\mu\phi(\rs).\pi)\})\\
  =\cup\{\{\pi\}\cup\delta_{F(k(\pi))}(\ee)\mid\pi\in \ee,\ k(\pi)\le\tau(\rs)
    ,\ p(\pi)\}~.
\end{multline*} 
\end{corollary}
\myproofbis{
Let $j\in\nat$. By the hypothesis on $\ee$ and
Corollary~\ref{cor:alternative} we have
\begin{align*}
  &\quad\alpha_\tau^{j+1}(\{\phi\sep\mu\in\Sigma_\tau\mid\phi\in
    \overline{\phi}_\tau(\ee),\ \mu\in\overline{\mu}(\ee),\ p(\mu\phi(\rs).\pi)\})\\
  &=\cup\left\{\{o.\pi\}\cup\alpha_{F(k(o.\pi))}^j(o.\phi\sep\mu)\left|
  \begin{array}{l}
    \phi\in\overline{\phi}_\tau(\ee),\ \mu\in\overline{\mu}(\ee)\\
    o=\mu\phi(\rs),\ p(o.\pi)
  \end{array}\right.\right\}\\
  &=\cup\left\{\{\pi\}\cup\alpha_{F(k(\pi))}^j(\phi'\sep\mu)\left|
  \begin{array}{l}
    \pi\in\ee,\ k(\pi)\le\tau(\rs),\ p(\pi)\\
    \phi'\in\overline{\phi}_{F(k(\pi))}(\ee),\ \mu\in\overline{\mu}(\ee)
  \end{array}\right.\right\}.
\end{align*}
Since $j$ is arbitrary we have
\begin{align*}
  &\alpha_\tau(\{\phi\sep\mu\in\Sigma_\tau\mid\phi\in
    \overline{\phi}_\tau(\ee),\ \mu\in\overline{\mu}(\ee)
    ,\ p(\mu\phi(\rs).\pi)\})\\
  &=\cup\left\{\{\pi\}\cup\alpha_{F(k(\pi))}\left(\left\{\phi'\sep\mu
    \left|\begin{array}{l}\phi'\in\overline{\phi}_{F(k(\pi))}(\ee)\\
    \mu\in\overline{\mu}(\ee)\end{array}\right.\right\}\right)\left|
    \begin{array}{l}
      \pi\in \ee,\ p(\pi)\\
      k(\pi)\le\tau(\rs)
    \end{array}\right.\right\},
\end{align*}
and the thesis follows by Corollary~\ref{cor:worst}.ii.
}

\mbox{}\\
\summary{Proposition \ref{prop:delta_fixpoints}.}
{Let $\delta(\tau)$ be an abstract garbage collector.
Then we have $\fp(\delta_\tau)=\codom(\alpha^\e_\tau)$ and
$\emptyset\in\fp(\delta_\tau)$. Moreover, if
$\mathtt{this}\in\domain(\tau)$,
then for every $X\subseteq\Sigma_\tau$ we have
$\alpha^\e_\tau(X)=\emptyset$ if and only if $X=\emptyset$.}
\myproof{Proposition}{prop:delta_fixpoints}{
We first prove that $\fp(\delta_\tau)=\codom(\alpha_\tau)$.
Let $X\subseteq\Sigma_\tau$ and $i\in\nat$. By Lemma~\ref{lem:small_step}
and monotonicity (Proposition~\ref{prop:delta_lco}) we have
\begin{align*}
  \alpha_\tau^i(X)&=\cup\{\alpha^i_\tau(\sigma)\mid\sigma\in X\}\\
  &=\cup\{\delta_\tau^i\alpha_\tau^i(\sigma)\mid\sigma\in X\}
    \subseteq\delta_\tau^i\alpha_\tau^i(X)\subseteq\delta_\tau
    \alpha_\tau^i(X)~.
\end{align*}
The converse inclusion $\alpha_\tau^i(X)\subseteq\delta_\tau\alpha_\tau^i(X)$
holds because
$\delta_\tau$ is reductive (Proposition \ref{prop:delta_lco}).
Then $\alpha_\tau^i(X)\in\fp(\delta_\tau)$.
Since $i$ is arbitrary we have
$\alpha_\tau(X)\in\fp(\delta_\tau)$.
Conversely, let $\ee\in\fp(\delta_\tau)$. Consider the set
of states constructed from the frames and memories in
Definition~\ref{def:worst} and let
\[
  X=\{\phi\sep\mu\in\Sigma_\tau\mid
      \phi\in\overline{\phi}_\tau(\ee),\ \mu\in\overline{\mu}(\ee)\}~.
\]
By Corollary~\ref{cor:worst}.ii and since $\ee\in\fp(\delta_\tau)$, we have
$\alpha_\tau(X)=\delta_\tau(\ee)=\ee$.

Since $\delta_\tau$ is reductive (Proposition \ref{prop:delta_lco}),
we have $\emptyset=\delta_\tau(\emptyset)$ \ie
$\emptyset\in\fp(\delta_\tau)$.

If $\mathtt{this}\in\dom(\tau)$, every
$\sigma\in\Sigma_\tau$ is such that $\alpha_\tau(\sigma)\not=\emptyset$,
since $\mathtt{this}$ cannot be unbound (Definition~\ref{def:concrete_states}).
Then $\alpha_\tau(X)=\emptyset$ if and only if $X=\emptyset$.%
}

The proof of Proposition \ref{prop:operations_e} requires some preliminary
results.

Corollary~\ref{cor:alphadelta} states that if we know that the approximation
of a set of concrete states $S$ is some $e\subseteq\Pi$, then we can conclude
that a better approximation of $S$ is $\delta(\ee)$. In other words,
garbage is never used in the approximation.
\begin{corollary}\label{cor:alphadelta}
Let $S\subseteq\Sigma_\tau$ and $\ee\subseteq\Pi$. Then
$\alpha_\tau(S)\subseteq\delta_\tau(\ee)$ if and only if
$\alpha_\tau(S)\subseteq \ee$.
\end{corollary}
\myproofbis{
Assume that $\alpha_\tau(S)\subseteq\delta_\tau(\ee)$. By reductivity
(Proposition \ref{prop:delta_lco}) we have $\alpha_\tau(S)\subseteq\ee$.
Conversely, assume that $\alpha_\tau(S)\subseteq\ee$. By
Proposition~\ref{prop:delta_fixpoints}
and monotonicity (Proposition~\ref{prop:delta_lco}) we have
$\alpha_\tau(S)=\delta_\tau\alpha_\tau(S)\subseteq\delta_\tau(\ee)$.}

Lemma~\ref{lem:codom} states that integer values, $\nil$
and the name of the variables
are not relevant to the definition of $\alpha$ (Definition~\ref{def:alpha}).
\begin{lemma}\label{lem:codom}
Let $\phi'\sep\mu\in\Sigma_{\tau'}$ and
$\phi''\sep\mu\in\Sigma_{\tau''}$ such that
$\codom(\phi')\cap\Loc=\codom(\phi'')\cap\Loc$.
Then
$\alpha_{\tau'}(\phi'\sep\mu)=\alpha_{\tau''}(\phi''\sep\mu)$.
\end{lemma}
\myproofbis{
From Definition~\ref{def:alpha}.
}

Lemma~\ref{lem:alpha_restrict} says that if we consider all the concrete states
approximated by some $e\subseteq\Pi$ and we restrict their frames, then
the resulting set of states is approximated by $\delta(e)$. In other words,
the operation $\delta$ garbage collects all objects that,
because of the restriction, are not longer reachable.
\begin{lemma}\label{lem:alpha_restrict}
Let $\mathit{vs}\subseteq\dom(\tau)$. Then
\[
  \alpha_{\tau|_{-\mathit{vs}}}(\{\phi|_{-\mathit{vs}}\sep
    \mu\mid\phi\sep\mu\in\Sigma_\tau\text{ and }
    \alpha_\tau(\phi\sep\mu)\subseteq \ee\})=\delta_{\tau|_
    {-\mathit{vs}}}(\ee)~.
\]
\end{lemma}
\myproofbis{
We have
\begin{align}
  &\quad\alpha_{\tau|_{-\mathit{vs}}}(\{\phi|_{-\mathit{vs}}\sep
    \mu\mid\phi\sep\mu\in\Sigma_\tau\text{ and }
    \alpha_\tau(\phi\sep\mu)\subseteq \ee\})\notag\\
  &=\alpha_{\tau|_{-\mathit{vs}}}(\{\phi|_{-\mathit{vs}}\sep
    \mu\in\Sigma_{\tau|_{-\mathit{vs}}}\mid
    \phi\sep\mu\in\Sigma_\tau\text{ and }
    \alpha_\tau(\phi\sep\mu)\subseteq \ee\})~,\elabel{eq:alpha_restrict1}
\end{align}
since if $\phi\sep\mu\in\Sigma_\tau$ then
$\phi|_{-\mathit{vs}}\sep\mu\in\Sigma_{\tau|_{-\mathit{vs}}}$.
We have that if $\alpha_\tau(\phi\sep\mu)\subseteq \ee$ then
$\alpha_{\tau|_{-\mathit{vs}}}(\phi|_{-\mathit{vs}}\sep\mu)\subseteq
\ee$. Hence \eref{eq:alpha_restrict1} is contained in $\ee$.
By Corollary~\ref{cor:alphadelta}, \eref{eq:alpha_restrict1} is also contained
in $\delta_{\tau|_{-\mathit{vs}}}(\ee)$.
But also the converse inclusion holds, since in
\eref{eq:alpha_restrict1} we can restrict the choice of $\phi\sep\mu
\in\Sigma_\tau$, so that \eref{eq:alpha_restrict1} contains
\begin{equation}\elabel{eq:alpha_restrict2}
  \alpha_{\tau|_{-\mathit{vs}}}\left(\left\{\phi|_{-\mathit{vs}}\sep
    \mu\in\Sigma_{\tau|_{-\mathit{vs}}}\left|
    \begin{array}{l}
      \phi\sep\mu\in\Sigma_\tau,\ \alpha_\tau(\phi\sep\mu)\subseteq \ee\\
      \phi\in\overline{\phi}_\tau(\ee),\ \mu\in\overline{\mu}(\ee)
    \end{array}\right.\right\}\right)~.
\end{equation}
By points ii and iii of
Lemma~\ref{lem:compatibility},~\eref{eq:alpha_restrict2}
is equal to
\begin{align*}
  &\quad\alpha_{\tau|_{-\mathit{vs}}}(\{\phi|_{-\mathit{vs}}\sep
    \mu\in\Sigma_{\tau|_{-\mathit{vs}}}\mid
    \phi\in\overline{\phi}_\tau(\ee),\ \mu\in\overline{\mu}(\ee)\})\\
  \text{(Definition~\ref{def:worst})}&=
    \alpha_{\tau|_{-\mathit{vs}}}(\{\phi\sep\mu\in
    \Sigma_{\tau|_{-\mathit{vs}}}\mid
    \phi\in\overline{\phi}_{\tau|_{-\mathit{vs}}}(\ee)
    \text{ and }\mu\in\overline{\mu}(\ee)\})\\
  \text{(Corollary~\ref{cor:worst}.ii)}&=\delta_{\tau|_{-\mathit{vs}}}(\ee)~.
\end{align*}
}

We are now ready to prove the correctness and optimality of the
abstract operations in Figure~\ref{fig:operations_e}.

\mbox{}\\
\summary{Proposition \ref{prop:e_insertion}.}
{The map $\alpha^\e_\tau$ (Definition~\ref{def:alpha}) is the abstraction
map of a Galois insertion from $\wp(\Sigma_\tau)$ to $\e_\tau$.}
\myproof{Proposition}{prop:operations_e}{
By the theory of abstract interpretation \cite{CousotC77}, given
$\ee\in\mathcal{E}_\tau$, the concretisation map induced
by the abstraction map of Definition~\ref{def:alpha} is
\[
  \gamma_\tau(\ee)=
    \{\sigma\in\Sigma_\tau\mid\alpha_\tau(\sigma)\subseteq \ee\}~.
\]
Moreover, the optimal abstract counterpart of a concrete operation
$\mathit{op}$ is $\alpha\mathit{op}\gamma$.

We consider every operation in Figure~\ref{fig:concrete_states} and we compute
the induced optimal abstract operation, which will always coincide with that
reported in Figure~\ref{fig:operations_e}.

Note that all the operations in Figure~\ref{fig:concrete_states} use
states in $\Sigma_\tau$ with $\mathtt{this}\in\dom(\tau)$
(Figure~\ref{fig:signatures}).
By Proposition~\ref{prop:delta_fixpoints}
we have $\gamma_\tau(\emptyset)=\emptyset$. Then the powerset extension of the
operations in Figure~\ref{fig:concrete_states} are strict on $\emptyset$.
The only exception is the second argument of $\mathsf{return}$, which is a
state whose frame is not required to contain \texttt{this}
(Figure~\ref{fig:signatures}). The operation
$\cup$ is not the powerset extension of an operation in Figure
\ref{fig:concrete_states}. Then it is not strict in general. Hence, in the
following, we will consider just the cases when the arguments of the abstract
counterparts of the operations in Figure~\ref{fig:concrete_states} are not
$\emptyset$ (except for the second argument of $\mathsf{return}$
and for $\cup$).

In this proof, we will use the following properties.
\begin{itemize}
\item[P1] If $\ee\in\mathcal{E}_\tau$, $\ee\not=\emptyset$ and $\mathtt{this}\in\dom
(\tau)$ then there exists $\pi\in \ee$ such that $k(\pi)\le\tau(\mathtt{this})$.
\item[P2] If $\ee\in\mathcal{E}_\tau$, $\ee\not=\emptyset$ and $\mathtt{this}\in\dom
(\tau)$ then there exists $\sigma\in\Sigma_\tau$ such that
$\alpha_\tau(\sigma)\subseteq \ee$.
\item[P3] $\alpha_\tau\gamma_\tau$ is the identity map.
\end{itemize}
P1 holds since $\ee=\delta_\tau(\ee)$ (Definition~\ref{def:property_e})
so that by Definition~\ref{def:delta}, we can conclude
that there exists such a $\pi$.
To see that P2 is a consequence of P1, let $\pi$ be
as defined in P1; then, letting $\sigma=[\mathtt{this}\mapsto l]\sep
[l\mapsto\pi\sep\init(F(k(\pi)))]$ for some $l\in\Loc$, we have
$\sigma \in \Sigma_\tau$. Moreover, by
Definition~\ref{def:alpha}, $\alpha_\tau(\sigma)=\{\pi\}\subseteq \ee$
so that P2 holds.
By Proposition~\ref{prop:e_insertion}, $\alpha_\tau$ is a Galois insertion
and hence, P3 holds.
\mbox{}\\
\proofoperation{nop}
\mbox{}\\
By P3 we have
\[
  \alpha_\tau(\mathsf{nop}_\tau(\gamma_\tau(\ee)))=
    \alpha_\tau\gamma_\tau(\ee)=\ee~.
\]
\proofoperation{get\_int,\ get\_null,\ get\_var}
\begin{align*}
  &\quad\alpha_{\tau[\rs\mapsto\integer]}
    (\mathsf{get\_int}^i_\tau(\gamma_\tau(\ee)))\\
  &=\alpha_{\tau[\rs\mapsto\integer]}(\{\phi[\rs\mapsto i]\sep\mu\mid
    \phi\sep\mu\in\gamma_\tau(\ee)\})\\
  \text{($*$)}
    &=\alpha_\tau(\{\phi\sep\mu\in\Sigma_\tau
    \mid\phi\sep\mu\in\gamma_\tau(\ee)\})
      =\alpha_\tau\gamma_\tau(\ee)=\ee~,
\end{align*}
where $*$ follows by Lemma~\ref{lem:codom}
since $\rs\not\in\dom(\tau)$. For the same reason, point $*$ follows
if $\rs$ is bound to $\nil$ or to some $\phi(v)$ with
$v\in\dom(\tau)$. Thus the proof above is also a proof
of the optimality of $\mathsf{get\_null}$ and of $\mathsf{get\_var}$.
\mbox{}\\
\proofoperation{expand}
\begin{align*}
  &\quad\alpha_{\tau[v\mapsto t]}(\mathsf{expand}^{v:t}_\tau(\gamma_\tau(\ee)))\\
  &=\alpha_{\tau[v\mapsto t]}(\{\phi[v\mapsto\init(t)]\sep\mu\mid
    \phi\sep\mu\in\gamma_\tau(\ee)\})\\
  \text{($*$)}
    &=\alpha_\tau(\{\phi\sep\mu\in\Sigma_\tau
    \mid\phi\sep\mu\in\gamma_\tau(\ee)\})
      =\alpha_\tau\gamma_\tau(\ee)=\ee~,
\end{align*}
where point $*$ follows by Lemma~\ref{lem:codom}, since $\init(t)\in\{0,\nil\}$
and $v\not\in\dom(\tau)$.
\mbox{}\\
\proofoperation{restrict}
\begin{align*}
  &\quad\alpha_{\tau|_{-\mathit{vs}}}(\mathsf{restrict}^\mathit{vs}
    _\tau(\gamma_\tau(\ee)))\notag\\
  &=\alpha_{\tau|_{-\mathit{vs}}}(\mathsf{restrict}^\mathit{vs}_\tau
    (\{\sigma\in\Sigma_\tau\mid\alpha_\tau(\sigma)\subseteq \ee\}))\notag\\
  &=\alpha_{\tau|_{-\mathit{vs}}}(\{\phi|_{-\mathit{vs}}\sep
    \mu\mid\phi\sep\mu\in\Sigma_\tau\text{ and }
    \alpha_\tau(\phi\sep\mu)\subseteq \ee\})\\
  \text{(Lemma~\ref{lem:alpha_restrict})}&=\delta_{\tau|_{-\mathit{vs}}}(\ee)~.
\end{align*}
\proofoperation{is\_null}
\begin{align*}
  &\quad\alpha_{\tau[\rs\mapsto\integer]}
    (\mathsf{is\_null}_\tau(\gamma_\tau(\ee)))\\
  &=\alpha_{\tau[\rs\mapsto\integer]}(\mathsf{is\_null}_\tau
    (\{\sigma\in\Sigma_\tau\mid\alpha_\tau(\sigma)\subseteq \ee\}))\\
  &=\alpha_{\tau[\rs\mapsto\integer]}\left(\left\{\phi[\rs\mapsto 1]\sep\mu
  \left|\begin{array}{l}
    \phi\sep\mu\in\Sigma_\tau\\
    \alpha_\tau(\phi\sep\mu)\subseteq\ee
  \end{array}\right.\right\}\right)\\
  \text{(Lemma~\ref{lem:codom})}
  &=\alpha_{\tau|_{-\rs}}(\{\phi|_{-\rs}\sep\mu\mid
    \phi\sep\mu\in\Sigma_\tau\text{ and }
    \alpha_\tau(\phi\sep\mu)\subseteq \ee\})\\
  \text{(Lemma~\ref{lem:alpha_restrict})}&=\delta_{\tau|_{-\rs}}(\ee)\\
  \text{(Definition~\ref{def:delta})}&=\delta_{\tau[\rs\mapsto\integer]}(\ee)~.
\end{align*}
\proofoperation{put\_var}
\begin{align}
  &\quad\alpha_{\tau|_{-\rs}}(\mathsf{put\_var}_\tau(\gamma_\tau(\ee)))\notag\\
  &=\alpha_{\tau|_{-\rs}}(\mathsf{put\_var}_\tau
    (\{\sigma\in\Sigma_\tau\mid\alpha_\tau(\sigma)\subseteq \ee\}))\notag\\
  &=\alpha_{\tau|_{-\rs}}(\{\phi[v\mapsto\phi(\rs)]|_{-\rs}\sep\mu
    \mid\phi\sep\mu\in\Sigma_\tau\text{ and }
    \alpha_\tau(\phi\sep\mu)\subseteq \ee\})~.\elabel{eq:putvar1}
\end{align}
Observe that $\codom(\phi[v\mapsto\phi(\rs)]|_{-\rs})=\codom(\phi|_{-v})$
so that,
by Lemmas~\ref{lem:codom} and~\ref{lem:alpha_restrict},
\eref{eq:putvar1} is equal to
\[
  \alpha_{\tau|_{-v}}(\{\phi|_{-v}\sep\mu
    \mid\phi\sep\mu\in\Sigma_\tau\text{ and }
    \alpha_\tau(\phi\sep\mu)\subseteq \ee\})
    =\delta_{\tau|_{-v}}(\ee)~.
\]
\proofoperation{call}
\begin{align*}
  &\quad
    \alpha_{P(\nu)|_{-\Out}}(\mathsf{call}_\tau^{\nu,v_1,\ldots,v_n}
    (\gamma_\tau(\ee)))\\
  &=\alpha_{P(\nu)|_{-\Out}}(\mathsf{call}_\tau
    ^{\nu,v_1,\ldots,v_n}
    (\{\sigma\in\Sigma_\tau\mid\alpha_\tau(\sigma)\subseteq \ee\}))\\
  &=\alpha_{P(\nu)|_{-\Out}}
    \left(\left\{\left.\left[\begin{array}{c}
      \iota_1\mapsto\phi(v_1),\\
      \vdots\\
      \iota_n\mapsto\phi(v_n),\\
      \mathtt{this}\mapsto\phi(\rs)\end{array}\right]\sep\mu
    \right|\begin{array}{l}
      \phi\sep\mu\in\Sigma_\tau\text{ and}\\
      \alpha_\tau(\phi\sep\mu)\subseteq \ee
    \end{array}\right\}\right)\\
  \text{($*$)}&=\alpha_{\tau|_{\{v_1,\ldots,v_n,\rs\}}}
    (\{\phi|_{\{v_1,\ldots,v_n,\rs\}}\sep\mu\mid
      \phi\sep\mu\in\Sigma_\tau,\ \alpha_\tau
        (\phi\sep\mu)\subseteq \ee\})\\
  \text{($**$)}&=\delta_{\tau|_{\{v_1,\ldots,v_n,\rs\}}}(\ee)~,
\end{align*}
where point $*$ follows by Lemma~\ref{lem:codom} and point $**$ follows
by Lemma~\ref{lem:alpha_restrict}.
\mbox{}\\
\proofoperation{is\_true,\ is\_false}
\begin{align*}
  &\quad\alpha_\tau(\mathsf{is\_true}_\tau(\gamma_\tau(\ee)))\\
  &=\alpha_\tau(\{\phi\sep\mu\in\gamma_\tau(\ee)\mid
    \phi(\rs)\ge 0\})\\
  \text{(Lemma~\ref{lem:codom})}&=\alpha_\tau\gamma_\tau(\ee)=\ee~.
\end{align*}
The optimality of $\mathsf{is\_false}$ follows by a similar proof.
\mbox{}\\
\proofoperation{new}
\mbox{}\\
Let $\kappa=k(\pi)$. Since $\rs\not\in\dom(\tau)$ we have
\begin{align}
  &\quad\alpha_{\tau[\rs\mapsto\kappa]}
    (\mathsf{new}^\pi_\tau(\gamma_\tau(\ee)))\notag\\
  &=\alpha_{\tau[\rs\mapsto\kappa]}(\mathsf{new}^\pi_\tau
    (\{\sigma\in\Sigma_\tau\mid\alpha_\tau(\sigma)\subseteq \ee\}))\notag\\
  &=\alpha_{\tau[\rs\mapsto\kappa]}
    \left(\left\{\begin{array}{c}
      \phi[\rs\mapsto l]\sep\\
      \sep\mu[l\mapsto\pi\sep\init(F(\kappa))
    ]\end{array}\left|\begin{array}{l}
      \phi\sep\mu\in\Sigma_\tau,\ \alpha_\tau
        (\phi\sep\mu)\subseteq \ee\\
      l\in\Loc\setminus\dom(\mu)
    \end{array}\right.\right\}\right)\notag\\
  &=\alpha_\tau(\{\phi\sep\mu\in\Sigma_\tau\mid
    \alpha_\tau(\phi\sep\mu)\subseteq \ee\})\cup\elabel{eq:new1}\\
  &\quad\cup\alpha_{[\rs\mapsto\kappa]}
    \left(\left\{
      \begin{array}{c}
        [\rs\mapsto l]\sep\\
        \sep[l\mapsto\pi\sep\init(F(\kappa))]
      \end{array}
      \left|\begin{array}{l}
        \phi\sep\mu\in\Sigma_\tau,\ \alpha_\tau
          (\phi\sep\mu)\subseteq \ee\\
        l\in\Loc\setminus\dom(\mu)
      \end{array}\right.
    \right\}\right).\elabel{eq:new2}
\end{align}
We have that \eref{eq:new1} is equal to $\ee$.
By P2 and Definition~\ref{def:alpha},
\eref{eq:new2} is equal to $\{\pi\}$.
\mbox{}\\
\proofoperation{=,\ +}
\begin{align*}
  &\quad\alpha_\tau(\mathsf{=}_\tau(\gamma_\tau(\ee_1))(\gamma_\tau(\ee_2)))\\
  &=\alpha_\tau(\{\mathsf{=}_\tau
    (\sigma_1)(\sigma_2)\mid
    \sigma_1\in\gamma_\tau(\ee_1),\ \sigma_2\in\gamma_\tau(\ee_2)\})\\
  \text{(P2)}&=\alpha_\tau(\{\sigma_2\mid\sigma_2\in\gamma_\tau(\ee_2)\})\\
  &=\alpha_\tau\gamma_\tau(\ee_2)=\ee_2~.
\end{align*}
The optimality of $\mathsf{+}$ follows by a similar proof.
\mbox{}\\
\proofoperation{return}
\mbox{}\\
Let $\tau'=\tau[\rs\mapsto P(\nu)(\Out)]$,
$\tau''=P(\nu)|_{\Out}$ and $L=\codom(\phi_1|_{-\rs})\cap\Loc$.
\begin{align}
  &\quad\alpha_{\tau'}(\mathsf{return}_\tau^\nu(\gamma_\tau(\ee_1))
    (\gamma_{\tau''}(\ee_2)))\notag\\
  &=\alpha_{\tau'}(\mathsf{return}_\tau^\nu
      (\{\sigma_1\in\Sigma_\tau\mid\alpha_\tau(\sigma_1)\subseteq \ee_1\})
    (\{\sigma_2\in\Sigma_{\tau''}\mid
      \alpha_{\tau''}(\sigma_2)\subseteq \ee_2\}))\notag\\
  &=\alpha_{\tau'}\left(\left\{
      \phi_1|_{-\rs}[\rs\mapsto\phi_2(\Out)]\sep\mu_2\left|
      \underbrace{\begin{array}{c}
        \phi_1\sep\mu_1\in\Sigma_\tau\\
        \phi_2\sep\mu_2\in\Sigma_{\tau''}\\
        \alpha_\tau(\phi_1\sep\mu_1)\subseteq \ee_1\\
        \alpha_{\tau''}(\phi_2\sep\mu_2)\subseteq \ee_2\\
        \mu_1=_L\mu_2
    \end{array}}_{Cond}\right.\right\}\right)\notag\\
  \text{($*$)}
    &=\alpha_{\tau|_{-\rs}}(\{\phi_1|_{-\rs}\sep\mu_2\mid Cond\})
    \cup\elabel{eq:unscope0}\\
  &\qquad\cup\alpha_{\tau''}(\{\phi_2\sep\mu_2\mid Cond\})\elabel{eq:unscope1}
\end{align}
where point $*$ follows by Lemma~\ref{lem:codom}.
Since $\alpha_{\tau''}(\phi_2\sep\mu_2)\subseteq \ee_2$, an upper
bound of \eref{eq:unscope1} is $\ee_2$. But $\ee_2$ is also a lower
bound of \eref{eq:unscope1} since, by Lemma~\ref{lem:compatibility}.iii,
a lower bound of \eref{eq:unscope1} is
\[
  \alpha_{\tau''}\left(\left\{\phi_2\sep\mu_2\left|\begin{array}{l}
      \phi_1\in\overline{\phi}_\tau(\ee_1),\ \mu_1\in\overline{\mu}(\ee_1)\\
      \phi_2\in\overline{\phi}_{\tau''}(\ee_2),\ \mu_2\in\overline{\mu}(\ee_2)
    \end{array}\right.\right\}\right)
\]
which by Corollary~\ref{cor:worst}.ii is equal to $\ee_2$. Note that the
condition $\mu_1=_L\mu_2$ is satisfied by Definition~\ref{def:worst}.

Instead \eref{eq:unscope0} is
\[
  \cup\left\{\{o.\pi\}\cup\alpha_{F(k(o.\pi))}(o.\phi\sep\mu_2)
    \left|\begin{array}{l}
    v\in\dom(\phi_1|_{-\rs})\\
    \phi_1|_{-\rs}(v)\in\Loc\\
    o=\mu_2\phi_1|_{-\rs}(v),\ Cond
    \end{array}\right.\right\}
\]
which, since $\mu_1=_L\mu_2$, is equal to
\begin{align}
  &\cup\left\{\{o.\pi\}\!\cup\!\alpha_{F(k(o.\pi))}
    (o.\phi\sep\mu_2)\left|\begin{array}{l}
      v\!\in\!\dom(\phi_1|_{-\rs})\\
      \phi_1|_{-\rs}(v)\!\in\!\Loc\\
      o=\mu_1\phi_1|_{-\rs}(v),\ Cond
    \end{array}\right.\right\}\notag\\
  \text{($*$)}\subseteq&
    \cup\left\{\{o.\pi\}\cup\delta_{F(k(o.\pi))}(\Pi)\left|\begin{array}{l}
      v\in\dom(\phi_1|_{-\rs}),\ \phi_1|_{-\rs}(v)\in\Loc\\
      o=\mu_1\phi_1|_{-\rs}(v),\ Cond
    \end{array}\right.\right\}\notag\\
  \text{($**$)}\subseteq&\cup\left\{\{\pi\}\cup\delta_{F(k(\pi))}(\Pi)\left|
    \begin{array}{l}
      \kappa\in\codom(\tau|_{-\rs})\cap\mathcal{K}\\
      \pi\in\ee_1,\ k(\pi)\le\kappa,\ Cond
    \end{array}\right.\right\}\notag\\
  \subseteq&\cup\left\{\{\pi\}\cup\delta_{F(k(\pi))}(\Pi)\left|
    \begin{array}{l}
      \kappa\in\codom(\tau|_{-\rs})\cap\mathcal{K}\\
      \pi\in \ee_1,\ k(\pi)\le\kappa
    \end{array}\right.\right\}~,\elabel{eq:unscope2}
\end{align}
where point $*$ follows by Lemma~\ref{lem:small_step} and
point $**$ holds since
$Cond$ requires that $\alpha_\tau(\phi_1\sep\mu_1)\subseteq
\ee_1$. But \eref{eq:unscope2} is also a lower bound of
\eref{eq:unscope0}, since \eref{eq:unscope0} contains
\begin{multline*}
  \alpha_{\tau|_{-\rs}}\left(\left\{\phi_1|_{-\rs}\sep\mu_2
    \in\Sigma_{\tau|_{-\rs}}\left|\begin{array}{l}
      \phi_1\in\overline{\phi}_\tau(\ee_1),\ \mu_1\in\overline{\mu}(\ee_1),\\
      \phi_2=\init(P(\nu)|_{\Out}),\ \mu_2\in\overline{\mu}(\Pi)
    \end{array}\right.\right\}\right)\\
  =\alpha_{\tau|_{-\rs}}(\{\phi\sep\mu
    \in\Sigma_{\tau|_{-\rs}}\mid
      \phi\in\overline{\phi}_{\tau|_{-\rs}}(\ee_1), \mu\in\overline{\mu}(\Pi)
    \})~,
\end{multline*}
which by Corollary~\ref{cor:worst}.i is equal to \eref{eq:unscope2}.
\mbox{}\\
\proofoperation{get\_field}
\mbox{}\\
Let $\tau'=\tau[\rs\mapsto F(\tau(\rs))(f)]$ and
$\tau''=[\rs\mapsto F(\tau(\rs))(f)]$. We have
\begin{align}
  &\quad\alpha_{\tau'}(\mathsf{get\_field}_\tau^f(\gamma_\tau(\ee)))\notag\\
  &=\alpha_{\tau'}(\mathsf{get\_field}_\tau^f(\{\phi\sep\mu\in
    \Sigma_\tau\mid\alpha_\tau(\phi\sep\mu)\subseteq \ee\}))\notag\\
  &=\alpha_{\tau'}\left(\left\{\phi|_{-\rs}[\rs\mapsto
    (\mu\phi(\rs)).\phi(f)]\sep\mu\left|\begin{array}{l}
      \phi\sep\mu\in\Sigma_\tau\\
      \phi(\rs)\not=\nil,\\
      \alpha_\tau(\phi\sep\mu)\subseteq \ee
    \end{array}\right.\right\}\right)\notag\\
  &=\alpha_{\tau|_{-\rs}}(\{\phi|_{-\rs}\sep\mu\mid
    \phi\sep\mu\in\Sigma_\tau,\ \phi(\rs)\not=\nil
    ,\ \alpha_\tau(\phi\sep\mu)\subseteq \ee\})\cup
    \notag\\
  &\quad\cup\alpha_{\tau''}\left(\left\{[\rs\mapsto
    (\mu\phi(\rs)).\phi(f)]\sep\mu\left|\begin{array}{l}
      \phi\sep\mu\in\Sigma_\tau\\
      \phi(\rs)\not=\nil,\\
      \alpha_\tau(\phi\sep\mu)\subseteq \ee
    \end{array}\right.\right\}\right)~.\elabel{eq:get_field3}
\end{align}
 \eref{eq:get_field3} is equal to
$\emptyset$ if $\{\pi\in \ee\mid k(\pi)\le
\tau(\rs)\}=\emptyset$, since in such a case the condition
$\phi(\rs)\not=\nil$ cannot be satisfied.
Since $\alpha_\tau(\phi\sep\mu)\subseteq \ee$, an upper bound of
\eref{eq:get_field3} is $\ee$. By Corollary~\ref{cor:alphadelta},
also $\delta_{\tau'}(\ee)$ is an upper bound of \eref{eq:get_field3}.
But it is also a lower bound of \eref{eq:get_field3}, since, from the
hypothesis on $\ee$ and from points ii and iii of
Lemma~\ref{lem:compatibility}, \eref{eq:get_field3} contains
\begin{align*}
  &\quad\alpha_{\tau|_{-\rs}}(\{\phi|_{-\rs}\sep\mu\in
    \Sigma_{\tau|_{-\rs}}\mid
    \phi\in\overline{\phi}_\tau(\ee),\ \mu\in\overline{\mu}(\ee)\})\cup\\
  &\quad\cup\alpha_{\tau''}(\{[\rs\mapsto
    (\mu\phi(\rs)).\phi(f)]\sep\mu\in\Sigma_{\tau''}\mid
      \phi\in\overline{\phi}_\tau(\ee),\ \mu\in\overline{\mu}(\ee)\})\\
  \text{($*$)}&=\alpha_{\tau|_{-\rs}}(\{\phi\sep\mu\in
    \Sigma_{\tau|_{-\rs}}\mid
    \phi\in\overline{\phi}_{\tau|_{-\rs}}(\ee),\ \mu
    \in\overline{\mu}(\ee)\})\cup\\
  &\quad\cup\alpha_{\tau''}(\{\phi\sep\mu\in\Sigma_{\tau''}\mid
      \phi\in\overline{\phi}_{\tau''}(\ee),\ \mu\in\overline{\mu}(\ee)\})\\
  \text{($**$)}&=\delta_{\tau|_{-\rs}}(\ee)\cup\delta_{\tau''}(\ee)=
    \delta_{\tau'}(\ee)~,
\end{align*}
where point $*$ follows by Definition~\ref{def:worst} and point $**$ follows by
Corollary~\ref{cor:worst}.ii.
\mbox{}\\
\proofoperation{lookup}
\begin{align}
  &\quad\alpha_\tau(\mathsf{lookup}^{m,\nu}_\tau(\gamma_\tau(\ee)))\notag\\
  &=\alpha_\tau(\mathsf{lookup}^{m,\nu}_\tau(\{\phi\sep\mu\in
    \Sigma_\tau\mid\alpha_\tau(\phi\sep\mu)\subseteq \ee\}))\notag\\
  &=\alpha_\tau\left(\left\{\phi\sep\mu\in\Sigma_\tau\left|
    \underbrace{\begin{array}{l}
      \alpha_\tau(\phi\sep\mu)\subseteq\ee,\ \phi(\rs)\not=\nil\\
      M(k((\mu\phi(\rs)).\pi))(m)=\nu
    \end{array}}_{Cond}\right.\right\}\right).\elabel{eq:lookup1}
\end{align}
Equation \eref{eq:lookup1} is equal to
$\emptyset$ if there is no $\pi\in \ee$ such that
$k(\pi)\le\tau(\rs)$ and $M(\pi)(m)=\nu$, because in such a case it is not
possible to satisfy the condition
$M(k((\mu\phi(\rs)).\pi))(m)=\nu$. Otherwise, it is equal to
\begin{align}
  &\alpha_{\tau|_{-\rs}}(\{\phi|_{-\rs}\sep\mu\mid
    \phi\sep\mu\in\Sigma_\tau,\ Cond\})\cup\elabel{eq:lookup2}\\
  &\cup\alpha_{\tau|_{\rs}}(\{\phi|_{\rs}\sep\mu\mid
    \phi\sep\mu\in\Sigma_\tau,\ Cond\})\elabel{eq:lookup3}~.
\end{align}
Since $Cond$ requires that
$\alpha_\tau(\phi\sep\mu)\subseteq \ee$, by Corollary
\ref{cor:alphadelta} an upper bound of \eref{eq:lookup2} is
$\delta_{\tau|_{-\rs}}(\ee)$. But it is also a lower
bound of \eref{eq:lookup2}, since a lower bound of \eref{eq:lookup2} is
\begin{align*}
  &\quad\alpha_{\tau|_{-\rs}}\left(\left\{\phi|_{-\rs}\sep\mu\in
    \Sigma_{\tau|_{-\rs}}\left|
    \begin{array}{l}
      \phi\in\overline{\phi}_\tau(\ee),\ \mu\in\overline{\mu}(\ee)\\
      \phi(\rs)\not=\nil\\
      M(k((\mu\phi(\rs)).\pi))(m)=\nu
    \end{array}\right.\right\}\right)\\
  \text{($*$)}&=\alpha_{\tau|_{-\rs}}(\{\phi\sep\mu\in
    \Sigma_{\tau|_{-\rs}}\mid
      \phi\in\overline{\phi}_{\tau|_{-\rs}}(\ee),\ \mu\in\overline{\mu}(\ee)\})\\
  \text{($**$)}&=\delta_{\tau|_{-\rs}}(\ee)~.
\end{align*}
Point $*$ follows from the hypothesis on $\ee$.
Point $**$ follows by Corollary~\ref{cor:worst}.ii.

Instead, \eref{eq:lookup3} is contained in
\begin{multline*}
  \alpha_{\tau|_{\rs}}\left(\left\{\phi|_{\rs}\sep\mu\in
    \Sigma_{\tau|_{\rs}}\left|\begin{array}{l}
      \phi\in\overline{\phi}_\tau(\ee),\ \mu
        \in\overline{\mu}(\ee)\\
      M(k((\mu\phi(\rs)).\pi))(m)=\nu\end{array}\right.\right\}\right)\\
  =\alpha_{\tau|_{\rs}}\left(\left\{\phi\sep\mu\in
    \Sigma_{\tau|_{\rs}}\left|\begin{array}{l}
      \phi\in\overline{\phi}_{\tau|_{\rs}}(\ee),\ \mu
        \in\overline{\mu}(\ee)\\
      M(k((\mu\phi(\rs)).\pi))(m)=\nu
    \end{array}\right.\right\}\right)~,
\end{multline*}
which, by Corollary~\ref{cor:worst2}, is
\[
  \cup\{\{\pi\}\cup\delta_{F(k(\pi))}(\ee)\mid
    \pi\in \ee,\ k(\pi)\le\tau(\rs),\ M(k(\pi))(m)=\nu\}~.
\]
\proofoperation{put\_field}
\begin{align}
  &\quad\alpha_{\tau|_{-\rs}}
    (\mathsf{put\_field}_{\tau,\tau'}(\gamma_\tau(\ee_1))
    (\gamma_\tau(\ee_2)))\notag\\
  &=\alpha_{\tau|_{-\rs}}(\mathsf{put\_field}_{\tau,\tau'}
    (\{\sigma_1\in\Sigma_\tau\mid\alpha_\tau(\sigma_1)\subseteq \ee_1\})
    \notag\\
  &\qquad\qquad\qquad
    (\{\sigma_2\in\Sigma_{\tau'}\mid\alpha_{\tau'}(\sigma_2)\subseteq \ee_2\}))
    \notag\\
  &=\alpha_{\tau|_{-\rs}}\!\!\left(\left\{\!\begin{array}{c}
    \phi_2|_{-\rs}\sep\mu_2[l\!\mapsto\!\mu_2\!(l).\pi\sep\\
      \sep\mu_2(l).\phi[f\mapsto\phi_2(\rs)]]
    \end{array}
    \left|\begin{array}{l}
      \phi_1\sep\mu_1\in\Sigma_\tau\\
      \phi_2\sep\mu_2\in\Sigma_{\tau'}\\
      \alpha_\tau(\phi_1\sep\mu_1)\subseteq\ee_1\\
      \alpha_{\tau'}(\phi_2\sep\mu_2)\subseteq\ee_2\\
      (l=\phi_1(\rs))\not=\nil\\
      \mu_1=_l\mu_2
    \end{array}\right.\right\}\!\right)\elabel{eq:put_field1}
\end{align}
which is $\emptyset$ if there is no $\pi\in \ee_1$ such that
$k(\pi)\le\tau(\rs)$, since in such a case the condition
$\phi_1(\rs)\not=\nil$ cannot be satisfied. Otherwise,
note that the operation $\mathsf{put\_field}$ copies the value of
$\phi_2(\rs)$, which is obviously reachable from $\phi_2$, inside
a field. Since $\alpha_{\tau'}(\phi_2\sep\mu_2)\subseteq \ee_2$,
we conclude that an upper bound of \eref{eq:put_field1} is
$\ee_2$. Then $\delta_{\tau|_{-\rs}}(\ee_2)$ is also an upper bound of
\eref{eq:put_field1} (Corollary~\ref{cor:alphadelta}). We show that it is
also a lower bound.
Let $\pi_1\in \ee_1$ be such that $k(\pi_1)\le\tau(\mathtt{this})$ (possible
for P1) and $\pi_2\in \ee_1$ be such that $k(\pi_2)\le\tau(\rs)$
(possible for the hypothesis on $\ee_1$). Let
$o_1=\pi_1\sep\init(F(k(\pi_1)))$ and $o_2=\pi_2\sep\init(F(k(\pi_2)))$. We
obtain the following lower bound of \eref{eq:put_field1} by choosing special
cases for $\phi_1$, $\mu_1$, $\phi_2$ and $\mu_2$:
\begin{equation}\elabel{eq:putfield2}
  \alpha_{\tau|_{-\rs}}\!\!\left(\left\{
    \begin{array}{c}
      \phi_2|_{-\rs}\!\sep\!\mu_2[l_2\!\mapsto\!\mu_2(l_2).\pi\sep\\
        \sep\mu_2(l_2).\phi[f\mapsto\phi_2(\rs)]]
    \end{array}
    \left|\begin{array}{l}
      \phi_1\!=\!\init(\tau)[\mathtt{this}\!\mapsto\!l_1,\rs\!\mapsto\!l_2]\\
      \phi_2\in\overline{\phi}_{\tau'}(\ee_2),\ \mu_2'\in\overline{\mu}(\ee_2)\\
      \phi_2\sep\mu_2'\in\Sigma_{\tau'}\\
      \mu_1=\mu_2=\mu_2'[l_1\mapsto o_1,l_2\mapsto o_2]\\
      l_1,l_2\in\Loc\setminus\dom(\mu_2'),\ l_1\not=l_2
    \end{array}\right.\right\}\right).
\end{equation}
Since $l_2$ is not used in $\phi_2$ nor in $\mu_2'$, \eref{eq:putfield2}
becomes
\begin{align*}
  &\quad\alpha_{\tau|_{-\rs}}\left(\left\{
    \phi_2|_{-\rs}\sep\mu_2\in\Sigma_{\tau|_{-\rs}}\left|\begin{array}{l}
      \phi_2\in\overline{\phi}_{\tau'}(\ee_2)\\
      \mu_2\in\overline{\mu}(\ee_2)
    \end{array}\right.\right\}\right)\\
  \text{(Definition~\ref{def:worst})}&=\alpha_{\tau|_{-\rs}}(\{
    \phi\sep\mu\in\Sigma_{\tau|_{-\rs}}\mid
      \phi\in\overline{\phi}_{\tau|_{-\rs}}(\ee_2),\ \mu\in\overline{\mu}(\ee_2)
    \})\\
  \text{(Lemma~\ref{lem:worst})}&=\delta_{\tau|_{-\rs}}(\ee_2)~.
\end{align*}
\proofoperation{\cup}
\mbox{}\\
By additivity (Proposition~\ref{prop:e_insertion}),
the best approximation of $\cup$ over $\wp(\Sigma_\tau)$ is $\cup$ over
$\wp(\Pi)$.}
\section{Proofs of Propositions~\ref{prop:xi_lco}, \ref{prop:xi_fixpoints},
\ref{prop:operations_er} and~\ref{prop:inclusion}
in Section~\ref{sec:erdomain}.}

\summary
{Proposition \ref{prop:xi_lco}.}
{The abstract garbage collector $\xi_\tau$ is an lco.}
\myproofbis{
By Definition~\ref{def:xi}, the map
$\xi_\tau$ is reductive and monotonic.
For idempotency, we have $\xi_\tau\xi_\tau(\bot)=\bot=\xi_\tau(\bot)$.
Let $s\in\Frame_\tau^\er\times\Memory^\er$.
If $\mathtt{this}\in\domain(\tau)$ and $\phi(\mathtt{this})=\emptyset$ then
$\xi_\tau\xi_\tau(s)=\bot=\xi_\tau(\bot)$. Otherwise, we prove that
$\rho_\tau\xi_\tau(s)=\rho_\tau(s)$, which entails the thesis
by Definition~\ref{def:xi}. We have
\begin{align*}
  \rho_\tau\xi_\tau(\phi\sep\mu)&=
    \rho_\tau(\phi\sep\cup\{\mu|_{\domain(F(k(\pi')))}\mid
    \pi'\in\rho_\tau(\phi\sep\mu)\}
    )\\
  &=\{\pi\in\phi(v)\mid v\in\domain(\tau),\ \tau(v)\in\mathcal{K}\}\ \cup\\
  &\qquad\cup\left\{\pi\in\mu(f)\left|
    \begin{array}{l}
      \pi'\in\rho_\tau(\phi\sep\mu),
        \ f\!\in\!\domain(F(k(\pi')))\\
      F(k(\pi'))(f)\in\mathcal{K}
    \end{array}\right.\right\}\\
  &=\rho_\tau(\phi\sep\mu).
\end{align*}}

To prove Proposition~\ref{prop:xi_fixpoints},
we need some preliminary definitions and results.

Let $s\in\Frame_\tau^\er\times\Memory^\er$. We define
frames and memories which use all possible creation points allowed by $s$.
\begin{definition}\label{def:er_worst}
Let $\phi\in\Frame_\tau^\er$, $\mu\in\Memory^\er$
and $l:\Pi\mapsto\Loc$ be one-to-one. We define
\begin{align*}
  \overline{\phi}_\tau&=\left\{\phi^\flat\in\Frame_\tau\left|
  \begin{array}{l}
    \text{for every $v\in\dom(\tau)$}\\
    \text{ if $\tau(v)=\integer$ then $\phi^\flat(v)=0$}\\
    \text{ if $\tau(v)\in\mathcal{K}$ and $\phi(v)=\emptyset$
      then $\phi^\flat(v)=\nil$}\\
    \text{ if $\tau(v)\in\mathcal{K}$ and $\phi(v)\not=\emptyset$
      then $\phi^\flat(v)\in l\phi(v)$}
  \end{array}\right.\right\}~,\\
  \overline{\mu}&=\left\{\mu^\flat\in\Memory\left|\begin{array}{l}
    \dom(\mu^\flat)=\codom(l),\ \mu^\flat(l(\pi))=\pi\sep\phi_\pi^\flat\\
    \text{with }\phi_\pi^\flat\in
      \overline{\mu}_{F(\pi)}\text{ for every $\pi\in\Pi$}
  \end{array}\right.\right\}~.
\end{align*}
\end{definition}
Lemma~\ref{lem:pre_small_er}
is needed in the proof of Lemma~\ref{lem:small_er}.
\begin{lemma}\label{lem:pre_small_er}
Let $\phi\in\Frame_\tau^\er$, $\mu\in\Memory^\er$,
$\phi^\flat\in\overline{\phi}_\tau$ and $\mu^\flat\in\overline{\mu}$.
Then
$\varepsilon_\tau(\phi^\flat\sep\mu^\flat)\subseteq\phi$.
\end{lemma}
\myproofbis{
For every $v\in\dom(\tau)$ we have
\begin{align*}
  \varepsilon_\tau(\phi^\flat\sep\mu^\flat)(v)&=\begin{cases}
    * & \text{if $\tau(v)=\integer$}\\
    \{(\mu^\flat\phi^\flat(v)).\pi\} & \text{if $\tau(v)\in\mathcal{K}$ and
      $\phi^\flat(v)\in\Loc$}\\
    \emptyset & \text{otherwise}
  \end{cases}\\
  \text{(Definition~\ref{def:er_worst})}&=\begin{cases}
    * & \text{if $\tau(v)=\integer$}\\
    \{\mu^\flat(l(\pi')).\pi\} & \text{if $\tau(v)\in\mathcal{K}$,
      $\phi^\flat(v)\in\Loc$, $\pi'\in\phi(v)$}\\
    \emptyset & \text{otherwise}
  \end{cases}\\
  &=\begin{cases}
    * & \text{if $\tau(v)=\integer$}\\
    \{\pi'\} & \text{if $\tau(v)\in\mathcal{K}$,
      $\phi^\flat(v)\in\Loc$, $\pi'\in\phi(v)$}\\
    \emptyset & \text{otherwise}
  \end{cases}\\
  &\subseteq\phi(v)~.
\end{align*}
}
We prove now some properties of the frames and memories of Definition
\ref{def:er_worst}.
\begin{lemma}\label{lem:small_er}
Let $\phi\in\Frame_\tau^\er$, $\mu\in\Memory^\er$,
$\phi^\flat\in\overline{\phi}_\tau$ and $\mu^\flat\in\overline{\mu}$.
Then
\vspace*{-1.5ex}
\begin{romanenumerate}
\item $\phi^\flat\sep\mu^\flat:\tau$;
\item $\phi^\flat\sep\mu^\flat\in\Sigma_\tau$ if and only if
  $\mathtt{this}\not\in\dom(\tau)$ or $\phi(\mathtt{this})\not=\emptyset$;
\item If $\phi^\flat\sep\mu^\flat\in\Sigma_\tau$ then
  $\alpha_\tau(\phi^\flat\sep\mu^\flat)\subseteq\phi\sep\mu$.
\end{romanenumerate}
\end{lemma}
\myproofbis{\mbox{}
\begin{romanenumerate}
\vspace*{-1ex}
\item Condition 1 of Definition~\ref{def:proptotau} holds since
$\codom(\phi^\flat)\cap\Loc\subseteq\codom(l)=\dom
(\mu^\flat)$. Moreover, if $v\in\dom(\phi^\flat)$ and
$\phi^\flat(v)\in\Loc$ then $\phi^\flat(v)\in l\phi(v)$. Thus there exists
$\pi\in\phi(v)$ with $(\mu^\flat\phi^\flat(v)).\pi=
\pi$ and such that
$k((\mu^\flat\phi^\flat(v)).\pi)=k(\pi)\le
\tau(v)$. Condition 2 holds since
if $o\in\codom(\mu^\flat)$ then $o.\phi=\phi_\pi^\flat$ for some
$\pi\in\Pi$. Since $\phi_\pi^\flat\in\overline{\mu}_{F(k(\pi))}$, reasoning
as above we have that $\phi_\pi^\flat$ is weakly $F(k(\pi))$-correct
\wrt $\mu^\flat$. Then $\phi^\flat\sep\mu^\flat:\tau$.
\item By point i, we know that $\phi^\flat\sep\mu^\flat:\tau$. From
Definition~\ref{def:concrete_states}, we have
$\phi^\flat\sep\mu^\flat\in\Sigma_\tau$ if
and only if $\mathtt{this}\not\in\dom(\tau)$ or
$\phi^\flat(\mathtt{this})\not=\nil$.
By Definition~\ref{def:er_worst}, the latter case holds if and only if
$\phi(\mathtt{this})\not=\emptyset$.
\item By Definition~\ref{def:er_abstraction} we have
\begin{align*}
  \alpha_\tau(\phi^\flat\sep\mu^\flat)&=\varepsilon_\tau(\phi^\flat\sep
    \mu^\flat)\sep\varepsilon_{\overline{\tau}}(\{\overline{o.\phi}
    \sep\mu^\flat
    \mid o\in O_\tau(\phi^\flat\sep\mu^\flat)\})\\
  \text{(Lemma~\ref{lem:pre_small_er})}&\subseteq\phi
    \sep\varepsilon_{\overline{\tau}}(\{\overline{o.\phi}\sep\mu^\flat
    \mid o\in O_\tau(\phi^\flat\sep\mu^\flat)\})~.
\end{align*}
By Definition~\ref{def:er_worst}, for every
$o\in O_\tau(\phi^\flat\sep\mu^\flat)$
we have $o.\phi\in\overline{\mu}_{F(k(\pi))}$
and hence $\overline{o.\phi}\subseteq\phi'$ with $\phi'\in\overline{\mu}
_{\overline{\tau}}$. Then we
have $\varepsilon_{\overline{\tau}}(\overline{o.\phi}
\sep\mu^\flat)\subseteq\varepsilon_{\overline{\tau}}(\phi'\sep\mu^\flat)$,
which by Lemma~\ref{lem:pre_small_er} is contained in $\mu$.
\end{romanenumerate}}
Lemma~\ref{lem:rhoo} states that, given an abstract state
$s$, if a creation point $\pi$ belongs to
$\rho^i(s)$ then there is a concrete state $\sigma$ from those
in Definition~\ref{def:er_worst} and an object in
$O^i(\sigma)$ created in $\pi$, and vice versa. In other words,
$\rho^i(s)$ collects all and only the creation points of the objects
which can ever be reached
in a concrete state approximated by $s$.
\begin{lemma}\label{lem:rhoo}
Let $\phi\in\Frame^\er_\tau$ be such that
if $\mathtt{this}\in\dom(\tau)$ then
$\phi(\mathtt{this})\not=\emptyset$, $\mu\in\Memory^\er$ and
$i\in\nat$. Then $\pi\in\rho_\tau^i(\phi\sep\mu)$ if and only if
there exist $\phi^\flat\in\overline{\phi}_\tau$ and $\mu^\flat\in
\overline{\mu}$ such that $\pi=o.\pi$ for a suitable
$o\in O_\tau^i(\phi^\flat\sep\mu^\flat)$.
\end{lemma}
\myproofbis{
We proceed by induction on $i$. If $i=0$ the result holds since
$\rho_\tau^0(\phi\sep\mu)=\emptyset$ and for every
$\phi^\flat\in\overline{\phi}_\tau$ and $\mu\in\overline{\mu}$ we
have $O_\tau^0(\phi^\flat\sep\mu^\flat)=\emptyset$.
Assume that it holds for a given $i\in\nat$. We have
$\pi\in\rho_\tau^{i+1}(\phi\sep\mu)$ if and only if
$\pi\in\phi(v)$ with $v\in\dom(\tau)$ (and hence $\tau(v)\in\mathcal{K}$) or
$\pi\in\rho^i_{F(k(\pi))}(\phi|_{\dom(F(k(\pi')))}\sep\mu)$ with
$v\in\dom(\tau)$ and $\pi'\in\phi(v)$ (and hence $\tau(v)\in\mathcal{K}$).
The first case holds if and only if $o.\pi=\pi$ with
$o=\mu^\flat\phi^\flat(v)$, $v\in\dom(\tau)$ and $\phi^\flat(v)\in\Loc$
for suitable $\phi^\flat\in\overline{\phi}_\tau$ and
$\mu^\flat\in\overline{\mu}$ (Definition~\ref{def:er_worst}).
By inductive hypothesis, the second case holds if and only if there exist
$\phi^\flat_1\in\overline{\phi}_{F(k(\pi'))}$ and
$\mu^\flat\in\overline{\mu}$ such that
$\pi=o.\pi$ for a suitable $o\in O_{F(k(\pi'))}^i(\phi^\flat_1\sep\mu^\flat)$,
if and only if (Definition~\ref{def:er_worst}) there exist
$\phi^\flat\in\overline{\phi}_\tau$ and
$\mu^\flat\in\overline{\mu}$ such that
$\pi=o.\pi$, $v\in\dom(\tau)$, $\phi^\flat(v)\in\Loc$,
$o'=\mu^\flat\phi^\flat(v)$
and $o\in O_{F(k(o'.\pi))}^i(o'.\phi\sep\mu^\flat)$.
Together, the first or the second case hold if and only if
there exist $\phi^\flat\in\overline{\phi}_\tau$ and
$\mu\in\overline{\mu}$ such that $o\in O_\tau^{i+1}(\phi^\flat\sep
\mu^\flat)$ and $o.\pi=\pi$ (Definition~\ref{def:reachability}).}

Lemma~\ref{lem:oo} says that the concrete states constructed through the frames
and memories of Definition~\ref{def:er_worst} represent a worst-case \wrt the
set of creation points of the objects reachable in every concrete state.
\begin{lemma}\label{lem:oo}
Let $\phi\sep\mu\in\Sigma_\tau$, $i\in\nat$
and $\phi^\#\sep\mu^\#=\alpha_\tau^\er(\phi\sep\mu)$.
If $o\in O_\tau^i(\phi\sep\mu)$ then there exist
$\phi^\flat\in\overline{\phi^\#}_\tau$ and $\mu^\flat\in\overline{\mu^\#}$ such
that $o'\in O_\tau^i(\phi^\flat\sep\mu^\flat)$ and $o'.\pi=o.\pi$.
\end{lemma}
\myproofbis{
We proceed by induction on $i$. We have $O_\tau^0(\phi\sep\mu)
=\emptyset$ and the result holds for $i=0$. Assume that it holds for a given
$i\in\nat$. Let $o\in O_\tau^{i+1}(\phi\sep\mu)$. We have
$o=\mu\phi(v)$ with $v\in\dom(\tau)$ and $\phi(v)\in\Loc$ or
$o\in O_{F(k(o'.\pi))}^i(o'.\phi\sep\mu)$ with
$v\in\dom(\tau)$, $\phi(v)\in\Loc$ and $o'=\mu\phi(v)$. In the first case,
we have $o.\pi\in\phi^\#(v)$ and there exist $\phi^\flat\in\overline{\phi^\#}
_\tau$ and $\mu^\flat\in\overline{\mu^\#}$ such that
$\mu^\flat\phi^\flat(v).\pi=\pi$ and the thesis follows by letting
$o'=\mu^\flat\phi^\flat(v)$.
In the second case, by inductive hypothesis we know that
there exist $\phi^\flat_1\in\overline{\phi^\#}_{F(k(o'.\pi))}$ and
$\mu^\flat\in\overline{\mu^\#}$ such that
$o''\in O_{F(k(o'.\pi))}^i(\phi^\flat_1\sep\mu^\flat)$,
$o''.\pi=o.\pi$, $v\in\dom(\tau)$, $\phi(v)\in\Loc$ and
$o'=\mu\phi(v)$ if and only if (Definitions~\ref{def:er_worst}
and~\ref{def:reachability})
there exist $\phi^\flat\in\overline{\phi^\#}_\tau$ and
$\mu^\flat\in\overline{\mu^\#}$ such that
$o''\in O_\tau^{i+1}(\phi^\flat\sep\mu^\flat)$
and $o''.\pi=o.\pi$.}

Lemma~\ref{lem:er_worst} gives an explicit definition of the abstraction of
the set of states constructed from the frames and memories of
Definition~\ref{def:er_worst}.
\begin{lemma}\label{lem:er_worst}
Let $\phi\in\Frame_\tau^\er$ and $\mu\in\Memory^\er$. Then
\[
  \alpha_\tau^\er(\{\phi^\flat\sep\mu^\flat\in\Sigma_\tau\mid
    \phi^\flat\in\overline{\phi}_\tau\text{ and }
    \mu^\flat\in\overline{\mu}\})=\xi_\tau(\phi\sep\mu)~.
\]
\end{lemma}
\myproofbis{
Let $A_\tau=\alpha_\tau^\er
(\{\phi^\flat\sep\mu^\flat\in\Sigma_\tau\mid
\phi^\flat\in\overline{\phi}_\tau\text{ and }
\mu^\flat\in\overline{\mu}\})$.
If $\mathtt{this}\in\dom(\tau)$ and $\phi(\mathtt{this})=\emptyset$, then
$A_\tau=\bot$ because of Lemma~\ref{lem:small_er}.ii. Moreover,
$\xi_\tau(\phi\sep\mu)=\bot$ (Definition
\ref{def:xi}). Otherwise, by Definition~\ref{def:er_worst} we have
\begin{align}
  A_\tau&=\epsilon_\tau\left(\left\{
    \phi^\flat\sep\mu^\flat\left|\begin{array}{l}
      \phi^\flat\in\overline{\phi}_\tau\\\mu^\flat\in\overline{\mu}
    \end{array}\right.\right\}\right)\sep
    \epsilon_{\overline{\tau}}\left(\left\{
    \overline{o.\phi}\sep\mu^\flat\left|\begin{array}{l}
      \phi^\flat\in\overline{\phi}_\tau,\ \mu^\flat\in\overline{\mu},\\
      o\in O_\tau(\phi^\flat\sep\mu^\flat)
    \end{array}\right.\right\}\right)\notag\\
  &=\phi\sep\epsilon_{\overline{\tau}}(\{
    \overline{\phi'}\sep\mu^\flat\mid
      \phi'\in\overline{\mu}_{F(k(o.\pi))},\ \phi^\flat\in\overline{\phi}_\tau
        ,\ \mu^\flat\in\overline{\mu}
        ,\ o\in O_\tau(\phi^\flat\sep\mu^\flat)
    \})\notag\\
  &=\phi\sep\epsilon_{\overline{\tau}}(\{
    \overline{\phi'}\sep\mu'\mid
      \phi'\!\in\overline{\mu}_{F(k(o.\pi))}
        ,\ \phi^\flat\!\in\overline{\phi}_\tau
        ,\ \mu',\mu^\flat\in\overline{\mu}
        ,\ o\in O_\tau(\phi^\flat\sep\mu^\flat)
    \})\elabel{eq:er_worst}
\end{align}
since $\epsilon_{\overline{\tau}}$ does not depend on the frames of the objects
in memory (Definition~\ref{def:extractor}).
By Lemma~\ref{lem:rhoo}, \eref{eq:er_worst} is equal to
\begin{align*}
  &\ \phi\sep\epsilon_{\overline{\tau}}(\{
    \overline{\phi'}\sep\mu'\mid
      \phi'\in\overline{\mu}_{F(k(\pi))},\ \mu'\in\overline{\mu}
        ,\ \pi\in\rho_\tau(\phi\sep\mu)
    \})\rangle\\
  =&\ \phi\sep\cup\{\mu|_{\dom(F(k(\pi)))}\mid\pi\in
    \rho_\tau(\phi\sep\mu)\}
    \cup\init(\overline{\tau})\\
  =&\ \xi_\tau(\phi\sep\mu)~.
\end{align*}}

We now prove Proposition~\ref{prop:xi_fixpoints}.
To do this, we will use the set of states
constructed from the frames and memories in Definition~\ref{def:er_worst}
to show that $\alpha^\er$ is onto.

\mbox{}\\
\summary
{Proposition \ref{prop:xi_fixpoints}.}
{Let $\xi_\tau$ be the abstract garbage collector of
Definition~\ref{def:xi}. Then $\fp(\xi_\tau)=\codom(\alpha_\tau^\er)$.}
\myproof{Proposition}{prop:xi_fixpoints}{
Let $X\subseteq\Sigma_\tau$. By Proposition~\ref{prop:xi_lco},
Lemmas~\ref{lem:oo} and~\ref{lem:er_worst} and
Definition~\ref{def:er_abstraction}, we have
\begin{align*}
  \alpha_\tau^\er(X)&=\cup\{\alpha_\tau^\er(\sigma)\mid\sigma\in X\}\\
  &\subseteq\cup\ \alpha_\tau^\er\left(\left\{\phi^\flat\sep\mu^\flat
      \in\Sigma_\tau\left|\begin{array}{l}
      \phi^\flat\in\overline{\alpha_\tau^\er(\sigma).\phi}_\tau\\
      \mu^\flat\in\overline{\alpha_\tau^\er(\sigma).\mu}\\
      \sigma\in X
    \end{array}\right.\right\}\right)\\
  &=\cup\{\xi_\tau\alpha_\tau^\er(\sigma)\mid\sigma\in X\}
    \subseteq\xi_\tau\alpha_\tau^\er(X)~.
\end{align*}
The converse inclusion holds since
$\xi_\tau$ is reductive (Proposition~\ref{prop:xi_lco}) and, hence
$\alpha_\tau^\er(X)\in\fp(\xi_\tau)$.
Conversely, let $s\in\fp(\xi_\tau)$ and
$X=\{\phi^\flat\sep\mu^\flat\in\Sigma_\tau\mid
\phi^\flat\in\overline{\phi}_\tau,\ \mu^\flat\in\overline{\mu}\}$.
By Lemma~\ref{lem:er_worst} and
since $s\in\fp(\xi_\tau)$, we have $\alpha_\tau^\er(X)=\xi_\tau(s)=s$.}

The proof of Proposition~\ref{prop:operations_er} requires some preliminary
results.

Corollary~\ref{cor:alphaxi} states that if we know that the approximation
of a set of concrete states $S$ is some $\phi\sep\mu$, then we can conclude
that a better approximation of $S$ is $\xi(\phi\sep\mu)$. In other words,
garbage is not used in the approximation.
\begin{corollary}\label{cor:alphaxi}
Let $S\subseteq\Sigma_\tau$, $\phi\in\Frame_\tau^\er$ and
$\mu\in\Memory^\er$. Then
$\alpha_\tau(S)\subseteq\xi_\tau(\phi\sep\mu)$ if and only if
$\alpha_\tau(S)\subseteq\phi\sep\mu$.
\end{corollary}
\myproofbis{
Assume that $\alpha_\tau(S)\subseteq\xi_\tau(\phi\sep\mu)$. By reductivity
(Proposition~\ref{prop:xi_lco}) we have $\alpha_\tau(S)\subseteq\phi\sep\mu$.
Conversely, assume that $\alpha_\tau(S)\subseteq\phi\sep\mu$. By 
Proposition~\ref{prop:xi_fixpoints} and monotonicity
(Proposition~\ref{prop:xi_lco}) we have
$\alpha_\tau(S)=\xi_\tau\alpha_\tau(S)\subseteq\xi_\tau(\phi\sep\mu)$.}

The following lemma will be used in the proof of
Proposition~\ref{prop:operations_er}. It states that the approximation of
a variable depends from the concrete value of that variable only, and that
the approximation of a memory is the same if the locations in
the frame do not change (although they may be bound to different variables).
\begin{lemma}\label{lem:er_codom}
Let $\phi'\sep\mu\in\Sigma_{\tau'}$ and
$\phi''\sep\mu\in\Sigma_{\tau''}$. Then
\begin{romanenumerate}
\item if $\phi'(v)= \phi''(v)$
for each $v\in\dom(\tau')\cap\dom(\tau'')$, then we have
$(\alpha_{\tau'}(\phi'\sep\mu)).\phi(v)=
(\alpha_{\tau''}(\phi''\sep\mu)).\phi(v)$;

\item if $\codom(\phi')\cap\Loc=\codom(\phi'')\cap\Loc$, then we have
  $(\alpha_{\tau'}(\phi'\sep\mu)).\mu=(\alpha_{\tau''}(\phi''\sep\mu)).\mu$.
\end{romanenumerate}
\end{lemma}
\myproofbis{
From Definition~\ref{def:er_abstraction}.}

Lemma~\ref{lem:alpha_restrict_er}
says that if we consider all the concrete states
approximated by some $\phi^\#\sep\mu^\#$ and we restrict their frames,
the resulting set of states is approximated by $\xi(\phi^\#\sep\mu^\#)$.
 In other words,
the operation $\xi$ garbage collects all objects that,
because of the restriction, are no longer reachable.
\begin{lemma}\label{lem:alpha_restrict_er}
Let $\mathit{vs}\subseteq\dom(\tau)$ and
$\phi^\#\sep\mu^\#\in\er_\tau$. Then
\[
  \alpha_{\tau|_{-\mathit{vs}}}\left(\left\{\phi|_{-\mathit{vs}}\sep
    \mu\left|\begin{array}{l}
    \phi\sep\mu\in\Sigma_\tau\\
    \alpha_\tau(\phi\sep\mu)\subseteq\phi^\#\sep\mu^\#
    \end{array}\right.\right\}\right)=\xi_{\tau|_
    {-\mathit{vs}}}(\phi^\#|_{-\mathit{vs}}\sep\mu^\#)~.
\]
\end{lemma}
\myproofbis{
We have
\begin{align}
  &\quad\alpha_{\tau|_{-\mathit{vs}}}(\{\phi|_{-\mathit{vs}}\sep
    \mu\mid\phi\sep\mu\in\Sigma_\tau\text{ and }
    \alpha_\tau(\phi\sep\mu)\subseteq\phi^\#\sep\mu^\#\})\notag\\
  &=\alpha_{\tau|_{-\mathit{vs}}}(\{\phi|_{-\mathit{vs}}\sep
    \mu\in\Sigma_{\tau|_{-\mathit{vs}}}\mid
    \phi\sep\mu\in\Sigma_\tau\text{ and }
    \alpha_\tau(\phi\sep\mu)\subseteq\phi^\#\sep\mu^\#\})~,
    \elabel{eq:alpha_restrict1_er}
\end{align}
since if $\phi\sep\mu\in\Sigma_\tau$ then
$\phi|_{-\mathit{vs}}\sep\mu\in\Sigma_{\tau|_{-\mathit{vs}}}$.
We have that, if $\alpha_\tau(\phi\sep\mu)\subseteq\phi^\#\sep\mu^\#$, then
$\alpha_{\tau|_{-\mathit{vs}}}(\phi|_{-\mathit{vs}}\sep\mu)\subseteq
\phi^\#|_{-\mathit{vs}}\sep\mu^\#$. Hence \eref{eq:alpha_restrict1_er} is
contained in $\phi^\#|_{-\mathit{vs}}\sep\mu^\#$.
By Corollary~\ref{cor:alphaxi}, the set~\eref{eq:alpha_restrict1_er}
is also contained in the set
$\xi_{\tau|_{-\mathit{vs}}}(\phi^\#|_{-\mathit{vs}}
\sep\mu^\#)$.
But also the converse inclusion holds, since in
\eref{eq:alpha_restrict1_er} we can restrict the choice of $\phi\sep\mu
\in\Sigma_\tau$, so that \eref{eq:alpha_restrict1_er} contains
\begin{equation}\elabel{eq:alpha_restrict2_er}
  \alpha_{\tau|_{-\mathit{vs}}}\left(\left\{\phi|_{-\mathit{vs}}\sep
    \mu\in\Sigma_{\tau|_{-\mathit{vs}}}\left|
    \begin{array}{l}
      \phi\sep\mu\in\Sigma_\tau,\ \alpha_\tau(\phi\sep\mu)\subseteq\phi^\#\sep
        \mu^\#\\
      \phi\in\overline{\phi^\#}_\tau,\ \mu\in\overline{\mu^\#}
    \end{array}\right.\right\}\right)~.
\end{equation}
By points ii and iii of Lemma~\ref{lem:compatibility},
\eref{eq:alpha_restrict2_er} is equal to
\begin{align*}
  &\quad\alpha_{\tau|_{-\mathit{vs}}}(\{\phi|_{-\mathit{vs}}\sep
    \mu\in\Sigma_{\tau|_{-\mathit{vs}}}\mid
    \phi\in\overline{\phi^\#}_\tau,\ \mu\in\overline{\mu^\#}\})\\
  \text{(Definition~\ref{def:er_worst})}&=
    \alpha_{\tau|_{-\mathit{vs}}}\left(\left\{\phi\sep\mu\in
    \Sigma_{\tau|_{-\mathit{vs}}}\left|\begin{array}{l}
      \phi\in\overline{(\phi^\#|_{-\mathit{vs}})}_{\tau|_{-\mathit{vs}}}\\
      \mu\in\overline{\mu^\#}
    \end{array}\right.\right\}\right)\\
  \text{(Lemma~\ref{lem:er_worst})}&=\xi_{\tau|_{-\mathit{vs}}}
    (\phi^\#|_{-\mathit{vs}}\sep\mu^\#)~.
\end{align*}}

We are now ready to prove the correctness and optimality of the
abstract operations in Figure~\ref{fig:operations_er}.

\mbox{}\\
\summary
{Proposition \ref{prop:operations_er}.}
{The operations in Figure~\ref{fig:operations_er} are the optimal counterparts
induced by $\alpha^\er$ of the operations in Figure~\ref{fig:concrete_states}
and of $\cup$.}
\myproof{proposition}{prop:operations_er}{
The strictness of the abstract operations (except $\cup$) follows by
reasoning as for the proof of strictness in
Proposition~\ref{prop:operations_e}. Note that
$\gamma_\tau(\bot)=\emptyset$ for all $\tau \in \Typing$ since,
by Definition~\ref{def:er_abstraction},
\[
   \gamma_\tau(\bot)=\{\sigma\in\Sigma_\tau\mid\alpha_\tau(\sigma)
\subseteq\bot\}=\{\sigma\in\Sigma_\tau\mid\alpha_\tau(\sigma)=\bot\}
=\emptyset.
\]
Hence $\mathsf{return}$ is also strict on both arguments.

We will use the corresponding versions of the properties P2 and P3
already used in the proof of Proposition~\ref{prop:operations_e}. They are
\begin{itemize}
\item[P2] If $\phi\sep\mu\in\mathcal{ER}_\tau$
then there exists $\sigma\in\Sigma_\tau$ such that
$\alpha_\tau(\sigma)\subseteq\phi\sep\mu$.
\item[P3] $\alpha_\tau\gamma_\tau$ is the identity map.
\end{itemize}
%

P2 holds since $\phi(\mathtt{this})\neq \emptyset$ so that
there exists $\pi\in\phi(\mathtt{this})$ and hence, letting
$\sigma=[\mathtt{this}\mapsto l]\sep [l\mapsto\pi\sep\init(F(k(\pi)))]$
for some $l\in\Loc$, we have $\sigma\in\Sigma_\tau$. Moreover,
\(
   \alpha_\tau(\sigma)=\phi^\bot[\mathtt{this}\mapsto\{\pi\}]\sep\mu^\bot
\subseteq\phi\sep\mu
\),
where $\phi^\bot$ and $\mu^\bot$ are the least elements of
$\Frame_\tau^\er$ and $\Memory^\er$, respectively.
By Proposition~\ref{prop:er_insertion}, $\alpha_\tau$ is a Galois insertion
and hence, P3 holds.

Most cases of the proof are similar to the corresponding cases in the proof
of Proposition~\ref{prop:operations_e}, provided we use Lemma
\ref{lem:er_codom} instead of Lemma~\ref{lem:codom},
Lemma~\ref{lem:alpha_restrict_er}
instead of Lemma~\ref{lem:alpha_restrict}, Definition~\ref{def:xi} instead
of Definition~\ref{def:delta}, and we modify
the syntax of the abstract elements. As an example, consider
\mbox{}\\
\proofoperation{get\_int,\ get\_null,\ get\_var}
\begin{align*}
  &\quad\alpha_{\tau[\rs\mapsto\integer]}
    (\mathsf{get\_int}^i_\tau(\gamma_\tau(\phi^\#\sep\mu^\#)))\\
  &=\alpha_{\tau[\rs\mapsto\integer]}(\{\phi'[\rs\mapsto i]\sep\mu'\mid
    \phi'\sep\mu'\in\gamma_\tau(\phi^\#\sep\mu^\#)\})\\
  \text{($*$)}&=\alpha_\tau(\{\phi'\sep\mu'\mid
    \phi'\sep\mu'\in\gamma_\tau(\phi^\#\sep\mu^\#)\}).\phi[\rs\mapsto *]\sep\\
  &\quad\sep\alpha_\tau(\{\phi'\sep\mu'\mid
    \phi'\sep\mu'\in\gamma_\tau(\phi^\#\sep\mu^\#)\}).\mu\\
  \text{(P3)}&=\phi^\#[\rs\mapsto *]\sep\mu^\#~.
\end{align*}
where point $*$ follows by Lemma~\ref{lem:er_codom} since
$\rs\not\in\dom(\tau)$ and $\tau[\rs\mapsto\integer](\rs)=\integer$.
The proof is similar for $\mathsf{get\_null}$ and $\mathsf{get\_var}$.

\mbox{}\\
Therefore, we only show the
cases which differ significantly from the corresponding case in
Proposition~\ref{prop:operations_e}.
\mbox{}\\
\proofoperation{is\_null}
\mbox{}\\
Let $A=\alpha_{\tau[\rs\mapsto\integer]}
(\mathsf{is\_null}_\tau(\gamma_\tau(\phi^\#\sep\mu^\#)))$. We have
\begin{align*}
  A&=\alpha_{\tau[\rs\mapsto\integer]}(\mathsf{is\_null}_\tau
    (\{\sigma\in\Sigma_\tau\mid\alpha_\tau(\sigma)\subseteq
    \phi^\#\sep\mu^\#\}))\\
  &=\alpha_{\tau[\rs\mapsto\integer]}(\{\phi[\rs\mapsto 1]\sep\mu\mid
    \phi\sep\mu\in\Sigma_\tau\text{ and }
    \alpha_\tau(\phi\sep\mu)\subseteq\phi^\#\sep\mu^\#\})~.
\end{align*}
By Lemma~\ref{lem:er_codom}.i we have
\begin{align*}
  A.\phi&=\phi^\#[\rs\mapsto *]\\
  \text{(Definition~\ref{def:xi})}&=\xi_{\tau[\rs\mapsto\integer]}
    (\phi^\#[\rs\mapsto *]\sep\mu^\#)~.
\end{align*}
Moreover, by Lemma~\ref{lem:er_codom}.ii we have
\begin{align*}
  A.\mu&=\alpha_{\tau|_{-\rs}}\left(\left\{\phi|_{-\rs}\sep\mu\left|
  \begin{array}{l}
    \phi\sep\mu\in\Sigma_\tau\\
    \alpha_\tau(\phi\sep\mu)\subseteq\phi^\#\sep\mu^\#
  \end{array}\right.\right\}\right).\mu\\
  \text{(Lemma~\ref{lem:alpha_restrict_er})}&=\xi_{\tau|_{-\rs}}
    (\phi^\#|_{-\rs}\sep\mu^\#).\mu\\
  \text{(Definition~\ref{def:xi})}&=\xi_{\tau[\rs\mapsto\integer]}
    (\phi^\#[\rs\mapsto *]\sep\mu^\#).\mu~.
\end{align*}
\mbox{}\\
\proofoperation{put\_var}
\mbox{}\\
Let $A=\alpha_{\tau|_{-\rs}}(\mathsf{put\_var}_\tau(\gamma_\tau(\phi^\#\sep
\mu^\#)))$. We have
\begin{align*}
  A&=\alpha_{\tau|_{-\rs}}(\mathsf{put\_var}_\tau
    (\{\sigma\in\Sigma_\tau\mid\alpha_\tau(\sigma)\subseteq\phi^\#\sep
    \mu^\#\}))\\
  &=\alpha_{\tau|_{-\rs}}\left(\left\{\phi[v\mapsto\phi(\rs)]|_{-\rs}\sep\mu
    \left|\begin{array}{l}
      \phi\sep\mu\in\Sigma_\tau\\
      \alpha_\tau(\phi\sep\mu)\subseteq\phi^\#\sep\mu^\#
    \end{array}\right.\right\}\right)~.
\end{align*}
By Lemma~\ref{lem:er_codom}.i we have
\begin{align*}
  A.\phi&=\phi^\#[v\mapsto\phi^\#(\rs)]|_{-\rs}\\
  \text{(Definition~\ref{def:xi})}&=\xi_{\tau|_{-\rs}}(\phi^\#[v\mapsto
    \phi^\#(\rs)]|_{-\rs}\sep\mu^\#).\phi~.
\end{align*}
Moreover, since
$\codom(\phi[v\mapsto\phi(\rs)]|_{-\rs})=\codom(\phi|_{-v})$, by Lemma
\ref{lem:er_codom}.ii we have
\begin{align*}
  A.\mu&=\alpha_{\tau|_{-v}}\left(\left\{\phi|_{-v}\sep\mu
    \left|\begin{array}{l}
      \phi\sep\mu\in\Sigma_\tau\\
      \alpha_\tau(\phi\sep\mu)\subseteq\phi^\#\sep\mu^\#
    \end{array}\right.\right\}\right).\mu\\
  \text{(Lemma~\ref{lem:alpha_restrict_er})}&=\xi_{\tau|_{-v}}
    (\phi^\#|_{-v}\sep\mu^\#).\mu\\
  \text{(Definition~\ref{def:xi})}&=\xi_{\tau|_{-\rs}}(\phi^\#[v\mapsto
    \phi^\#(\rs)]|_{-\rs}\sep\mu^\#).\mu~.
\end{align*}
\proofoperation{call}

\mbox{}\\
Let $p=P(\nu)|_{-\Out}$.
Recall that $\dom(p)=\{\iota_1,\ldots,\iota_n,\mathtt{this}\}$.
Let $\tau_{'}=\tau[v_1\mapsto\iota_1,\ldots,v_n\mapsto\iota_n,\rs\mapsto
\mathtt{this}]$ and $\phi^\#_{'}=\phi^\#[v_1\mapsto\iota_1,\ldots,
v_n\mapsto\iota_n,\rs\mapsto\mathtt{this}]$. We have
\begin{align*}
  &\quad
    \alpha_p(\mathsf{call}_\tau^{\nu,v_1,\ldots,v_n}
    (\gamma_\tau(\phi^\#\sep\mu^\#)))\\
  &=\alpha_p(\mathsf{call}_\tau^{\nu,v_1,\ldots,v_n}
    (\{\sigma\in\Sigma_\tau\mid\alpha_\tau(\sigma)\subseteq
    \phi^\#\sep\mu^\#\}))\\
  &=\alpha_p\left(\left\{\left.\left[\begin{array}{c}
      \iota_1\mapsto\phi(v_1),\\
      \vdots\\
      \iota_n\mapsto\phi(v_n),\\
      \mathtt{this}\mapsto\phi(\rs)\end{array}\right]\sep\mu
    \right|\begin{array}{l}
      \phi\sep\mu\in\Sigma_\tau\text{ and}\\
      \alpha_\tau(\phi\sep\mu)\subseteq\phi^\#\sep\mu^\#
    \end{array}\right\}\right)\\
  \text{(Lemma~\ref{lem:er_codom})}
    &=\alpha_p(\{\phi|_p\sep\mu\mid\phi\sep\mu\in\Sigma_{\tau_{'}}\text{ and }
    \alpha_{\tau_{'}}(\phi\sep\mu)\subseteq\phi^\#_{'}\sep\mu^\#\})\\
  \text{(Lemma~\ref{lem:alpha_restrict_er})}&=\xi_p(\phi^\#_{'}|_p\sep\mu^\#)\\
  &=\xi_p\left(\left[\begin{array}{l}
    \iota_1\mapsto\phi^\#(v_1),\\
    \vdots\\
    \iota_n\mapsto\phi^\#(v_n),\\
    \mathtt{this}\mapsto\phi^\#(\rs)
    \end{array}\right]\sep\mu^\#\right)~.
\end{align*}
\mbox{}\\
\proofoperation{new}
\mbox{}\\
Let $\kappa=k(\pi)$ and $A=\alpha_{\tau[\rs\mapsto\kappa]}
(\mathsf{new}^\pi_\tau(\gamma_\tau(\phi^\#\sep\mu^\#)))$.
Since $\rs\not\in\dom(\tau)$ we have
\begin{align*}
  A&=\alpha_{\tau[\rs\mapsto\kappa]}(\mathsf{new}^\pi_\tau
    (\{\sigma\in\Sigma_\tau\mid\alpha_\tau(\sigma)\subseteq\phi^\#\sep
    \mu^\#\}))\\
  &=\alpha_{\tau[\rs\mapsto\kappa]}
    \left(\left\{\begin{array}{c}
      \phi[\rs\mapsto l]\sep\\
      \sep\mu[l\mapsto\pi\sep\init(F(\kappa))
    ]\end{array}\left|\begin{array}{l}
      \phi\sep\mu\in\Sigma_\tau\\
      \alpha_\tau(\phi\sep\mu)\subseteq\phi^\#\sep\mu^\#\\
      l\in\Loc\setminus\dom(\mu)
    \end{array}\right.\right\}\right)~.
\end{align*}
By Lemma~\ref{lem:er_codom}.i we have
\begin{align*}
  A.\phi&=\alpha_\tau(\{\phi\sep\mu\mid\phi\sep\mu\in\Sigma_\tau
    \text{ and }\alpha_\tau(\phi\sep\mu)\subseteq\phi^\#\sep\mu^\#\}).\phi
    [\rs\mapsto\{\pi\}]\\
  &=\alpha_\tau\gamma_\tau(\phi^\#\sep\mu^\#).\phi[\rs\mapsto\{\pi\}]\\
  \text{(P3)}&=\phi^\#[\rs\mapsto\{\pi\}]~.
\end{align*}
The newly created object $o=\pi\sep\init(F(\kappa))$ has its fields
bound to $\nil$:
$o.\phi(f)=\init(F(\kappa))(f)\in\{0,\nil\}$ for every $f\in\dom(o.\phi)$.
Hence it does not contribute to the memory component $A.\mu$ and by
Lemma~\ref{lem:er_codom}.ii we have
\begin{align*}
  A.\mu&=\alpha_\tau(\{\phi\sep\mu\mid\phi\sep\mu\in\Sigma_\tau
    \text{ and }\alpha_\tau(\phi\sep\mu)\subseteq\phi^\#\sep\mu^\#\}).\mu\\
  &=\alpha_\tau\gamma_\tau(\phi^\#\sep\mu^\#).\mu\\
  \text{(P3)}&=\mu^\#~.
\end{align*}
\mbox{}\\
\proofoperation{return}
\mbox{}\\
Let $\tau'=\tau[\rs\mapsto P(\nu)(\Out)]$,
$\tau''=P(\nu)|_{\Out}$ and $L=\codom(\phi_1|_{-\rs})\cap\Loc$.
\begin{align*}
  &\quad\alpha_{\tau'}(\mathsf{return}_\tau^\nu(\gamma_\tau(\phi_1^\#\sep
    \mu_1^\#))(\gamma_{\tau''}(\phi_2^\#\sep\mu_2^\#)))\\
  &=\alpha_{\tau'}(\mathsf{return}_\tau^\nu
    (\{\sigma_1\in\Sigma_\tau\mid\alpha_\tau(\sigma_1)\subseteq\phi_1^\#
    \sep\mu_1^\#\})\\
  &\qquad(\{\sigma_2\in\Sigma_{\tau''}\mid
    \alpha_{\tau''}(\sigma_2)\subseteq\phi_2^\#\sep\mu_2^\#\}))\\
  &=\alpha_{\tau'}\left(\!\left\{
      \phi_1|_{-\rs}[\rs\mapsto\phi_2(\Out)]\sep\mu_2\left|
      \underbrace{\begin{array}{c}
        \phi_1\sep\mu_1\in\Sigma_\tau\\
        \phi_2\sep\mu_2\in\Sigma_{\tau''}\\
        \alpha_\tau(\phi_1\sep\mu_1)\subseteq\phi_1^\#\sep\mu_1^\#\\
        \alpha_{\tau''}(\phi_2\sep\mu_2)\subseteq\phi_2^\#\sep\mu_2^\#\\
        \mu_1=_L\mu_2
    \end{array}}_{Cond}\right.\right\}\right)\\
  \text{($*$)}
    &=\underbrace{\alpha_{\tau|_{-\rs}}(\{\phi_1|_{-\rs}\sep\mu_2\mid Cond\})
    }_A\cup\underbrace{\alpha_{\tau''}(\{\phi_2\sep\mu_2\mid Cond\})
    [\Out\mapsto\rs]}_B
\end{align*}
where point $*$ follows by Definition~\ref{def:er_abstraction}.
Since $\alpha_{\tau''}(\phi_2\sep\mu_2)\subseteq\phi_2^\#\sep\mu_2^\#$, we
have $B\subseteq\phi_2^\#[\Out\mapsto\rs]\sep\mu_2^\#$.
But the converse inclusion holds also, since by Lemma
\ref{lem:small_er}.iii we have
\[
  B\supseteq\alpha_{\tau''}\left(\left\{\phi_2\sep\mu_2\left|\begin{array}{l}
    \phi_1\in\overline{\phi_1^\#}_\tau,\ \mu_1\in\overline{\mu_1^\#}\\
    \phi_2\in\overline{\phi_2^\#}_{\tau''},\ \mu_2\in\overline{\mu_2^\#}
    \end{array}\right.\right\}\right)[\Out\mapsto\rs]
\]
which by Lemma~\ref{lem:er_worst} is equal to $\phi_2^\#[\Out
\mapsto\rs]\sep\mu_2^\#$. Note that the
condition $\mu_1=_L\mu_2$ is satisfied by Definition~\ref{def:er_worst}.
Since $\dom(\tau'')=\{\Out\}$, we conclude that
$B=[\rs\mapsto\phi_2^\#(\Out)]\sep\mu_2^\#$.

With regard to $A$, we have
\begin{equation}\elabel{eq:er_return}
\begin{split}
  A&\supseteq\alpha_{\tau|_{-\rs}}\left\{\phi_1|_{-\rs}\sep\mu_2\left|
    \begin{array}{l}
    Cond,\ \phi_1\in\overline{\phi_1^\#}_\tau,\ \mu_1\in\overline{\mu_1^\#}\\
    \phi_2=\init(\tau''),\ \mu_2\in\overline{\mu^\top}
  \end{array}\right.\right\}\\
  \text{(Lemma~\ref{lem:small_er})}&=\alpha_{\tau|_{-\rs}}
    \{\phi_1|_{-\rs}\sep\mu_2\mid
      \phi_1\in\overline{\phi_1^\#}_\tau,\ \mu_2\in\overline{\mu^\top}\}\\
  \text{(Lemma~\ref{lem:er_worst})}&=\xi_{\tau|_{-\rs}}(\phi_1^\#|_{-\rs}
    \sep\mu^\top)~.
\end{split}
\end{equation}
Moreover, for every $v\in\dom(\tau|_{-\rs})$ such that
$\tau(v)\in\mathcal{K}$, we have
\begin{align*}
  A.\phi(v)&=\{o.\pi\mid
    \phi_1|_{-\rs}(v)\in\Loc,\ o=\mu_2\phi_1|_{-\rs}(v),\ Cond\}\\
  \text{(since $\mu_1=_L\mu_2$)}&=\{o.\pi\mid
    \phi_1|_{-\rs}(v)\in\Loc,\ o=\mu_1\phi_1|_{-\rs}(v),\ Cond\}\\
  &\subseteq\left\{(\mu_1\phi_1(v)).\pi\left|\begin{array}{l}
    \phi_1(v)\in\Loc
      ,\ \phi_1\sep\mu_1\in\Sigma_\tau\\
    \alpha_\tau(\phi_1\sep\mu_1)
      \subseteq\phi_1^\#\sep\mu_1^\#\end{array}\right.\right\}\\
  &=(\alpha_\tau\gamma_\tau(\phi_1^\#\sep\mu_1^\#)).\phi(v)\\
  \text{(P1)}&=\phi_1^\#(v)~.
\end{align*}
We conclude that $A.\phi\subseteq\phi_1^\#|_{-\rs}$. Moreover, we have
$A.\mu\subseteq\mu^\top$. Hence $A\subseteq\phi_1^\#|_{-\rs}\sep\mu^\top$
and, by Corollary~\ref{cor:alphaxi},
$A\subseteq\xi_{\tau|_{-\rs}}(\phi_1^\#|_{-\rs}\sep\mu^\top)$.
Together with \eref{eq:er_return}, this proves that
$A=\xi_{\tau|_{-\rs}}(\phi_1^\#|_{-\rs}\sep\mu^\top)$.
\mbox{}\\
\proofoperation{get\_field}
\mbox{}\\
Let $\tau'=\tau[\rs\mapsto(F\tau(\rs))(f)]$ and
$A=\alpha_{\tau'}(\mathsf{get\_field}_\tau^f(\gamma_\tau(\phi^\#\sep
\mu^\#)))$. We have
\begin{align*}
  A&=\alpha_{\tau'}(\mathsf{get\_field}_\tau^f(\{\phi\sep\mu\in
    \Sigma_\tau\mid\alpha_\tau(\phi\sep\mu)\subseteq\phi^\#\sep\mu^\#\}))\\
  &=\alpha_{\tau'}\left(\left\{\phi[\rs\mapsto
    (\mu\phi(\rs)).\phi(f)]\sep\mu\left|\begin{array}{l}
      \phi\sep\mu\in\Sigma_\tau\\
      \phi(\rs)\not=\nil\\
      \alpha_\tau(\phi\sep\mu)\subseteq\phi^\#\sep\mu^\#
    \end{array}\right.\right\}\right)
\end{align*}
which is $\bot$ when $\phi^\#(\rs)=\emptyset$, since in such a case the
condition $\phi(\rs)\not=\nil$ cannot be satisfied. Assume then that we
have $\phi^\#(\rs)\not=\emptyset$ and let $f'=(\mu\phi(\rs)).\phi(f)$. We
conclude that
\begin{align*}
  A&\supseteq\alpha_{\tau'}\left(\left\{\phi[\rs\mapsto
    f']\sep\mu\left|\begin{array}{l}
      \phi\sep\mu\in\Sigma_\tau\\
      \phi(\rs)\not=\nil\\
      \alpha_\tau(\phi\sep\mu)\subseteq\phi^\#\sep\mu^\#\\
      \phi\in\overline{\phi^\#}_\tau,\ \mu\in\overline{\mu^\#}
    \end{array}\right.\right\}\right)\\
  \text{(Definition~\ref{def:er_worst})}&=\alpha_{\tau'}\left(\left\{
    \phi[\rs\mapsto f']\sep\mu\left|\begin{array}{l}
      \phi\sep\mu\in\Sigma_\tau\\
      \alpha_\tau(\phi\sep\mu)\subseteq\phi^\#\sep\mu^\#\\
      \phi\in\overline{\phi^\#}_\tau,\ \mu\in\overline{\mu^\#}
    \end{array}\right.\right\}\right)\\
  \text{(Lemma~\ref{lem:small_er})}&=\alpha_{\tau'}(\{
    \phi[\rs\mapsto f']\sep\mu\mid
    \phi\in\overline{\phi^\#}_\tau,\ \mu\in\overline{\mu^\#}\})\\
  \text{(Definition~\ref{def:er_worst})}&=\alpha_{\tau'}(\{
    \phi\sep\mu\mid
    \phi\in\overline{\phi^\#[\rs\mapsto\mu(f)]}_{\tau'},\ \mu
    \in\overline{\mu^\#}\})\\
  \text{(Lemma~\ref{lem:er_worst})}&=\phi^\#[\rs\mapsto\mu(f)]\sep\mu^\#~.
\end{align*}
We prove that the converse inclusion also holds.
Let $x=(\mu\phi(\rs)).\phi(f)$. If
$x\in\Loc$, the object $\mu(x)$ is reachable by construction from
$\phi(\rs)$. Hence we have
\begin{align*}
  A.\mu&\subseteq\alpha_\tau(\{\phi\sep\mu\mid
    \phi\sep\mu\in\Sigma_\tau,\ \phi(\rs)\not=\nil,\ \alpha_\tau
    (\phi\sep\mu)\subseteq\phi^\#\sep\mu^\#\}).\mu\\
  &\subseteq\alpha_\tau(\{\phi\sep\mu\mid
    \phi\sep\mu\in\Sigma_\tau,\ \alpha_\tau(\phi\sep\mu)
    \subseteq\phi^\#\sep\mu^\#\}).\mu\\
  &=\alpha_\tau\gamma_\tau(\phi^\#\sep\mu^\#).\mu\\
  \text{(P3)}&=\mu^\#~.
\end{align*}
If $\phi(\rs)\not=\nil$ then $o=\mu\phi(\rs)\in O_\tau(\phi\sep\mu)$ and
$\varepsilon_{\overline{\tau}}(\overline{o.\phi}\sep\mu)\subseteq\mu^\#$
(Definition~\ref{def:er_abstraction}). Hence, if $(\mu\phi(\rs)).\phi(f)
\not=\nil$ then we have that
$((\mu\phi(\rs)).\phi(f)).\pi\in\mu^\#(f)$. By Lemma
\ref{lem:er_codom} we conclude that
\[
  A.\phi\subseteq\phi^\#[\rs\mapsto\mu^\#(f)]~.
\]
\proofoperation{lookup}
\mbox{}\\
Let $A=\alpha_\tau(\mathsf{lookup}^{m,\nu}_\tau(\gamma_\tau
(\phi^\#\sep\mu^\#)))$. We have
\begin{align*}
  A&=\alpha_\tau(\mathsf{lookup}^{m,\nu}_\tau(\{\phi\sep\mu\in
    \Sigma_\tau\mid\alpha_\tau(\phi\sep\mu)\subseteq\phi^\#\sep\mu^\#\}))\\
  &=\alpha_\tau\left(\left\{\phi\sep\mu\in\Sigma_\tau\left|
    \begin{array}{l}
      \alpha_\tau(\phi\sep\mu)\subseteq\phi^\#\sep\mu^\#\\
      \phi(\rs)\not=\nil,\ M((\mu\phi(\rs)).\pi)(m)=\nu
    \end{array}\right.\right\}\right)~.
\end{align*}
We have $A=\bot$ if there is no $\pi\in\phi^\#(\rs)$ such that
$M(\pi)(m)=\nu$, because in such a case the condition
$M((\mu\phi(\rs)).\pi)(m)=\nu$ cannot be satisfied. Otherwise we have
\begin{align*}
  A&\supseteq\alpha_\tau\left(\left\{\phi\sep\mu\in\Sigma_\tau\left|
    \begin{array}{l}
      \alpha_\tau(\phi\sep\mu)\subseteq\phi^\#\sep\mu^\#\\
      \phi(\rs)\not=\nil\\
      M(k((\mu\phi(\rs)).\pi))(m)=\nu\\
      \phi\in\overline{\phi^\#}_\tau
        ,\ \mu\in\overline{\mu^\#}
    \end{array}\right.\right\}\right)\\
  \text{(Definition~\ref{def:xi})}&=\alpha_\tau
    \left(\left\{\phi\sep\mu\in\Sigma_\tau\left|
    \begin{array}{l}
      \alpha_\tau(\phi\sep\mu)\subseteq\phi^\#\sep\mu^\#\\
      M(k((\mu\phi(\rs)).\pi))(m)=\nu\\
      \phi\in\overline{\phi^\#}_\tau
        ,\ \mu\in\overline{\mu^\#}
    \end{array}\right.\right\}\right)\\
  \text{(Lemma~\ref{lem:small_er}.iii)}&=\alpha_\tau
    \left(\left\{\phi\sep\mu\in\Sigma_\tau\left|
    \begin{array}{l}
      M(k((\mu\phi(\rs)).\pi))(m)=\nu\\
      \phi\in\overline{\phi^\#}_\tau
        ,\ \mu\in\overline{\mu^\#}
    \end{array}\right.\right\}\right)\\
  \text{(Definition~\ref{def:er_worst})}&=\alpha_\tau
    \left(\left\{\phi\sep\mu\in\Sigma_\tau\left|
    \begin{array}{l}
      (\mu\phi(\rs)).\pi\in S,\\
      \phi\in\overline{\phi^\#}_\tau
        ,\ \mu\in\overline{\mu^\#}
    \end{array}\right.\right\}\right)
\end{align*}
where $S=\{\pi\in\phi^\#(\rs)\mid M(k(\pi))(m)=\nu\}$. By Definition
\ref{def:er_worst} we have
\begin{align*}
  A&\supseteq\alpha_\tau(\{\phi\sep\mu\in\Sigma_\tau\mid
    \phi\in\overline{\phi^\#[\rs\mapsto S]}_\tau
    ,\ \mu\in\overline{\mu^\#}\})\\
  \text{(Lemma~\ref{lem:er_worst})}&=\xi_\tau(\phi^\#[\rs\mapsto S]\sep
    \mu^\#)~.
\end{align*}
We prove that also the converse inclusion holds. Note that if
$\alpha_\tau(\phi\sep\mu)\subseteq\phi^\#[\rs\mapsto S]\sep\mu^\#$ and
$\phi(\rs)\not=\nil$ then $M(k((\mu\phi(\rs)).\pi))(m)=\nu$. Hence we have
\begin{align*}
  A&\subseteq\alpha_\tau\left(\left\{
    \phi\sep\mu\in\Sigma_\tau\left|\begin{array}{l}
      \alpha_\tau
       (\phi\sep\mu)\subseteq\phi^\#[\rs\mapsto S]\sep\mu^\#,\\
      \phi(\rs)\not=\nil\end{array}\right.\right\}\right)\\
  \text{(Definition~\ref{def:er_abstraction})}&=
    \alpha_\tau(\{\phi\sep\mu\in\Sigma_\tau\mid\alpha_\tau
    (\phi\sep\mu)\subseteq\phi^\#[\rs\mapsto S]\sep\mu^\#\})\\
  &=\alpha_\tau\gamma_\tau(\phi^\#[\rs\mapsto S]\sep\mu^\#)\\
  \text{(P3)}&=\phi^\#[\rs\mapsto S]\sep\mu^\#~.
\end{align*}
\proofoperation{put\_field}
\mbox{}\\
Let $\tau''=\tau|_{-\rs}$ and $A=\alpha_{\tau''}
(\mathsf{put\_field}_{\tau,\tau'}(\gamma_\tau(\phi_1^\#\sep\mu_1^\#))
(\gamma_{\tau'}(\phi_2^\#\sep\mu_2^\#)))$. We have
\begin{align*}
  A&=\alpha_{\tau''}(\mathsf{put\_field}_{\tau,\tau'}
    (\{\sigma_1\in\Sigma_\tau\mid\alpha_\tau(\sigma_1)\subseteq\phi_1^\#
    \sep\mu_1^\#\})\\
  &\qquad(\{\sigma_2\in\Sigma_{\tau'}\mid\alpha_{\tau'}(\sigma_2)\subseteq\phi_2^\#
    \sep\mu_2^\#\}))\\
  &=\alpha_{\tau''}\left(\left\{\begin{array}{c}
    \phi_2|_{-\rs}\sep\mu_2[l\mapsto\mu_2(l).\pi\sep\\
      \sep\mu_2(l).\phi[f\mapsto\phi_2(\rs)]]
    \end{array}
    \left|\begin{array}{l}
      \phi_1\sep\mu_1\in\Sigma_\tau,\\
      \phi_2\sep\mu_2\in\Sigma_{\tau'}\\
      \alpha_\tau(\phi_1\sep\mu_1)\subseteq\phi_1^\#\sep\mu_1^\#\\
      \alpha_{\tau'}(\phi_2\sep\mu_2)\subseteq\phi_2^\#\sep\mu_2^\#\\
      (l=\phi_1(\rs))\not=\nil,\\
      \mu_1=_l\mu_2
    \end{array}\right.\right\}\right)
\end{align*}
which is $\bot$ if $\phi_1^\#(\rs)=\emptyset$, since in such a
case the condition $\phi_1(\rs)\not=\nil$ cannot be satisfied.
Assume then that $\phi_1^\#(\rs)\not=\emptyset$. If no creation point
in $\phi_1^\#(\rs)$ occurs in $\phi_2^\#|_{-\rs}\sep\mu_2^\#$ then
$\mu_2(l)\not\in O_{\tau''}(\phi_2|_{-\rs}\sep\mu_2)$. Hence the update
of the content of $l$ does not contribute to $\alpha_{\tau''}$
(Definition~\ref{def:er_abstraction}) and we have
\[
  A=\alpha_{\tau''}\left(\left\{
    \phi_2|_{-\rs}\sep\mu_2
    \left|\begin{array}{l}
      \phi_1\sep\mu_1\in\Sigma_\tau,\ \phi_2\sep\mu_2\in\Sigma_{\tau'}\\
      \alpha_\tau(\phi_1\sep\mu_1)\subseteq\phi_1^\#\sep\mu_1^\#\\
      \alpha_{\tau'}(\phi_2\sep\mu_2)\subseteq\phi_2^\#\sep\mu_2^\#\\
      (l=\phi_1(\rs))\not=\nil,\ \mu_1=_l\mu_2
    \end{array}\right.\right\}\right)~.
\]
Let $\phi_2\sep\mu_2\in\Sigma_{\tau'}$ be such that
$\alpha_{\tau'}(\phi_2\sep\mu_2)\subseteq\phi_2^\#\sep\mu_2^\#$. By P2, we can
always find $\phi_1\sep\mu_1\in\Sigma_\tau$ such that
$\alpha_\tau(\phi_1\sep\mu_1)\subseteq\phi_1^\#\sep\mu_1^\#$.
By the hypothesis $\phi_1^\#(\rs)\not=\emptyset$
we can assume that $(l=\phi_1(\rs))\not=\nil$.
If $\mu_1=_l\mu_2$ does not hold,
we can assume that $l\not\in\dom(\mu_2)$ (up to renaming).
Let $o=\mu_1(l)$. We define $\mu_2'=\mu_2[l\mapsto o.\pi\sep\init(k(o.\pi))]$.
We have $\phi_2\sep\mu_2'\in\Sigma_{\tau'}$ and, since
the extra location $l$ does not contribute to $\alpha_{\tau'}$, we have
$\alpha_{\tau'}(\phi_2\sep\mu_2)=\alpha_{\tau'}(\phi_2\sep\mu_2')$
and $\alpha_{\tau''}(\phi_2|_{-\rs}\sep\mu_2)=
\alpha_{\tau''}(\phi_2|_{-\rs}\sep\mu_2')$.
Moreover, $\mu_1=_l\mu_2$ holds by construction. We conclude that
the constraints on $\phi_1\sep\mu_1$ and
the constraint $\mu_1=_l\mu_2$ do not contribute to $A$, and we have
\begin{align*}
  A&=\alpha_{\tau''}\left(\left\{
    \phi_2|_{-\rs}\sep\mu_2\left|\begin{array}{l}
      \phi_2\sep\mu_2\in\Sigma_{\tau'},\\
      \alpha_{\tau'}(\phi_2\sep\mu_2)\subseteq\phi_2^\#\sep\mu_2^\#
  \end{array}\right.\right\}\right)\\
  \text{(Corollary~\ref{cor:alphaxi})}&=\xi_{\tau''}(\phi_2^\#|_{-\rs}
    \sep\mu_2^\#)~.
\end{align*}

Otherwise, since
the objects reachable from $\phi_2(\rs)$ belong to the set
$O_{\tau'}(\phi_2\sep\mu_2)$, by Lemma~\ref{lem:er_codom} we have
\begin{align*}
  A&\subseteq\alpha_{\tau''}\left(\left\{\begin{array}{c}
    \phi_2|_{-\rs}\sep\mu_2[l\mapsto\mu_2(l).\pi\sep\\
      \sep\mu_2(l).\phi[f\mapsto\phi_2(\rs)]]
    \end{array}
    \left|\begin{array}{l}
      \phi_2\sep\mu_2\in\Sigma_{\tau'},\\
      \alpha_{\tau'}
        (\phi_2\sep\mu_2)\subseteq\phi_2^\#\sep\mu_2^\#\\
      l\in\dom(\mu_2),\\
      f\in\dom(F(k(\mu_2(l).\pi)))
    \end{array}\right.\right\}\right)\\
  &\subseteq\phi_2^\#|_{-\rs}\sep
    \mu_2^\#[f\mapsto\mu_2^\#(f)\cup\phi_2^\#(\rs)]~.
\end{align*}
By Corollary~\ref{cor:alphaxi} we conclude that
$A\subseteq\xi_{\tau''}(\phi_2^\#|_{-\rs}\sep
\mu_2^\#[f\mapsto\mu_2^\#(f)\cup\phi_2^\#(\rs)])$.

We prove the converse inclusion now.
Since we assume that there is a $\pi\in\phi_1^\#(\rs)$ which occurs in
$\phi_2^\#|_{-\rs}\sep\mu_2^\#$, then we can find
$\phi_1\sep\mu_1\in\Sigma_\tau$ with
$\alpha_\tau(\phi_1\sep\mu_1)\subseteq\phi_1^\#\sep\mu_1^\#$
and $\phi_2\sep\mu_2\in\Sigma_{\tau'}$ with
$\alpha_{\tau'}(\phi_2\sep\mu_2)\subseteq\phi_2^\#\sep\mu_2^\#$
such that $\phi_1(\rs)=l$, $\mu_1(l).\pi=\pi$ and
$\mu_1=_l\mu_2$. Note that $\phi_2(\rs)$ is only constrained
by $\alpha_{\tau'}(\phi_2\sep\mu_2)\subseteq\phi_2^\#\sep\mu_2^\#$
that is, $\phi_2(\rs)$ can range over all $\phi_2^\#(\rs)$.
Moreover, by the existence of $\pi$ we can assume that $l$ is
reachable in $\phi_2\sep\mu_2$ that is
$\mu_2(l)\in O_{\tau''}(\phi_2|_{-\rs}\sep\mu_2)$.
We conclude that
\begin{equation}\elabel{eq:putfield6}
  A.\mu(f)\supseteq\phi_2^\#(\rs).
\end{equation}
Moreover, given again
$\phi_1\sep\mu_1\in\Sigma_\tau$ with
$\alpha_\tau(\phi_1\sep\mu_1)\subseteq\phi_1^\#\sep\mu_1^\#$
and $\phi_2\sep\mu_2\in\Sigma_{\tau'}$ with
$\alpha_{\tau'}(\phi_2\sep\mu_2)\subseteq\phi_2^\#\sep\mu_2^\#$
and $(l=\phi_1(\rs))\not=\nil$, the condition
$\mu_1=_l\mu_2$ can be made true by renaming $l$ into $l'$ in
$\phi_2\sep\mu_2$ (if $l$ occurs there) and
extending $\mu_2$ with an unreachable $l$ bound to $\mu_1(l)$. We conclude that
we can always find $\phi_1\sep\mu_1$ and
$\phi_2\sep\mu_2$ such that
$\alpha_\tau(\phi_1\sep\mu_1)\subseteq\phi_1^\#\sep\mu_1^\#$,
$\alpha_{\tau'}(\phi_2\sep\mu_2)\subseteq\phi_2^\#\sep\mu_2^\#$,
$(l=\phi_1(\rs))\not=\nil$, $\mu_1=_l\mu_2$ and
$l$ is not reachable from $\phi_2\sep\mu_2$:
$\mu_2(l)\not\in O_{\tau''}(\phi_2|_{-\rs}\sep\mu_2)$.
As a consequence and by using P2, we have
\begin{align}
  A&\supseteq\alpha_{\tau''}\left(\left\{
    \phi_2|_{-\rs}\sep\mu_2\left|\begin{array}{l}
      \phi_2\sep\mu_2\in\Sigma_{\tau'},\\
      \alpha_{\tau'}(\phi_2\sep\mu_2)\subseteq\phi_2^\#\sep\mu_2^\#
  \end{array}\right.\right\}\right)\notag\\
  \text{(Corollary~\ref{cor:alphaxi})}&=\xi_{\tau''}(\phi_2^\#|_{-\rs}
    \sep\mu_2^\#)~.\elabel{eq:putfield7}
\end{align}
By merging~\eref{eq:putfield6} and~\eref{eq:putfield7} we conclude that
\begin{equation}\elabel{eq:putfield8}
  A\supseteq\xi_{\tau''}(\phi_2^\#|_{-\rs}
    \sep\mu_2^\#)\cup(\phi_\bot\sep\mu_\bot[f\mapsto\phi_2^\#(\rs)])
\end{equation}
where $\phi_\bot$ maps all variables to $\emptyset$ and
$\mu_\bot$ maps all fields to $\emptyset$.
We still have to prove that in the equation above we can move
$\phi_2^\#(\rs)$ inside the garbage collector $\xi_{\tau''}$.
But this is true since by Figure~\ref{fig:signatures} we know that
$f$ is a field of $F(\tau(\rs))$ so that $f$ is a field of
the objects created at
the creation point $\pi\in\phi_1^\#(\rs)$ which we assume to occur in
$\phi_2^\#|_{-\rs}\sep\mu_2^\#$. Hence $\xi_{\tau''}$ cannot
garbage collect the set $\phi_2^\#(\rs)$ bound to $f$. In conclusion,
\eref{eq:putfield8} becomes
\[
  A\supseteq\xi_{\tau''}(\phi_2^\#|_{-\rs}
    \sep\mu_2^\#[f\mapsto\mu_2^\#(f)\cup\phi_2^\#(\rs)]).
\]

\mbox{}\\
\proofoperation{\cup}
\mbox{}\\
By additivity (Proposition \ref{prop:er_insertion}),
the best approximation of $\cup$ over $\wp(\Sigma_\tau)$ is
(pointwise) $\cup$ over $\er$.
}

The proof of Proposition~\ref{prop:inclusion} needs the following result
that $\theta_\tau(e)$ is an element of $\er_\tau$ and
approximates exactly the same concrete states as $e$.

\begin{lemma}\label{lem:inclusion}
Let $\sigma\in\Sigma_\tau$ and $\ee\in\e_\tau$.
Then $\theta_\tau(\ee)\in\er_\tau$. Moreover,
$\alpha^\e_\tau(\sigma)\subseteq\ee$ if and only if
$\alpha^\er_\tau(\sigma)\subseteq\theta_\tau(\ee)$.
\end{lemma}
\myproofbis{
We have $\theta_\tau(\ee)\in\er_\tau$ by idempotency of $\xi_\tau$ (Proposition
\ref{prop:xi_lco}) and Definition~\ref{def:er_domain}.

Let $\alpha^\e_\tau(\sigma)\subseteq\ee$ and $v\in\domain(\tau)$. If
$\tau(v)=\integer$, then $\varepsilon_\tau(\sigma)(v)=*=\vartheta_\tau(\ee)(v)$.
If $\tau(v)\in\mathcal{K}$, then every $\pi\in\varepsilon_\tau(\sigma)(v)$
is such that $k(\pi)\le\tau(v)$ (Definitions~\ref{def:extractor}
and~\ref{def:proptotau}). Moreover,
$\pi=\mu(l).\pi$ for some $l\in\codom(\phi)\cap\Loc$
(Definition~\ref{def:extractor}). Hence $\pi\in\alpha_\tau^\e(\sigma)$
(Definition~\ref{def:alpha}), and $\pi\in\ee$. By
Definition~\ref{def:implementation} we conclude that
$\pi\in\vartheta_\tau(\ee)(v)$.
Hence $\alpha_\tau^\er(\sigma).\phi=\varepsilon_\tau(\sigma)\subseteq
\vartheta_\tau(\ee)$. Let now $f\in\domain(\widetilde{\tau})$. If
$\widetilde{\tau}(f)=\integer$, then $\varepsilon_{\widetilde{\tau}}
(\{\widetilde{o.\phi}\sep\mu\mid o\in O_\tau(\sigma)\})(f)=*
=\vartheta_{\widetilde{\tau}}(\ee)(f)$. If $\widetilde{\tau}(f)\in\mathcal{K}$,
then every $\pi\in\varepsilon_{\widetilde{\tau}}(\{\widetilde{o.\phi}\sep\mu
\mid o\in O_\tau(\sigma)\})(f)$ is such that
$k(\pi)\le\widetilde{\tau}(f)$ (Definitions~\ref{def:extractor}
and~\ref{def:proptotau}). Moreover, $\pi=\mu(l).\pi$ for some
$l\in\codom(o.\phi)\cap\Loc$ with $o\in O_\tau(\sigma)$
(Definition~\ref{def:extractor}). Hence $\pi\in\alpha_\tau^\e(\sigma)$
(Definition~\ref{def:alpha}), and $\pi\in\ee$. By
Definition~\ref{def:implementation} we conclude that
$\pi\in\vartheta_{\widetilde{\tau}}
(\ee)(f)$. Hence $\alpha_{\widetilde{\tau}}^\er(\sigma).\mu=
\varepsilon_{\widetilde{\tau}}(\{\widetilde{o.\phi}\sep\mu\mid
o\in O_\tau(\sigma)\})\subseteq\vartheta_{\widetilde{\tau}}(\ee)$.
In conclusion, we have $\alpha_\tau^\er(\sigma)\subseteq
\vartheta_\tau(\ee)\sep\vartheta_{\widetilde{\tau}}(\ee)$.
Since $\xi_\tau$ is monotonic (Proposition~\ref{prop:xi_lco}) and by
Proposition~\ref{prop:xi_fixpoints}, we have
$\alpha_\tau^\er(\sigma)\subseteq\xi_\tau(\vartheta_\tau(\ee)\sep
\vartheta_{\widetilde{\tau}}(\ee))=\theta_\tau(\ee)$.

Conversely, let $\alpha_\tau^\er(\sigma)\subseteq\theta_\tau(\ee)$.
Let $\pi\in\alpha_\tau^\e(\sigma)$. By Definition~\ref{def:alpha}
we have $\pi=o.\pi$ with $o\in O_\tau(\sigma)$.
By Definition~\ref{def:er_abstraction} we have
$\pi\in\alpha_\tau^\er(\sigma).\phi(v)$ for some $v\in\domain(\tau)$
or $\pi\in\alpha_\tau^\er(\sigma).\mu(f)$ for some
$f\in\domain(\widetilde{\tau})$,
and hence $\pi\in\theta_\tau(\ee).\phi(v)$, in the first case, or
$\pi\in\theta_\tau(\ee).\mu(f)$, in the second case.
In both cases, by Definition~\ref{def:implementation} we have $\pi\in\ee$.
Thus
$\alpha_\tau^\e(\sigma)\subseteq\ee$.}

We can now prove that every element of $\e$ represents the same
set of concrete states as an element of $\er$.

\mbox{}\\
\summary
{Proposition~\ref{prop:inclusion}.}
{Let $\gamma_\tau^\e$ and $\gamma_\tau^\er$ be the
concretisation maps induced by the abstraction maps of
Definitions~\ref{def:alpha} and~\ref{def:er_abstraction}, respectively.
Then $\gamma_\tau^\e(\e_\tau)\subseteq\gamma_\tau^\er(\er_\tau)$.}
\myproofbis{
By Lemma~\ref{lem:inclusion}, for any $\ee\in\e_\tau$, we have
\begin{align*}
  \gamma_\tau^\e(\ee)&=\{\sigma\in\Sigma_\tau\mid\alpha_\tau^\e(\sigma)
    \subseteq\ee\}\\
  &=\{\sigma\in\Sigma_\tau\mid\alpha_\tau^\er(\sigma)
    \subseteq\theta_\tau(\ee)\}=\gamma_\tau^\er(\theta_\tau(\ee)).
\end{align*}
Since this holds for all $\ee \in \e_\tau$,
we have the thesis.
}

} 
{} 
\end{article}
\end{document}